\documentclass[numberedappendix]{emulateapj}
\def\arcs{$''$}

\def\sfrd{\,{\rm M_\odot\,yr^{-1}\,Mpc^{-3}}}
\def\HI{\hbox{H~$\scriptstyle\rm I\ $}}

\begin{document}
\title{Galaxies at $z\sim6$: The $UV$ Luminosity Function
and Luminosity Density from 506 HUDF, HUDF-Ps, and GOODS $i$-dropouts}
\author{R.J. Bouwens$^{3}$, G.D. Illingworth$^{3}$, J.P. Blakeslee$^{4}$,
        M. Franx$^{5}$}
\affil{1 Based on observations made with the NASA/ESA Hubble Space
Telescope, which is operated by the Association of Universities for
Research in Astronomy, Inc., under NASA contract NAS 5-26555. These
observations are associated with programs \#9803.}
\affil{2 Observations have been carried out using the Very Large
Telescope at the European Southern Observatory (ESO) Paranal Observatory under program ID: LP168.A-0485.}
\affil{3 Astronomy Department, University of California, Santa Cruz,
CA 95064}
\affil{4 Department of Physics \& Astronomy, Washington State
University, Pullman, WA 99164-2814}
\affil{5 Leiden Observatory, Postbus 9513, 2300 RA Leiden,
Netherlands.}

\lefthead{BOUWENS ET AL.\ }
\righthead{$z\sim6$ UV LF}

\begin{abstract}
We have detected 506 $i$-dropouts ($z\sim6$ galaxies) in deep,
wide-area HST ACS fields: HUDF, enhanced GOODS, and HUDF-Parallel ACS
fields (HUDF-Ps).  The contamination levels are $\lesssim8$\% (i.e.,
$\gtrsim92$\% are at $z\sim6$).  With these samples, we present the
most comprehensive, quantitative analyses of $z\sim6$ galaxies yet and
provide optimal measures of the $UV$ luminosity function (LF) and
luminosity density at $z\sim6$, and their evolution to $z\sim3$.  We
redetermine the size and color evolution from $z\sim6$ to $z\sim3$.
Field-to-field variations (cosmic variance), completeness, flux, and
contamination corrections are modelled systematically and
quantitatively.  After corrections, we derive a rest-frame continuum
UV ($\sim1350\AA$) LF at $z\sim6$ that extends to $M_{1350,AB}$ $\sim
-17.5$ ($0.04L_{z=3}^{*}$).  There is strong evidence for evolution of
the LF between $z\sim6$ and $z\sim3$, most likely through a
brightening ($0.6\pm0.2$ mag) of $M^{*}$ (at 99.7\% confidence) though
the degree depends upon the faint--end slope.  As expected from
hierarchical models, the most luminous galaxies are deficient at
$z\sim6$.  Density evolution ($\phi^*$) is ruled out at $>$99.99\%
confidence.  Despite large changes in the LF, the luminosity density
at $z\sim6$ is similar ($0.82\pm0.21\times$) to that at $z\sim3$.
Changes in the mean UV color of galaxies from $z\sim6$ to $z\sim3$
suggest an evolution in dust content, indicating the true evolution is
substantially larger: at $z\sim6$ the star formation rate density is
just $\sim30$\% of the $z\sim3$ value.  Our UV luminosity function is
consistent with $z\sim6$ galaxies providing the necessary UV flux to
reionize the universe.

\end{abstract}
\section{Introduction}

The deep $z_{850}$-band capabilities of the Hubble Space Telescope
(HST) Advanced Camera for Surveys (ACS) greatly enhanced the ability
of astronomers to identify and observe galaxies at $z\sim6$.  Flux in
the $z_{850}$-band can be contrasted with flux in the $i_{775}$-band,
allowing for identification of $z\sim6$ $i_{775}$-dropouts.  Early
studies revealed that $i$-dropouts were both smaller (Bouwens et al.\
2003b; Stanway et al.\ 2004b; Bouwens et al.\ 2006b, hereafter, B06b)
and less numerous than dropouts at lower redshifts (Stanway et al.\
2003; Bouwens et al.\ 2003b; Dickinson et al.\ 2004; Stanway et al.\
2004b; B06b).  However, since much of the early work was at bright
magnitudes ($z_{850,AB}\lesssim27$), it was still quite unclear from
these studies how this population extended to fainter magnitudes or
lower surface brightnesses.

With the availability of significantly deeper $i$ and $z$ data from
the Hubble Ultra Deep Field (HUDF; Beckwith et al.\ 2006) and
the HUDF-Ps (Bouwens et al.\ 2004b), this situation is largely
changed.  There are already a number of papers that take advantage of
this depth to comment on the faint-end slope (Bouwens et al.\ 2004a;
Bunker et al.\ 2004, hereafter, BSEM04; Yan \& Windhorst 2004b), the
rest-frame UV colors (Stanway et al.\ 2005; Yan et al.\ 2005), and the
surface brightness distribution at $z\sim6$ (BSEM04; Bouwens et al.\
2004b).  These data have also provided us with some new insight into
the long standing question of how the universe was reionized.  Some
authors (e.g., Yan \& Windhorst 2004b; Lehnert \& Bremer 2003) have
argued that it is largely the faint galaxies that were instrumental in
this process, while others have emphasized the role that possible
evolution in metallicity or the initial mass function (IMF) may have
on the process (Stiavelli et al.\ 2004b).  Finally, other groups
(e.g., BSEM04) have even questioned whether the observed galaxy
population is sufficient to reionize the universe at all.

While providing many interesting initial results, there were a number
of limitations to these early analyses.  Some (e.g., BSEM04)
restricted themselves to a bright limit (in their analyses of the two
most notable data sets) to minimize the importance of incompleteness,
flux, or contamination corrections.  Other analyses did not calculate
the selection volume for their survey self-consistentally from the
observed UV colors (but rather assumed a simple $z=5.5-6.5$ top-hat
selection window: e.g., Yan \& Windhorst 2004b).  Moreover, none of
these early studies made a detailed account of the uncertainties in
their LF determinations or made an attempt to correct for
field-to-field variations, which can be substantial ($\sim35$\% rms)
for single 11.3 arcmin$^2$ ACS Wide-Field Camera (WFC) fields
(Somerville et al.\ 2004; Bouwens et al.\ 2004a; BSEM04).  Correcting
for these variations is important for ensuring that a consistent
normalization is used at bright, intermediate, and faint magnitudes
and thus the derived luminosity function (LF) is not compromised.
Particularly important in this regard are the implications for the
faint-end slope and the number of lower luminosity galaxies.  Such
objects have the potential to provide the necessary UV flux to
reionize the universe (Lehnert \& Bremer 2003; Yan \& Windhorst
2004b).

The purpose of this paper is to redress many of these limitations and
provide a systematic analysis of $i$-dropouts from some of the
deepest, widest area surveys available for study.  We consider fields
at three different depths.  The two wide-area GOODS fields ($\sim$160
arcmin$^2$ each), here enhanced to include the extensive supernova
search data, form the backbone of our probe, providing important
statistics at the bright end while controlling for field-to-field
variations.  At the faint end, there is the Hubble Ultra Deep Field
(HUDF; 11 arcmin$^2$), which in addition to constraining the faint-end
slope allows us to quantify the incompleteness, flux biases, and
contamination in our shallower probes.  Finally, at intermediate
magnitudes, we have the two HUDF-Ps (17 arcmin$^2$ in total), which
provide an important bridge between our faintest and brightest fields.
Together these three data sets provide a good measure of the
$i$-dropout surface density over a 5 mag baseline, from
$z_{850,AB}\sim24.5$ to 29.5.

This paper is structured as follows.  We begin with a description of
the data (\S2), describe our selection criteria (\S3), and then
compile an $i$-dropout sample in the HUDF.  We use the color
information to make inferences about the contamination rate, intrinsic
colors, and overall redshift distribution.  We then proceed to an
analysis of our shallower fields and incorporate the data from those
fields into our $i$-dropout probe, deriving the $i$-dropout surface
density from $z_{850,AB}\sim24.5$ to 29.5.  In \S4, we compare the
present probe with previous catalogs and surface density
determinations.  In \S5, we use this surface density to derive a LF in
the rest-frame UV ($\sim1350\AA$) and compare it with the LF derived
at $z\sim3$ (Steidel et al.\ 1999).  Finally, we discuss these
results, comment on the likely physical implications (\S6), and
conclude (\S7).  We make use of appendices to develop some key
technical issues, while not interrupting the flow of the paper.  Where
necessary, we use the ``concordance cosmology''
$(\Omega_{M},\Omega_{\Lambda},h)=(0.3,0.7,0.7)$.  We note that the
results are not very dependent on the details of the cosmology and
that $M^*$ and $\phi^*$ change by $\lesssim12$\% (\S5) when expressed
in terms of the one year WMAP measurements
$(\Omega_{M},\Omega_{\Lambda},h)=(0.24,0.76,0.73$; Spergel et al.\
2003).

\section{Observations}

As noted above, the present analysis leverages data sets of three
different depths to obtain a fairly optimal measure of the number
densities of $i$-dropouts over a 5 mag baseline.  Table 1 provides a
summary of these data sets.

\begin{deluxetable*}{cccc}
\tablewidth{5.5in}
\tablecaption{Observational Data.}
\tablehead{
\colhead{} & \colhead{Detection Limits\tablenotemark{a}} & \colhead{PSF FWHM} & \colhead{Areal Coverage}\\
\colhead{Passband} & \colhead{(10$\sigma$)} & \colhead{(arcsec)} & \colhead{(arcmin$^2$)}}
\startdata
\multicolumn{4}{c}{HUDF} \\
$B_{435}$ & 29.6 & 0.09 & 11.2 \\
$V_{606}$ & 30.0 & 0.09 & 11.2 \\
$i_{775}$ & 29.9 & 0.09 & 11.2 \\
$z_{850}$ & 29.2 & 0.10 & 11.2 \\
$J_{110}$ & 26.9 & 0.33 & 5.8 \\
$H_{160}$ & 26.7 & 0.37 & 5.8 \\
\multicolumn{4}{c}{} \\
\multicolumn{4}{c}{HUDF-Ps} \\
$B_{435}$ & 28.9 & 0.09 & 17.0\tablenotemark{b} \\
$V_{606}$ & 29.2 & 0.09 & 17.0\tablenotemark{b} \\
$i_{775}$ & 28.8 & 0.09 & 17.0\tablenotemark{b} \\
$z_{850}$ & 28.5 & 0.10 & 17.0\tablenotemark{b} \\
\multicolumn{4}{c}{} \\
\multicolumn{4}{c}{GOODS fields} \\
$B_{435}$ & 28.2 & 0.09 & 316 \\
$V_{606}$ & 28.4 & 0.09 & 316 \\
$i_{775}$ & 27.7 & 0.09 & 316 \\
$z_{850}$ & 27.5 & 0.10 & 316 \\
$J$ & $\sim25$ & $\sim$0.45$''$ & 131\\
$K_s$ & $\sim24.5$ & $\sim$0.45$''$ & 131\\
\enddata
\tablenotetext{a}{$0.2''$-diameter aperture for the ACS data,
$0.6''$-diameter aperture for NICMOS data, and $0.8''$-diameter for
ISAAC data.}
\tablenotetext{b}{A significant fraction of the area from the HUDF-Ps
was not used because it did not meet our minimal S/N requirements
(\S2.2).  The area used is tabulated here.}
\end{deluxetable*}

\subsection{ACS HUDF}

The $B_{435}V_{606}i_{775}z_{850}$ images used for this analysis are
the v1.0 reductions of the HUDF (Beckwith et al.\ 2006), binned
on a 0.03\arcs$\,$pixel scale.  While the observations cover $\sim$12
arcmin$^2$, our search area was restricted to the deepest 11.2
arcmin$^2$.  The zeropoints used for these images are the latest
values from the continuing ACS calibrations (Sirianni et al.\ 2005).
Photometry performed using these zeropoints was offset slightly to
account for the estimated Galactic absorption $E(B-V)=0.007$
(Schlegel, Finkbeiner, \& Davis 1998).  The $10\sigma$ limits for
these images were 29.6, 30.0, 29.9, and 29.2, respectively, in
0.2\arcs-diameter aperture.  Point-spread functions (PSFs) were
0.09-0.10\arcs$\,$FWHM.

Extremely deep Near-Infrared Camera and Multi-Object Spectrometry
(NICMOS) coverage is available over a portion of the HUDF (5.76
arcmin$^2$; Thompson et al.\ 2005).  That program included eight
orbits in the NIC3 $J_{110}$ filter and eight orbits in the NIC3
$H_{160}$ filter over nine separate pointings, for a total of 144
orbits.  The pointings were arranged in a $3\times3$ grid, each
separated by 45\arcs.  Although there is some variation in depth
across the mosaic, typical $5\sigma$ limits for the images were 27.6
and 27.4 in the $J_{110}$ and $H_{160}$ passbands (0.6\arcs-diameter
aperture), respectively.  Our reduction of the NICMOS data was a
slight improvement on that initially made available with the treasury
release and was made possible by more exact position matching with the
HUDF $z_{850}$-band image.  This reduction is described in more detail
in Thompson et al.\ (2005).  The resulting NIC3 PSFs had FWHMs of
0.33\arcs$\,$and 0.37\arcs$\,$ in the $J_{110}$ and $H_{160}$ bands,
respectively.  The $F110W$ and $F160W$ zero points used are those
recently determined by STScI (de Jong et al.\ 2006; see also Coe et
al.\ 2006).  These zeropoints are offset by $-0.16$ and $-0.04$ (de
Jong et al.\ 2006) from those previously made available by the Space
Telescope Science Institute (STScI; 2004 June).

\subsection{HUDF ACS Parallels}

The two HUDF-Ps were taken in parallel to the HUDF NICMOS observations
(GO-9803: Thompson et al.\ 2005).  Each field consists of 72 orbits of
ACS observations (9 orbits of $B_{435}$, 9 orbits of $V_{606}$, 18
orbits of $i_{775}$, 27 orbits of $z_{850}$, and 9 orbits of G800L)
that reaches nearly $\sim1$ mag deeper than the original 5-epoch ACS
GOODS observations.  They also reach fainter ($\sim0.2-0.4$ mag) than
the WFPC2 HDF-N (Williams et al.\ 1996) and HDF-S (Williams et al.\
2000).  Processing of the data included alignment, background
subtraction, cosmic ray rejection, and drizzling onto a 0.03\arcs$\,$
grid, and was performed by the ``Apsis'' pipeline (Blakeslee et al.\
2003).  Artifacts in the original exposures such as satellite trails
or the ``figure eight'' patterns (resulting from scattered light off
the internal dewars) were explicitly masked out before drizzling the
images together.  The reductions of these fields used in this paper
are different from those described in several previous publications
(Blakeslee et al.\ 2004; Bouwens et al.\ 2004a).  Our principal reason
for this was to bin the data on a very similar 0.03\arcs-pixel scale
to that available for the ACS GOODS fields (\S2.3) and the HUDF
(Beckwith et al.\ 2006).  The similar pixel scale made it
straightforward to degrade the deeper data to the quality of the
shallower data and therefore estimate quantities like the
completeness, flux biases, and contamination rate (see Appendix C).

To maximize depth, we combined the ACS parallel data (Thompson et al.\
2005) with overlapping ACS WFC exposures from the CDF-S GOODS
(Giavalisco et al.\ 2004a), GEMS (Rix et al.\ 2004), and SNe search
programs (A. Riess et al.\ 2006, in preparation).  Incorporating the
latter data resulted in modest increases in the mean depth of our
images ($+0.2$ mag).  Only regions having exposure times in excess of
5 orbits, 11 orbits, and 18 orbits in the $V_{606}$, $i_{775}$, and
$z_{850}$ bands, respectively, are considered in our selection (or
equivalently their $10\sigma$ depths were required to exceed 28.9,
28.6, and 28.1 in the $V_{606}$, $i_{775}$, and $z_{850}$ bands,
respectively, in a 0.2\arcs-diameter apertures).  This corresponded to
10.0 arcmin$^2$ in the first HUDF-Parallel [hereafter, referred to as
  HUDFP1] and 7.0 arcmin$^2$ in the second [hereafter, referred to as
  HUDFP2].  The $10\sigma$ depths for the deepest portion of these
parallels were 28.9, 29.2, 28.8, and 28.5 in the $B_{435}$, $V_{606}$,
$i_{775}$, and $z_{850}$ bands, respectively, in 0.2\arcs-diameter
apertures ($\sim$0.7-1.1 mags less deep than the HUDF).

\subsection{ACS GOODS}

The current analysis makes use of our own reductions of the ACS data
available over the two GOODS fields ($\sim160$ arcmin$^2$).  Though a
public reduction of the data over this area was available (i.e., the
GOODS version 1.0 reduction: Giavalisco et al.\ 2004a), it did not
include the significant amounts of ACS data taken over these fields
after the initial 398-orbit GOODS campaign.  These include 195 orbits
of $V_{606}i_{775}z_{850}$ data taken for additional SNe searches
(A. Riess et al.\ 2006, in preparation; S. Perlmutter et al.\ 2006, in
preparation), $\gtrsim100$ orbits of $z_{850}$-band data for SNe
follow-up (A. Riess et al.\ 2006, in preparation; S. Perlmutter et
al.\ 2005, in preparation), $\sim40$ orbits of overlapping $V_{606}$
and $z_{850}$ data from the GEMS program (Rix et al.\ 2004), and 128
orbits of $B_{435}V_{606}i_{775}z_{850}$ data over the ACS parallels
to the HUDF NICMOS field (Thompson et al.\ 2005).  These data
substantially enhance the GOODS version 1.0 data set, and should
largely be included in the GOODS version 2.0 release.  Instead of
waiting for the release, we carried out our own reduction.  Similar to
our handling of the HUDF-Parallel ACS fields, we processed the ACS
data with our ``Apsis'' pipeline (Blakeslee et al.\ 2003).  They were
drizzled onto the same astrometric grid as the images (35 individual
8k x 8k frames) which made up the v1.0 reductions of the two GOODS
fields (Giavalisco et al.\ 2004a).  These images--and our own
reductions--were done on a 0.03\arcs $\,$pixel scale very similar to
the HUDF.  The approximate $10\sigma$ depths of those data were 28.2,
28.4, 27.7, and 27.5 in the $B_{435}$, $V_{606}$, $i_{775}$, and
$z_{850}$ bands, respectively.  These data reach nearly $\sim0.15$ mag
and $\sim0.4$ mag deeper in the $i_{775}$ and $z_{850}$ bands,
respectively, than the GOODS v1.0 reductions.

One complication with the analysis of the two GOODS fields is the
notable variation in the depth.  The extensive overlap regions between
adjacent exposures in the ACS tiling ($\sim7$ arcmin$^2$ for each
field) are appreciably deeper ($\sim0.4$ mag), the many outer regions
($\sim30$ arcmin$^2$ for each field) only covered by three epochs of
data (5 epochs including the SNe search data) are shallower ($\sim0.3$
mag), and other regions of these fields with missing exposures (e.g.,
due to guide star acquisition problems) also are shallower (Giavalisco
et al.\ 2004a).  As we demonstrated in an earlier study on the
$i$-dropouts in the RDCS1252-2927 field (Bouwens et al.\ 2003b), such
variations can have a dramatic impact on the number of $i$-dropouts
selected (changing the numbers by factors of $\sim1.8$ for just
$\sim$0.4 mag alterations in depth), and therefore any selection of
$i$-dropouts off the undegraded GOODS images requires an accurate
accounting for these variations.  This could be done, for example, by
laying down objects at random positions across the GOODS mosaic and
then attempting to recover them.  Instead of adopting this more
involved approach, we took a simpler route, degrading the entire frame
to a uniform S/N and ignoring regions below this S/N in the object
selection.  Our procedure for executing the degradation is detailed in
Appendix B.  The threshold we settled on was $0.1$ mag brighter than
that obtained with a 2.5, 3.5, and 9-orbit exposure in the $V_{606}$,
$i_{775}$, and $z_{850}$ bands, respectively, and was chosen as a
compromise between depth and area.  This threshold is equivalent to
10$\sigma$ depths of 28.3, 27.5, and 27.4 in the $V_{606}$, $i_{775}$,
and $z_{850}$ bands, respectively.  Throughout this work, this is what
we mean when we refer to the S/N levels of the GOODS fields (or to
deeper fields degraded to GOODS depth).

We also made use of the Infrared Spectrometer and Array Camera (ISAAC)
$JK_s$ data for the CDF-S GOODS field (B. Vandame et al.\ 2006, in
preparation) to better estimate the contamination from lower redshift
interlopers.  The data consist of 21 separate $\sim$3-4 hr
$2.5'\times2.5'$ ISAAC exposures in the $J$ ($\sim1.25\mu m$) and
$K_s$ ($\sim2.16\mu m$) bands that reach $\sim$25.7 and $\sim$25 AB
magnitudes ($5\sigma$), respectively, in a 0.8\arcs-diameter aperture.
The entire mosaic covers 131 arcmin$^2$ or about $\sim85$\% of the ACS
GOODS area.  B. Vandame et al.\ (2006, in preparation) estimated the
seeing for the frames to range from 0.31\arcs$\,$to 0.66\arcs, with a
median value of 0.46\arcs.  Zero points for the individual ISAAC
frames were derived by matching photometry of $\sim$50 stars on each
frame with the shallower SOFI (Arnouts et al.\ 2001) and Two Micron
All Sky Survey (2MASS; Skrutskie et al.\ 1997) images.

As a check on the zero points, we performed photometry on the $\sim20$
$z\sim0.4-1.0$ E/S0s in each ISAAC frame (ACS $BViz$ + ISAAC $JK_s$
bands) and then fit a spectral energy distribution (SED) to the six
optical-infrared fluxes.  While our $K_s$-band fluxes are in excellent
agreement with the fit results, we noticed that our $J$-band fluxes
were generally $\sim0.1$ mag fainter than expected.  Since B. Vandame
et al.\ (2006, in preparation) noted a similar $\sim0.1$ mag faintward
offset relative to the photometry of the K20 survey (Cimatti et al.\
2002), we took this offset to be real and offset the $J$-band
zeropoints quoted by B. Vandame et al.\ (2006, in preparation) by 0.1
mag.  No such shifts were applied to the $K_s$-band fluxes.  Similarly,
the seeing estimates obtained for different ISAAC images (B. Vandame
et al.\ 2006, in preparation) were examined and compared with our own
estimates.  In general, the FWHMs we obtained were $\sim0.02$\arcs$\,$
to $\sim0.05$\arcs$\,$ larger than the B. Vandame et al.\ (2006, in
preparation) estimates.  We elected to apply our estimates throughout
in determining the optical-infrared colors of objects in the CDF-South
GOODS field (i.e., Appendix D4.1).  Given that our only use of $J$ and
$K_s$ photometry in this study is for quantifying contamination, these
adjustments should have no large effect on the other quantities
derived here.

\section{Analysis}

Our procedure for doing object detection and photometry is identical
to that detailed in a number of previous publications by our group
(e.g., Bouwens et al. 2003a; Bouwens et al.\ 2006a, hereinafter,
B06a).  SExtractor (Bertin \& Arnouts 1996) was run in double-image
mode, with the $z_{850}$-band images used for object detection and the
other images used as the measurement images.  The infrared
coverage--although superior in probing beyond the break--was not used
in the detection procedure because (1) the signal-to-noise ratio (S/N)
and resolution of these images were in general much poorer than the
$z_{850}$-band images and (2) these images--where available--tended to
be very inhomogeneous in nature.  Photometry was done using two scaled
Kron (1980) apertures, the smaller ones to measure colors and the
larger ones to convert these colors to total magnitudes.  Small
corrections were applied to the total magnitudes ($0.1$ mag to the
$B_{435}$, $V_{606}$, and $i_{775}$ bands and $0.125$ mag in the
$z_{850}$ band: Sirianni et al.\ 2005) to account for the flux that
falls outside these apertures (typically $\sim0.8''$ in diameter).
Optical-infrared colors were obtained by degrading the optical images
to the same PSF as the coincident infrared image and then measuring
the flux in an aperture that maximized the S/N (typically
0.8\arcs-1.4\arcs-diameter apertures).

One minor issue in the construction of our $i$-dropout catalogs was
the choice of the SExtractor deblending parameter.  A small value for
this parameter minimizes blending with foreground sources, but also
causes many of the more clumpy $i$-dropouts to split into multiple
pieces.  Conversely, a large value for this parameter largely avoids
such splitting, but results in more blending with foreground sources.
After extensive testing, we opted to use a larger value for the
deblending parameter (i.e., DEBLEND\_MINCONT = 0.15) than the defaults
(i.e., DEBLEND\_MINCONT = 0.005).  Although this results in a greater
degree of blending (e.g., 17\% of $i$-dropouts are blended with
foreground objects in the HUDF vs. 11\% using much smaller deblending
parameters: Appendix D1), it should avoid splitting
physically-associated systems into multiple pieces--which would result
in small systematic errors.  Corrections can be made for these
additional incompleteness levels (see Appendix D1).\footnote{Even
better results could have been obtained here, if there was some source
detection and photometry software available that had been designed to
take advantage of color information in source deblending.  Since
dropouts have highly unique colors, it would be fairly straightforward
to distinguish clumps that make up one of these objects from other
foreground objects.  SExtractor currently only uses the detection
image for this process and does not consider color information.}  To
ensure that the object blending was reasonable, a detailed visual
inspection was performed on each of the objects in our samples (\S3.2;
\S3.4).  No objects were found that included any obvious contribution
from foreground sources.  The highly unique colors of dropout sources
made this check a fairly unambiguous process.

\subsection{$i$-dropout Selection}

As in several previous publications on this subject (Stanway et al.\
2003; Bouwens et al.\ 2004a; Dickinson et al.\ 2004; B06a),
$i$-dropouts are selected using a simple $(i_{775}-z_{850})_{AB}$ cut.
At intermediate magnitudes ($24<z_{850,AB}<27$), such cuts have
already been shown to be quite efficient at isolating objects with
blue $z-J$ colors indicative of $z\sim6$ starbursts (Stanway et al.\
2003; Bouwens et al.\ 2003b; Dickinson et al.\ 2004; Stanway et al.\
2005; B06a).  Our choice of a more inclusive
$(i_{775}-z_{850})_{AB}>1.3$ criterion rather than the $>1.4$ and
$>1.5$ criteria used in previous work (Bouwens et al.\ 2003; Bouwens
et al.\ 2004a; B06a) was motivated by our desire to maximize the size
of our sample.  While this also results in a somewhat higher
contamination rate (Appendix D4), an increasing amount of data is now
available, both in the IR (Thompson et al.\ 2005; Vandame et al.\
2006, in preparation) and with the ACS GRISM (Pirzkal et al.\ 2004;
Malhotra et al.\ 2005) to better constrain the contamination.  Note
that in computing the $i-z$ color for selection, we set the $i$-band
flux to its $2\sigma$ upper limit in the case of a non detection.  In
addition to our $i-z>1.3$ criteria, we also required that objects have
$(V_{606}-z_{850})_{AB}$ colors redder than 2.8 or be non-detections
($<2\sigma$) in the $V_{606}$-band to exclude lower-redshift
interlopers.  Appendix A provides a justification for the
$(V_{606}-z_{850})_{AB}$ color cut by comparing it with a number of
intrinsically red galaxies uncovered in the CDF-South
(Table~2).  To guard against spurious sources that come
in the form of low-surface brightness variations in the background
(Appendix D4.4), we required that objects in the HUDF be at least
$3.5\sigma$ detections in a 0.3$\arcsec$-diameter aperture.  The
detection requirement was increased to 4 and 4.5 $\sigma$ for the
HUDF-Ps and GOODS fields, respectively, to cope with the likely larger
non-Gaussian signatures present in the smaller exposure stacks that
comprise these data.  Point sources brighter than some fiducial
$z_{850}$-band magnitude (26.8 for GOODS fields, 27.5 for the HUDF-Ps,
and 28.4 for the HUDF) were removed at this stage (point sources were
defined to have SExtractor stellarity parameters $>0.75$).  Faintward
of these fiducial limits, point sources could no longer be reliably
identified (their contribution was treated as a contamination fraction
and estimated statistically: see Appendix D4.3).
Table~3 contains a list of all objects excluded as
stars.  Finally, we carefully inspected all of our candidate
$i$-dropouts to ensure that they did not arise from diffraction spikes
around stars or the extended low-surface brightness wings around
ellipticals.

\begin{deluxetable*}{cccccccc}
\tablewidth{5.5in}
\tablecolumns{8}
\tabletypesize{\footnotesize}
\tablecaption{A sample of intrinsically red $(i_{775}-z_{850})_{AB}>1,
(z_{850}-J)_{AB}>0.8$, likely low-redshift objects identified in the
CDF-S GOODS field that may serve as interlopers for our $i$-dropout
samples (Figure~A1, Figure~D3).\tablenotemark{a}}
\tablehead{
\colhead{} & \colhead{} & \colhead{} & \colhead{} &
\colhead{} & \colhead{} & \colhead{} & \colhead{$r_{hl}$}\\
\colhead{R.A.} & \colhead{Decl.} & \colhead{$z$} & \colhead{$i-z$} &
\colhead{$V-z$} & \colhead{$z-J$} & \colhead{$z-K_s$} & \colhead{(arcsec)}
}
\startdata
03:32:35.63 & -27:43:10.1 & 21.97$\pm$0.01 & 1.1 & 2.7 & 1.0 & 1.7 & 0.30\\
03:32:39.41 & -27:54:11.8 & 22.17$\pm$0.01 & 1.3 & 2.8 & 1.1 & 2.2 & 0.58\\
03:32:25.07 & -27:52:49.3 & 22.28$\pm$0.01 & 1.1 & 2.7 & 0.9 & 1.9 & 0.33\\
03:32:17.15 & -27:52:32.0 & 22.46$\pm$0.01 & 1.0 & 2.6 & 0.9 & 1.8 & 0.28\\
03:32:25.76 & -27:43:47.3 & 22.49$\pm$0.02 & 1.1 & 2.4 & 1.1 & 2.2 & 0.60\\
03:32:43.93 & -27:42:32.4 & 22.72$\pm$0.01 & 1.2 & 3.2 & 1.6 & 2.4 & 0.35\\
\enddata 

\tablenotetext{a}{Table~2 is published in its entirety
in the electronic version of the Astrophysical Journal.  A portion is
shown here for guidance regarding its form and content.  Objects in
this table are most likely low-redshift interlopers and were helpful
in optimizing our selection cuts to avoid contamination.  Half-light
radii are estimated from the growth curve.  Limits on colors are
$2\sigma$.  The $z_{850}-J$ and $z_{850}-K_s$ colors were measured
with respect to the $J+K_s$ ISAAC imaging (\S2.3: B. Vandame et al.\
2006, in preparation).  Right ascension and declination use the
J2000.0 equinox; units of right ascension are hours, minutes, and
seconds, and units of declination are degrees, arcminutes, and
arcseconds.}
\end{deluxetable*}

\begin{deluxetable*}{cccccccccc}
\tablewidth{5.5in}
\tablecolumns{10}
\tabletypesize{\footnotesize}
\tablecaption{$i-z>1.3$ pointlike (stellar) sources not included in our $i$-dropout compilation.\tablenotemark{a}}
\tablehead{
\colhead{} & \colhead{} & \colhead{} & \colhead{} & \colhead{} & \colhead{} & 
\colhead{} & \colhead{} & \colhead{$r_{hl}$}\\
\colhead{Object ID} & \colhead{R.A.} & \colhead{Decl.} & \colhead{$z_{850}$} & \colhead{$i - z$} & \colhead{$z - J$} & 
\colhead{$z - K_s$} & \colhead{S/G} & \colhead{(arcsec)}}
\startdata
HDFN-6581218516 & 12:36:58.12 &  62:18:51.6 & 23.02$\pm$0.01 & 1.5 & --- & --- & 0.93 & 0.09 \\
CDFS-2295952287 & 03:32:29.59 & -27:52:28.7 & 23.05$\pm$0.01 & 1.4 & 1.0 & 0.9 & 0.98 & 0.08 \\
HDFN-7340515534 & 12:37:34.05 &  62:15:53.4 & 23.17$\pm$0.01 & 1.5 & --- & --- & 0.94 & 0.09 \\
CDFS-2192345455 & 03:32:19.23 & -27:45:45.5 & 23.30$\pm$0.01 & 1.3 & 0.8 & 1.0 & 0.96 & 0.08 \\
CDFS-2181947466 & 03:32:18.19 & -27:47:46.6 & 23.60$\pm$0.01 & 1.4 & 1.1 & 1.3 & 0.99 & 0.08 \\
HDFN-6388514511 & 12:36:38.85 &  62:14:51.1 & 23.89$\pm$0.01 & 1.7 & --- & --- & 0.92 & 0.09 \\
\enddata
\tablenotetext{a}{Table~3 is published in its entirety in the
electronic version of the Astrophysical Journal.  A portion is shown
here for guidance regarding its form and content.  Similar comments to
Table~6 apply.  A ``b'' superscript denotes the same object.  All
limits are 2$\sigma$.  ``S/G'' denotes the SExtractor stellarity
parameter, where 0 is for an extended object and 1 is for a point
source.  Right ascension and declination use the J2000.0 equinox;
units of right ascension are hours, minutes, and seconds; units of
declination are degrees, arcminutes, and arcseconds.}
\end{deluxetable*}

\subsection{$i$-dropouts in the HUDF}

Applying the above selection criteria to the HUDF results in a sample
of 122 $i$-dropouts.  Objects range in magnitude from
$z_{850,AB}=25.0$ to 29.4 (the $8\sigma$ limit).  At $z\sim6$, this
corresponds to $0.04-2.2$ times the characteristic rest-frame UV
luminosity at $z\sim3$ (Steidel et al.\ 1999).  Table~4
summarizes the positions, magnitudes, $i-z$ colors, sizes,
stellarities, $z-J_{110}$ colors, and $J_{110}-H_{160}$ colors of
different objects in our HUDF $i$-dropout sample.
$V_{606}i_{775}z_{850}$ color cutouts are provided in Figure 1 for the
brightest 28 $i$-dropouts from the HUDF.

The deeper optical and infrared imaging available in the central
region of the HUDF allow us to extend our knowledge of the
contamination rate from low-redshift interlopers (e.g., dusty/evolved
$z\sim1-3$ objects) to fainter magnitudes ($z_{850,AB}\gtrsim27$) than
has been previously possible.  While there have already been several
studies using these data to argue that this contamination is small
(Yan \& Windhorst 2004b; Stanway et al.\ 2005), the present selection
pushes slightly deeper.  As in our analysis of the $i$-dropouts in
RDCS1252-2927 and the CDF-S GOODS field (Bouwens et al.\ 2003b; B06a),
we consider the canonical $(i_{775}-z_{850})$ versus $(z_{850} -
J_{110})$ color-color plot (Figure~2).  It is
immediately apparent that the contamination rate from low-redshift
interlopers is low.  Only two of the 43 $i$-dropouts observed to
$z_{850,AB}\sim28.7$ had $z-J_{110}$ colors inconsistent with the
expected position of $z\sim6$ starbursts in color-color space
(\textit{shaded orange region}), suggesting a very low ($\sim$5\%)
contamination rate for the sample as a whole.  Splitting the sample
across several magnitude bins, we can obtain a magnitude-dependent
contamination fraction (Table~D7).

\begin{figure}
\includegraphics[width=3.2in]{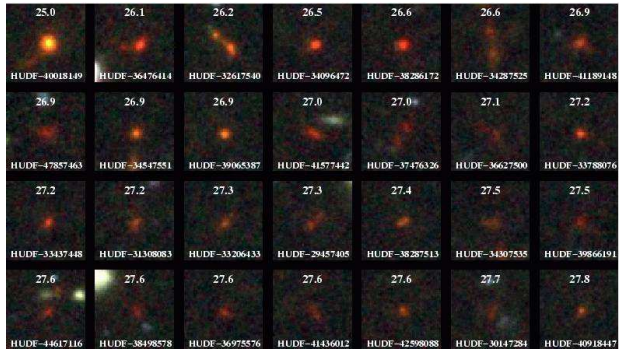}
\caption{Postage stamps ($V_{606} i_{775} z_{850}$ color images) of
the brightest 24 $i_{775}$-dropouts from the HUDF.  Objects are ordered
in terms of their $z_{850}$-band magnitude.  The $z_{850}$-band
magnitudes and object IDs are shown above and below each object,
respectively.  Each postage stamp is 3.0\arcs$\,$in size.  These high
S/N images show definitive evidence for assymetries, mergers, and
other interactions--similar to that seen at lower redshifts
($z\sim2-5$).}
\end{figure}

\begin{figure}
\includegraphics[width=3.4in]{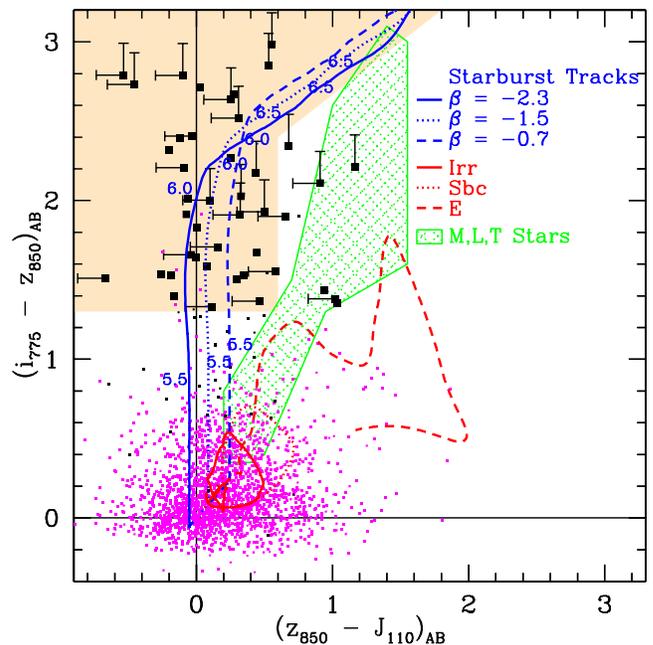}
\caption{The $i_{775}-z_{850}$/$z_{850}-J_{110}$ color-color diagram
showing the photometry of objects (\textit{squares}) in the HUDF NICMOS
footprint.  Objects undetected ($<2\sigma$) in the $V_{606}$ band or
whose $(V_{606}-z_{850})_{AB}$ colors are redder than 2.8 (see
Appendix A) are shown as filled black squares.  Other objects in the
photometric sample are displayed as magenta dots.  The tracks made by
starbursts of various UV spectral slopes $\beta$ are plotted here as a
function of redshift to indicate the position of likely high-redshift
$i$-dropouts (\textit{blue lines}).  For contrast, similar tracks have
been included for a number of low-redshift templates to show the
position of possible contaminants (\textit{red lines}) along with the
colors for early stellar types M0-T7 (Knapp et al.\ 2004; Geballe et
al.\ 2002; Leggett al.\ 2002) (\textit{green hatched region}).
Although a $(i_{775}-z_{850})_{AB}>1.3$ criterion is used to select
the $i$-dropout sample, almost all $i-z>1.3$ objects have very blue
$(z_{850}-J_{110})_{AB}$ colors, as expected for bona-fide $5.5<z<7$
star-forming objects. (The shaded orange region shows the expected
position of high-redshift objects and is used to estimate the
contamination rate; see \S3.2.)  This suggests that our optically
selected sample has a very low contamination rate
($\lesssim5$\%).}
\end{figure}

\begin{deluxetable*}{cccccccccc}
\tablewidth{5.5in}
\tablecolumns{9}
\tabletypesize{\footnotesize}
\tablecaption{HUDF $i$-dropout sample.\tablenotemark{a}}
\tablehead{
\colhead{} &
\colhead{} & \colhead{} &
\colhead{} & \colhead{} & \colhead{} & \colhead{} & \colhead{} & \colhead{$r_{hl}$}\\
\colhead{Object ID} &
\colhead{R.A.} & \colhead{Decl.} &
\colhead{$z_{850}$} & \colhead{$i - z$} & \colhead{$z - J$} & \colhead{$J - H$} & \colhead{S/G} & \colhead{(arcsec)}}
\startdata
HUDF-40018149 & 03:32:40.01 & -27:48:14.9 & 24.99$\pm$0.01 & 1.6 & 0.0 & $-$0.1 & 0.03 & 0.16 \\
HUDF-36476414 & 03:32:36.47 & -27:46:41.4 & 26.08$\pm$0.02 & 2.4 & $-$0.1 & 0.4 & 0.03 & 0.19 \\
HUDF-32617540 & 03:32:32.61 & -27:47:54.0 & 26.21$\pm$0.03 & 1.4 & --- & --- & 0.03 & 0.24 \\
HUDF-34096472 & 03:32:34.09 & -27:46:47.2 & 26.50$\pm$0.02 & 2.2 & --- & --- & 0.06 & 0.12 \\
HUDF-38286172 & 03:32:38.28 & -27:46:17.2 & 26.56$\pm$0.03 & 2.7 & 0.0 & 0.0 & 0.06 & 0.12 \\
HUDF-34287525 & 03:32:34.28 & -27:47:52.5 & 26.58$\pm$0.05 & 1.5 & 0.3 & $-$0.2 & 0.00 & 0.31 \\
\enddata

\tablenotetext{a}{Table~4 is published in its entirety
in the electronic version of the Astrophysical Journal.  A portion is
shown here for guidance regarding its form and content.  All
magnitudes are AB.  Right ascension and declination use the J2000.0
equinox; units of right ascension are hours, minutes, and seconds;
units of declination are degrees, arcminutes, and arcseconds.  All
limits are 2$\sigma$.  ``S/G'' denotes SExtractor stellarity
parameter, where 0 indicates an extended object and 1 indicates a
point source (objects with S/G $>$ 0.75 and $z_{850,AB}<28.4$ are
taken to be stars and thus not included here).  The term ``faint''
means that an object was not detected ($>2\sigma$) in either of the
passbands for a measured color and thus is not quoted.  Typical errors
on the $i-z$, $z-J$, and $J-H$ colors are 0.3, 0.2-0.3, and 0.3-0.4
mag, respectively.}
\end{deluxetable*}

\subsection{Rest-frame UV Colors and Redshifts}

The $izJH$ photometry available for the HUDF can also be used to
estimate both the rest-frame UV colors and redshifts for sample
objects.  The $z-J$/$J-H$ color-color diagram, in particular, serves
as a useful starting point because at $z\gtrsim5.9$ it provides a
fairly unique mapping onto redshift and rest-frame UV color
(Figure~3).  Here we only include $i$-dropouts to a
limiting magnitude of $H_{160,AB}\sim27.7$.  Faintward of this, there
are substantial errors on the $J$ and $H$ photometry for individual
objects, and hence it is only possible to estimate the average colors
for this population.  We obtain these colors by stacking the
$i$-dropouts in two different faint magnitude intervals
$27.4<z_{850,AB}<27.9$ and $27.9<z_{850,AB}<28.9$.  Despite concerns
about possible errors in the NICMOS zero points (\S2.1), the position
of the data is such as to suggest moderately blue rest-frame UV
colors, i.e., $\beta=-2.0$, and a mean redshift somewhat below 6.

\begin{figure}
\includegraphics[width=3.4in]{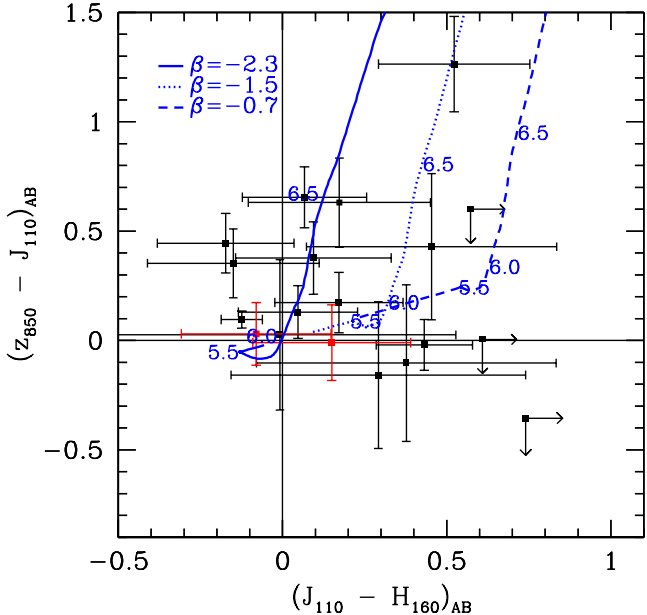}
\caption{The $(z_{850}-J_{110})$/$(J_{110}-H_{160})$ color-color
diagram showing the photometry of all 14 bright $H_{160,AB}<27.7$
objects from the HUDF with deep NICMOS coverage.  Each of these objects
was required to be at least a $2\sigma$ detection in the $H_{160,AB}$
band.  To make these measurements on even fainter dropouts, we
included stacked photometry for objects in the magnitude bins
$27.4<z_{850,AB}<27.9$ and $27.9<z_{850,AB}<28.9$ (\textit{red
squares}; \S3.3).  Model tracks are as in Figure~2.
Errors are $1\sigma$ while the limits are $2\sigma$.  The bright
objects shown here are all clearly resolved and therefore non-stellar.
The large fraction of objects with $J_{110}-H_{160}$ colors of $<0.3$
suggests a reasonably blue $\beta\sim-2.0$ population.  Although a
majority of the objects have $z_{850}-J_{110}$ colors suggestive of a
pile-up at the lower redshift end of the window, i.e., $z\lesssim6$, a
few objects have $z_{850}-J_{110}$ colors consistent with being at
higher redshift, i.e., $z\gtrsim6.1$ (see also Malhotra et al.\
2005).}
\end{figure}

Although illustrative, Figure~3 does not provide us with a
very useful way of quantifying the mean properties of our sample, such
as the redshift and rest-frame UV slope.  To accomplish this, a better
approach is to consider the distribution of $(z_{850}-J_{110})_{AB}$
and $(J_{110}-H_{160})_{AB}$ colors.  This is analogous to the
modeling we did in our previous work on $U$, $B$, and $V$-dropout
samples from the Hubble Deep Field (HDF) and GOODS fields (B06a).  The
$(z_{850}-J_{110})_{AB}$ colors (reddened by the Ly$\alpha$ forest)
are most useful for inferences about the mean redshift of the sample
while the $(J_{110}-H_{160})_{AB}$ colors (redward of the break) are
most useful for inferences about the mean rest-frame UV color of the
sample.  A schematic illustration of this is provided in the top
panels of Figure~4, where the predicted
$(z_{850}-J_{110})_{AB}$ and $(J_{110}-H_{160})_{AB}$ colors are shown
as a function of the UV continuum slope $\beta$ (annotated) and
redshift (\textit{right vertical axis}).  Using the blue lines as a
guide, the $(z_{850}-J_{110})_{AB}$ colors of $i$-dropouts observed in
the HUDF (\textit{shaded histogram}: selected to have
$H_{160,AB}\lesssim27.7$) suggest that these objects are predominantly
at $z\sim5.5-6$.  The $(J_{110}-H_{160})_{AB}$ colors indicate a mean
$UV$ continuum slope $\beta$ of $\sim -2.0$.

We can make these inferences more rigorous by performing some
simulations.  To make the simulations as realistic as possible, we
project a HUDF $B_{435}$-dropout sample (Bouwens et al.\ 2004b) to
$z\sim5-7$ and scale the sizes of individual $B$-dropouts as
$(1+z)^{-1.1}$ (for fixed luminosity).  This scaling is derived in
\S3.7 using the current data sets and is in good agreement with
previous measurements (Bouwens et al.\ 2004a,b; B06b; Ferguson et al.\
2004).  The actual simulations are executed using our well-tested
cloning software (Bouwens et al.\ 1998a,b; Bouwens et al.\ 2003a,b;
B06b), which handles the artificial redshifting and reselection of
individual objects.  $B$-dropouts are distributed in redshift
(assuming no clustering) according to the product of their individual
$1/V_{max}$ and the available cosmological volume.  Here, three mean
rest-frame UV slopes are assumed for the simulations: $\beta=-2.2$,
$-1.8$, and $-1.4$.  A $1\sigma$ scatter of 0.5 (in the UV slope
$\beta$) is assumed for each.  The results of the simulations are
shown in the bottom panels of Figure~4 (\textit{black
lines}) and compared with the observations.  It seems clear that the
observed $(J_{110}-H_{160})_{AB}$ colors (\textit{histogram}) can be
best fit by a model with a mean $\beta$ of $-2.0$, somewhere in
between the $\beta=-2.2$ and $\beta=-1.8$ model results
(\textit{dotted line}).  All models however yield a tail toward red
$(z_{850}-J_{110})_{AB}$ colors that does not occur in the
observations (\textit{histogram}).  This suggests a deficit of
$i$-dropouts at the higher redshift end of the $z\sim5.5-7$ selection
window.  To model this, we assumed that the space density of
$i$-dropouts in the HUDF was a strong function of redshift, i.e.,
$e^{-(z-5.5)}$, while adopting the best-fitting mean-frame UV slope
$\beta$ found above ($-2.0$).  The results are shown in
Figure~4 as a solid purple line and provide a rough fit
to the median colors.  We note that very similar conclusions have come
from the GRAPES program (Malhotra et al.\ 2005), where even better
redshift measurements are possible from the GRISM data.  Malhotra et
al.\ (2005) demonstrated that the majority of bright
($z_{850,AB}\lesssim28$) $i$-dropouts in the HUDF (15 out of 23
objects) are at $z\sim5.9\pm0.2$.

\begin{figure}
\includegraphics[width=3.5in]{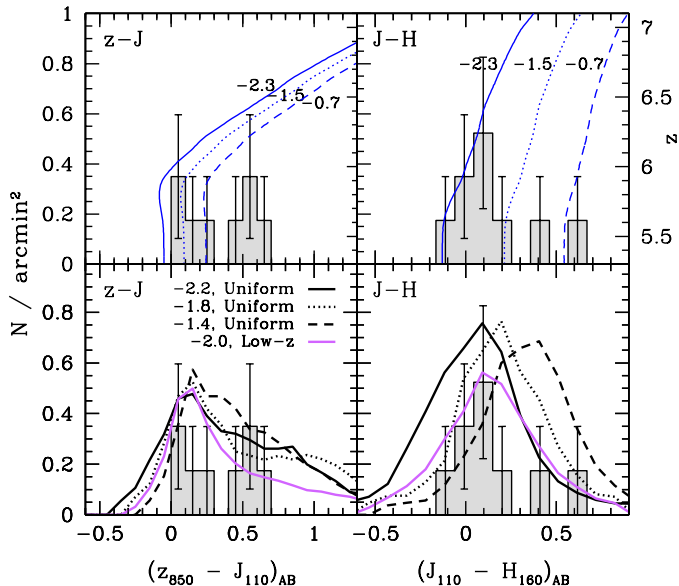}
\caption{\textit{Top panels}; The observed $(z_{850}-J_{110})_{AB}$
and $(J_{110}-H_{160})_{AB}$ color distributions (\textit{histogram
with} $1\sigma$ \textit{Poisson errors}).  Overplotted is the color as
a function of redshift (right-hand axis) for a range of $\beta's$:
$\beta=-0.7$ (\textit{dashed blue line}), $\beta=-1.5$ (\textit{dotted
blue line}), and $\beta=-2.3$ (\textit{solid blue line}).  The binned
colors were derived from $i$-dropouts in the HUDF with
$H_{160,AB}\lesssim27.8$.  \textit{Bottom panels}; The observed
$(z_{850}-J_{110})_{AB}$ and $(J_{110}-H_{160})_{AB}$ color
distributions (\textit{histogram}) vs. that recovered from simulations
assuming three different mean rest-frame UV slopes $\beta = -2.2$
(\textit{solid black line}), $\beta=-1.8$ (\textit{dotted black
line}), and $\beta=-1.4$ (\textit{dashed black line}).  To model the
deficit of objects with large $(z_{850}-J_{110})$ colors, we also
included the results from a simulation in which the object density was
proportional to $e^{-(z-5.5)}$ (assuming a mean $\beta$ of $-2.0$).
Together the $z-J_{110}$ (providing information on Ly$\alpha$ forest
attenuation) and $J_{110}-H_{160}$ (constraining the spectral slope)
colors place good constraints on the rest-frame UV slope $\beta$ and
redshift.  These results suggest that most $i$-dropouts in the HUDF
NICMOS footprint lie below $z\sim 6.2$ with a mean $\beta$ of $-2.0$.
More details can be found in \S3.3.}
\end{figure}

\subsection{$i$-dropouts in the GOODS/HUDF-P fields}

To control for field-to-field variations and to add numbers at bright
and intermediate magnitudes (where statistics in the HUDF are poor),
it was useful to incorporate the HUDF results with those derived from
the shallower HUDF-Ps and GOODS fields.  The selection of dropouts
from these fields was performed using nearly identical selection
criteria to that used for the HUDF (\S3.2).  Sixty-eight and 332
dropouts were found in the HUDF-Ps and GOODS fields, respectively
(Tables~5-6).  These are
significantly more dropouts ($\sim2-5$ times) than were found in our
initial studies on these fields (Bouwens et al.\ 2004a; B06a) and is
due to our slightly more inclusive selection criteria ($i-z>1.3$
rather than $i-z>1.4$), better pixelization (0.03\arcs$\,$rather than
0.05\arcs), greater depths ($0.2$ mag fainter for the HUDF-Ps and
$0.4$ mag fainter for the GOODS fields), and larger areas probed (an
additional $\sim40$ arcmin$^2$ for the GOODS fields).  Our total
$i$-dropout sample (from all three data sets) has 506 individual
objects (16 of the total 522 dropouts from these three fields are
found in both our GOODS and HUDF/HUDF-Ps catalogs and so are only
counted once).

\begin{deluxetable*}{cccccccc}
\tablewidth{5.5in}
\tablecolumns{8}
\tabletypesize{\footnotesize}
\tablecaption{HUDF-Ps $i$-dropout sample.}
\tablehead{
\colhead{} &
\colhead{} & \colhead{} &
\colhead{} & \colhead{} & \colhead{} & \colhead{$r_{hl}$}\\
\colhead{Object ID} &
\colhead{R.A.} & \colhead{Decl.} &
\colhead{$z_{850}$} & \colhead{$i - z$} & \colhead{S/G} & \colhead{(arcsec)}}
\startdata
HUDFP1-2494954244 & 03:32:49.49 & -27:54:24.4 & 26.08$\pm$0.06 & 1.4 & 0.00 & 0.21 \\
HUDFP2-2064148469 & 03:32:06.41 & -27:48:46.9 & 26.11$\pm$0.06 & 1.7 & 0.01 & 0.19 \\
HUDFP1-2439856440 & 03:32:43.98 & -27:56:44.0 & 26.32$\pm$0.04 & 1.6 & 0.38 & 0.10 \\
HUDFP1-2483955541 & 03:32:48.39 & -27:55:54.1 & 26.63$\pm$0.05 & 1.6 & 0.35 & 0.09 \\
HUDFP1-2427156555 & 03:32:42.71 & -27:56:55.5 & 26.77$\pm$0.09 & 2.0 & 0.00 & 0.17 \\
HUDFP1-2394754149 & 03:32:39.47 & -27:54:14.9 & 26.88$\pm$0.10 & 1.3 & 0.01 & 0.15 \\
\enddata

\tablenotetext{a}{Table~5 is published in its entirety
in the electronic version of the Astrophysical Journal.  A portion is
shown here for guidance regarding its form and content.  Right
ascension and declination use the J2000.0 equinox; units of right
ascension are hours, minutes, and seconds; units of declination are
degrees, arcminutes, and arcseconds.  All magnitudes are AB.  All
limits are 2$\sigma$.  ``S/G'' denotes SExtractor stellarity
parameter, where 0 indicates an extended object and 1 indicates a
point source (objects with S/G $>$ 0.75 and $z_{850,AB}<27.5$ are
taken to be stars and thus not included here).  Typical errors on the
$i-z$ colors are 0.3 mag.}
\end{deluxetable*}

\begin{deluxetable*}{ccccccccc}
\tablewidth{5.5in}
\tablecolumns{9}
\tabletypesize{\footnotesize}
\tablecaption{GOODS $i$-dropout sample.\tablenotemark{a}}
\tablehead{
\colhead{} &
\colhead{} & \colhead{} &
\colhead{} & \colhead{} & \colhead{} & \colhead{} & \colhead{} & \colhead{$r_{hl}$}\\
\colhead{Object ID} &
\colhead{R.A.} & \colhead{Decl.} &
\colhead{$z_{850}$} & \colhead{$i - z$} & \colhead{$z - J$} & \colhead{$z - K_s$} & \colhead{S/G} & \colhead{(arcsec)}
}
\startdata
CDFS-2256155487 & 03:32:25.61 & -27:55:48.7 & 24.51$\pm$0.02 & 1.6 & -0.1 & -1.0 & 0.37 & 0.11 \\
HDFN-5426312091 & 12:35:42.63 &  62:12:09.1 & 25.15$\pm$0.06 & 1.5 & --- & --- & 0.02 & 0.23 \\
CDFS-2400148141 & 03:32:40.01 & -27:48:14.1 & 25.17$\pm$0.04 & 1.6 & 0.0 & $<$-0.4 & 0.36 & 0.12 \\
CDFS-2331939491 & 03:32:33.19 & -27:39:49.1 & 25.29$\pm$0.06 & 2.4 & --- & --- & 0.02 & 0.21 \\
CDFS-2237840378 & 03:32:23.78 & -27:40:37.8 & 25.34$\pm$0.07 & 1.6 & --- & --- & 0.01 & 0.22 \\
CDFS-2334852466\tablenotemark{*} & 03:32:33.48 & -27:52:46.6 & 25.37$\pm$0.08 & 1.4 & 1.2 & 2.5 & 0.00 & 0.29 \\
\enddata

\tablenotetext{a}{Table~6 is published in its entirety in the
electronic version of the Astrophysical Journal.  A portion is shown
here for guidance regarding its form and content.  Right ascension and
declination use the J2000.0 equinox; units of right ascension are
hours, minutes, and seconds; units of declination are degrees,
arcminutes, and arcseconds.  All magnitudes are AB.  Right ascension
and declination use the J2000 equinox.  All limits are 2$\sigma$.
``S/G'' denotes the SExtractor stellarity parameter, where 0 is for an
extended object and 1 is for a point source (objects with S/G $>$ 0.75
and $z_{850,AB}<26.8$ are taken to be stars and thus not included
here).  Objects denoted with an asterisk have $z_{850}-K_s$ colors
which suggest they are low-redshift interlopers (Figure~D3).}
\end{deluxetable*}

\subsection{Corrections for Depth}

The properties of all our $i$-dropouts samples are summarized in
Table~7.  To put these samples together to obtain a
single measure of the $i$-dropout surface density, we must account for
the sizeable effect of survey depth.  A simple illustration of this
can be found in the top panel of Figure~5, which
contrasts $i$-dropouts selected from the HUDF, HUDF-Ps, and GOODS
fields.  Although incompleteness is clearly the dominant effect in the
observed differences, other selection and measurement biases also play
a role.  We relegate a detailed discussion of these biases to Appendix
D.  However, it is useful to give a brief summary here of the main
corrections.

\begin{deluxetable}{lrrrr}
\tablewidth{3in}
\tablecolumns{5}
\tabletypesize{\footnotesize}
\tablecaption{Summary of $i$-dropout samples.\tablenotemark{a}}
\tablehead{
\colhead{} & \colhead{Area} &
\colhead{} & \colhead{} & \colhead{}\\
\colhead{Sample} & \colhead{(arcmin$^{2}$)} &
\colhead{No.} & \colhead{Mag. Limit\tablenotemark{a}} & \colhead{$L_{z=3} ^{*}$\tablenotemark{b}}}
\startdata
CDFS GOODS & 166\tablenotemark{*} & 181 & $z \sim 27.9$ & 0.17 \\
HDFN GOODS & 150 & 151 & $z \sim 27.9$ & 0.17 \\
HUDFP1 & 10 & 54\tablenotemark{$\dagger$} & $z \sim 28.6$ & 0.09 \\
HUDFP2 & 7 & 14\tablenotemark{$\dagger$} & $z \sim 28.6$ & 0.09 \\
HUDF & 11 & 122\tablenotemark{$\dagger$} & $z \sim 29.4$ & 0.04 \\
\enddata
\tablenotetext{*}{Due to our inclusion of the ACS parallels to the HUDF
 NICMOS field in our reductions of the CDF-S GOODS field (\S2.3),
 the total area available there for $i$-dropout searches exceeded that
 available in the HDF-N GOODS field.}
\tablenotetext{a}{The magnitude limit is the $\sim$8$\sigma$ detection
 limit for objects in a 0.2\arcs-diameter aperture.}
\tablenotetext{b}{Magnitude limit in units of $L_{z=3}^{*}$ (Steidel et al.\ 1999).}
\tablenotetext{$\dagger$}{7, 7, and 2 $i$-dropouts from our HUDF, HUDFP1, and
HUDFP2 catalogs, respectively, also occur in our CDFS GOODS catalog.}
\end{deluxetable}

We divide these corrections into completeness, flux, and contamination
corrections.  These corrections allow an approximate conversion from
the surface densities measured in our shallower data to their
equivalent surface densities if measured with HUDF quality data.  Our
first set of corrections, the completeness corrections (Appendix D1),
makes up for the fact that our shallower surveys preferentially miss
the larger, lower surface brightness fraction of galaxies in any given
magnitude interval.  In general, these corrections tend to be small
($\lesssim10$\%) except near the magnitude limit of the data, where
they can be $\gtrsim50$\%.  For the GOODS data, these corrections
enter at $z_{850,AB}\gtrsim26.8$ and for the HUDF-Ps data, they enter
at $z_{850,AB}\gtrsim27.5$.  Table D3 show the
results of the simulations.  As with other results in this section,
these were obtained by degrading the deeper data to the depths of the
shallower data and repeating the selection.

The purpose of our second set of corrections, the flux corrections
(Appendix D2), was to compensate for the fact that our shallower
surveys may estimate lower fluxes for objects than would be measured
in deeper exposures.  Here, the corrections proved to be relatively
small ($\sim0.1-0.2$ mags) except for those objects near the magnitude
limit (Figure~D2) where some brightening was observed
(0.3 mag).  This brightening appeared to be the result of a Malmquist
bias.  The first and second corrections were implemented using the
estimated transfer functions (Appendix D3:
Tables~D5 and D6).

Finally, our third set of corrections (Appendix D4) was used to
subtract out the likely contamination rate for our different samples.
We included a variety of different sources of contamination in this
estimate: low-mass stars (Figure~D5), intrinsically-red lower redshift
interlopers (Table~D7), objects that entered our sample due to
photometric scatter (Tables~D8-D9), and finally spurious objects.  In
general, all sources of contamination were small and never contributed
more than 15\% of the objects in any given magnitude
interval.\footnote{The recent findings from the GRAPES team (Malhotra
et al.\ 2005) are consistent with these contamination estimates.  For
an $(i_{775}-z_{850})_{AB}>1.3$ selection (where a spectrum could be
unambiguously extracted), the GRAPES team found that only one out of
15 objects was a contaminant (i.e., a $z_{850,AB}\sim25.4$ star).  In
the current HUDF selection (\S3.2), this object was rejected as a
star.}

\begin{figure}
\includegraphics[width=3.2in]{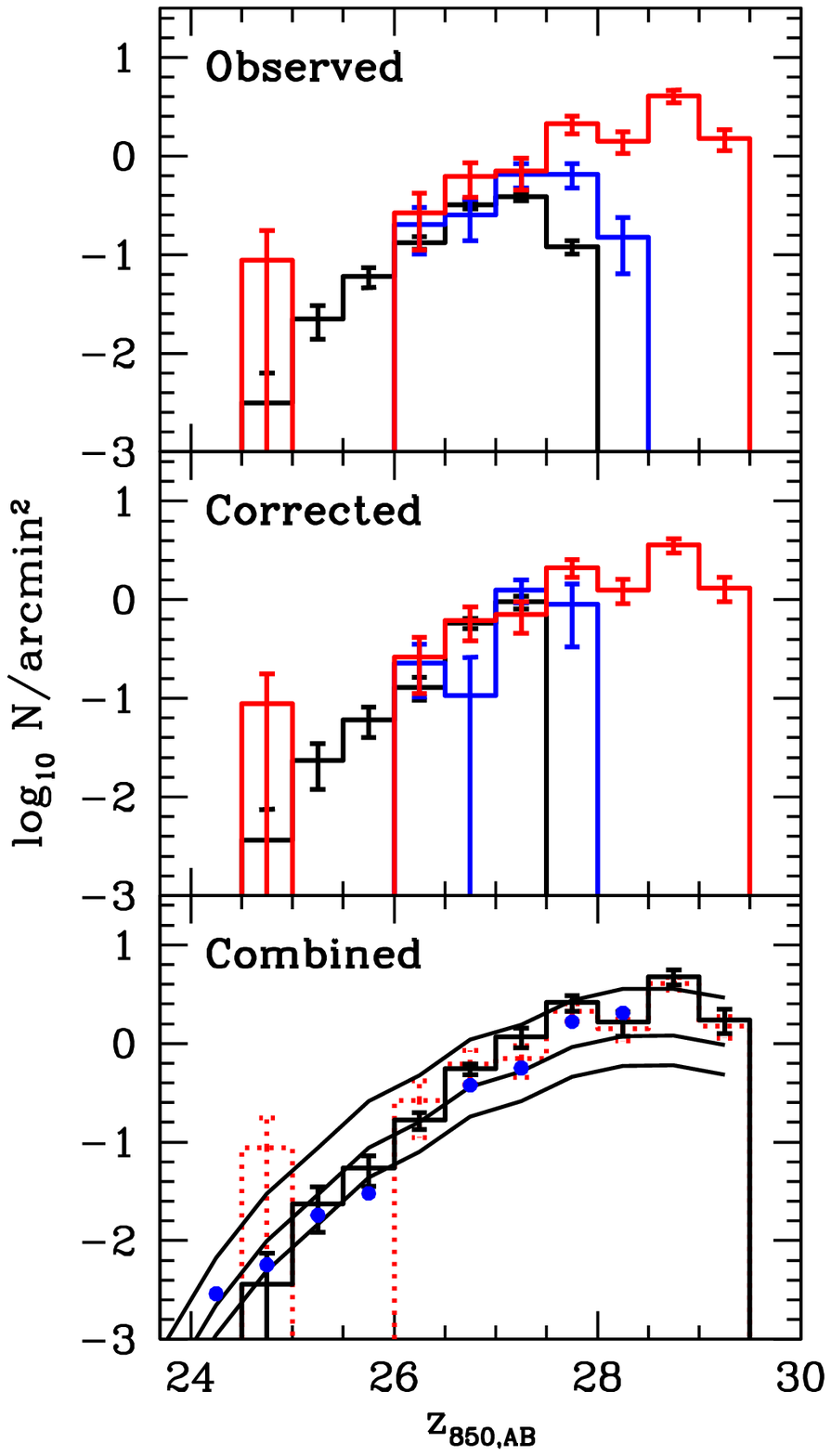}
\caption{Surface densities (per 0.5 mag interval) of $i$-dropouts
observed at three different depths: GOODS (\textit{black histogram}),
HUDF-Ps (\textit{blue histogram}), and HUDF (\textit{red histogram}).
Errors are the $1\sigma$ Poissonian uncertainties.  The top panel
presents the uncorrected surface densities.  The middle panel presents
these same surface densities, but corrected to the same level of
completeness and flux biases as are present in the HUDF (Appendix D)
and with the contaminants removed (note that incompleteness in the
HUDF [or at HUDF depths] is accounted for using the selection volume,
at a later stage: Figure~8).  The bottom panel shows
the cumulative surface density obtained by combining these fields
through a maximum likelihood procedure (\S3.8).  The predicted surface
densities of $i$-dropouts assuming the Steidel et al.\ (1999) LF, the
Steidel et al.\ (1999) LF divided by 3, and the Steidel et al.\ (1999)
LF divided by 6 are shown with the three solid black curves in the
bottom panel.  The uncorrected HUDF counts are also included in the
bottom panel as the dotted red histogram.  The equivalent differential
counts of BSEM04 are included in this panel as blue circles (from
their Figure 10).  The no-evolution ($z\sim3$) predictions exceed the
observed counts by factors of $\approx6$ at the bright end
($z_{850,AB}\lesssim26$: Stanway et al.\ 2003, 2004b; Dickinson et
al.\ 2004), $\approx3$ at more intermediate magnitudes
($z_{850,AB}\sim26-27$), and $\lesssim2$ at faint end
($z_{850,AB}\gtrsim27$).  The effect of depth on the extracted counts
is obvious in the top panel.  A detailed quantification of the
relevant biases (selection and measurement) is provided in
Table~D3 and
Figure~D2.}
\end{figure}

\subsection{Field-to-Field Variations}

The effective normalization of the luminosity function is expected to
show significant variations as a function of position and environment
($\sim35$\% rms for single ACS pointings).  This is the result of
large-scale structure (loosely referred to as ``cosmic variance'').
Since the goal of these studies is to derive a luminosity function
that is representative of the cosmic average, our challenge is to
remove these variations.  A simple averaging of the $i$-dropouts from
the different fields is not appropriate since the fields differ in the
magnitude ranges they probe.  One would have no guarantee with such a
procedure that the average normalization obtained at brighter
magnitudes is the same as that obtained at fainter magnitudes, thus
allowing for discontinuities in the normalization.  This could impact
the shape of the derived LF (see Appendix E).

To remove these differences, it is necessary to estimate the relative
normalizations of $i$-dropouts in our different survey fields.  We do
this by degrading the deeper data to the depths of the shallower
survey fields and then comparing the surface densities of $i$-dropouts
derived.  To maximize the significance, the present comparisons are
done in two stages: (1) comparing the HUDF against the HUDF-Ps and (2)
comparing the deeper three fields (HUDF + HUDF-Ps) against the GOODS
fields.  The overall normalization for our data sets is set by the
mean of the two GOODS fields, which--sampling the largest comoving
volume--should provide our best estimate of the cosmic average.

For the first stage, $i$-dropouts in the HUDF are normalized relative
to $i$-dropouts in our deepest three fields.  The normalization factor
is determined by degrading the HUDF to the same S/N level as the two
parallels and then comparing the number of dropouts in the fields.
This degradation was performed 10 times and the S/N (weight maps) of
both parallels were matched on a pixel-by-pixel basis (as in Appendix
C and Appendix D1).  Our findings are shown in
Table~8 and point to the HUDF having a
normalization similar to the first parallel (50.2 vs. 43.6), but
substantially higher than that of the second parallel (27.8
vs. 11.4).\footnote{Note that more objects are found in the
degradation of the HUDF to the depth and area of the first parallel
than the second.  This is due to the slightly larger depth and area
for that field (due to a greater overlap with exposures from the GOODS
fields).}  Taken together, this suggests that the HUDF is 16$\pm$24\%
overdense relative to the mean of the HUDF-Ps, or 10$\pm$15\% overdense
relative to the cosmic average defined by the three
fields.\footnote{For example, the first stage normalization factor
(1.10$\pm$0.15) quoted for the HUDF can be calculated from the numbers
given in Table~8 as
$3(35.7)/(11.4+50.2+35.7)\sim1.10$ where $35.7$ is the average number
of dropouts found in the degradations of the HUDF to the depth of the
parallels, i.e., $(43.6+27.8)/2\sim35.7$.}  These fields also enable
us to comment on the observed field-to-field variations, which appear
to be $\sim$46\% rms on $\sim11$ arcmin$^2$ scales.  This is
consistent with the $\sim35$\% rms variations one obtains assuming a
$\Lambda$CDM power spectrum, $\Delta z = 0.7$ selection window, pencil
beam geometry, and bias of 4, which is appropriate (Mo \& White 1996)
for objects of number density $\sim10^{-3}$ Mpc$^{-3}$ probed by these
fields (Figure~10).

\begin{deluxetable}{lrr}
\tablewidth{3.3in}
\tabletypesize{\footnotesize}
\tablecaption{Number of $i$-dropouts found in the two HUDF-Ps and in
the HUDF degraded to the same depths.}
\tablehead{ \colhead{} & \multicolumn{2}{c}{Number of dropouts}\\
\colhead{Field} & \colhead{HUDFP1} & \colhead{HUDFP2}}
\startdata
HUDFP1 & 50.2\tablenotemark{$\dagger$} & -- \\
HUDFP2 & -- & 11.4\tablenotemark{$\dagger\ddagger$} \\
HUDF & 43.6 & 27.8\tablenotemark{*} \\
\enddata
\tablenotetext{$\dagger$}{These numbers have been corrected for the
expected contamination from low-redshift objects scattering into our
sample ($\sim3$ per field, see Table~D8).}
\tablenotetext{$\ddagger$}{Note that no comparable deficit in $B$ or
$V$-dropouts is found in HUDFP2 relative to other fields (e.g., the HUDF
or HUDFP1), suggesting that the apparent underabundance of $i$-dropouts here
is not related to the reduction or processing of the data (or any
bright stars in the foreground).}
\tablenotetext{*}{The depth and
selection area in the second parallel were smaller than that of the
first due to a lesser overlap with GOODS.  As a result, degradations
of the HUDF to the depth of the first parallel revealed more objects
than degradations to the depth of the second.}
\end{deluxetable}

For the second stage, the normalization of the deeper three fields is
adjusted to match that of the GOODS fields.  As before, the
normalization factor is estimated by degrading the HUDF and HUDF-Ps to
the S/N level of the GOODS fields and extracting $i$-dropout samples
using selection criteria identical to that used for GOODS.  The
results of these experiments are shown in
Table~9, and it is clear that the average surface
density derived from the three deeper fields
($0.69\pm0.15\,\,\textrm{arcmin}^{-2}$) is somewhat lower
($0.70\pm0.16$ times) than that found in both GOODS fields ($0.99\pm
0.06\,\,\textrm{arcmin}^{-2}$).  $0.70\pm0.16$ is the second stage
normalization factor.  Interestingly enough, the surface density of
$i$-dropouts is 9\% $\pm$ 13\% larger in the CDF-S GOODS field than it
is in the HDF-N GOODS field.  However, this is not inconsistent with
the sort of variations expected from cosmic variance ($\pm$20\%) in
fields of this size (160 arcmin$^2$; Somerville et al.\ 2004).
Multiplying the first and second stage factors together, we arrive at
an overall normalization factor for the HUDF and HUDF-Ps.  These factors
are summarized in Table~10 under the ``Two Stage''
column.

\begin{deluxetable}{lc}
\tablewidth{3in}
\tablecolumns{2}
\tabletypesize{\footnotesize}
\tablecaption{Surface Densities of $i$-dropouts from different fields
at GOODS depth.}
\tablehead{
\colhead{} & \colhead{Surface Density}\\
\colhead{Field} & \colhead{(arcmin$^{-2}$)}}
\startdata
HDFN GOODS & 0.94$\pm$0.08\tablenotemark{$\dagger$} \\
CDFS GOODS & 1.03$\pm$0.08\tablenotemark{$\dagger$} \\
HUDFP1 & 0.78$\pm$0.28 \\
HUDFP2 & 0.33$\pm$0.21 \\
HUDF & 0.97$\pm$0.29 \\
\enddata 
\tablenotetext{$\dagger$}{These surface densities have been corrected
for the expected contamination rate from low-redshift objects
scattering into our sample ($0.06$ contaminants arcmin$^{-2}$, see
Tables~D8 and D9).}
\end{deluxetable}

As an alternative to this procedure, the normalization of our deeper
fields can be derived by comparing directly with the surface density
of $i$-dropouts found at GOODS depth (Table~9).
Using the above results (i.e., Table~9), we
derive a normalization of $0.98\pm0.30$ and $0.56\pm0.18$ for the HUDF
and HUDF-Ps fields, respectively.  These values are compiled under the
``One Stage'' column in Table~10.  While consistent,
they are of slightly lower significance than our estimates made with
the two stage procedure.  We adopt the results of the two stage
procedure as our final estimate of the relative normalization and take
the reciprocal of this normalization as our adjustment factor.

\begin{deluxetable}{lrrc}
\tablewidth{0pt}
\tabletypesize{\footnotesize}
\tablecaption{Adjustments made to the $i$-dropout surface densities
from the different fields used in this study.}
\tablehead{ \colhead{} & \multicolumn{2}{c}{Relative Normalization}\\
\colhead{Field} & \colhead{Two Stage\tablenotemark{a}} & \colhead{One
Stage\tablenotemark{b}} & \colhead{Adjustment Factor\tablenotemark{c}}}
\startdata
HUDFPs & $0.67\pm0.16$ & $0.56\pm0.18$ & 1.50 \\
HUDF & $0.77\pm0.20$ & $0.98\pm0.30$ & 1.30 \\
GOODS & 1.0 (fixed) & 1.0 (fixed) & 1.00 \\
\enddata
\tablenotetext{a}{The two stage normalization (\S3.6) is obtained by
comparing the surface densities of $i$-dropouts in a field with those
of the two GOODS fields.  This is a two stage process, in which the
normalization of a given field is first tied to the deepest three
fields (Table~8) and these fields, in turn, are
tied to the two GOODS fields (Table~9).  The
final normalization factor is then the product of the normalization
factors derived from these two comparisons, e.g.,
$(1.10\pm0.15)(0.70\pm0.16)=0.77\pm0.20$ for the HUDF (see \S3.6).  The
two stage normalization has the advantage of a larger overlap between
the different surveys being tied together.  This overlap translates into
smaller uncertainties in the overall normalization factors (estimated
assuming Poissonian errors).}
\tablenotetext{b}{The one stage normalization (\S3.6) is obtained by
comparing the surface densities of $i$-dropouts in a field with the
average of that found in the two GOODS fields
(Table~9), e.g., $(0.97\pm0.29\,
\textrm{arcmin}^{-2})/((0.94\pm0.08\,\textrm{arcmin}^{-2}+1.03\pm0.08
\,\textrm{arcmin}^{-2})/2)\sim 0.98\pm0.30$ for the HUDF.}
\tablenotetext{c}{The adopted adjustment factor is equal to the
reciprocal of the normalization relative to GOODS.  We use the two
stage normalizations because of their smaller uncertainties.}
\end{deluxetable}

\subsection{Dependence of Galaxy Size on Redshift}

Data from the HUDF, HUDF-Ps, and GOODS fields also allow us to revisit
our analyses on the physical sizes of galaxies at $z\sim6$ and how
these sizes compare with those at latter times.  Previously, we had
carried out our analyses using each of the above fields separately
(Bouwens et al.\ 2004b; Bouwens et al.\ 2004a; B06b).  With the
combined data set, we can significantly improve this analysis.  For
this paper, these sizes are important for modeling the selection
effects of our $i$-dropout samples.  Similarly to our previous work,
we model the sizes of $i$-dropouts in all three samples using
different size scalings $(1+z)^{-m}$ ($m$ = 0, 1, and 2) of a
$z\sim2.5$ HDF-N + HDF-S $U$-dropout sample (B06a).  We project
objects from this sample to higher redshift ($z\sim5-7$) using our
cloning software, add them to noise frames, and then reselect them in
exactly the same way as the observed samples.  Only galaxies 1 mag
brightward of the selection limits are considered for our comparisons
(to avoid being dominated by selection effects).\footnote{As shown in
Figure~7, the observed $z_{850}$-band magnitudes can
correspond to a wide range of absolute magnitudes.  This may make it
more challenging to measure size evolution at $z\sim6$ using a
fixed-magnitude $i$-dropout sample.  Fortunately, this should not bias
the size evolution measured here since we have included all of these
effects in our simulations.}  Figure~6 for the
$i$-dropouts from all three fields.  Here it is evident that the
typical half-light radius for $i$-dropouts at $z_{850,AB}\sim27$ is
0.8 kpc (after correction for the PSF).  Relative to the sizes of
objects at lower redshift, the $(1+z)^{0}$ and $(1+z)^{-2}$ scalings
seem to nicely bracket the observed range.  To derive a more precise
estimate, we rely on comparisons between the mean half-light radii
obtained from the observations and simulations.  Interpolating between
our simulation results, our best-fit values for the size-evolution
exponent $m$ are $1.2\pm0.4$, $1.0\pm0.5$, and $1.0\pm0.4$ for the
HUDF, HUDF-Ps, and GOODS fields, respectively.  Combining the results
from the three fields to obtain a single scaling (and thus assuming
that this redshift scaling is luminosity independent) yields
$m=1.1\pm0.3$.  This is in good agreement with several previous
determinations: $m=[0.8,2.0]_{1\sigma}$ (B06b), $m=1.57 _{-0.50}
^{+0.53}$ (Bouwens et al.\ 2004a), $m=0.94_{-0.19} ^{+0.25}$ (Bouwens
et al.\ 2004b), and the Ferguson et al.\ (2004) $H(z)^{-1}$ size
scaling, which is equivalent to $m = 1.47$ over the redshift range
$2.5<z<6$.

\begin{figure*}
\epsscale{1.0}
\begin{center}
\includegraphics[width=7.0in]{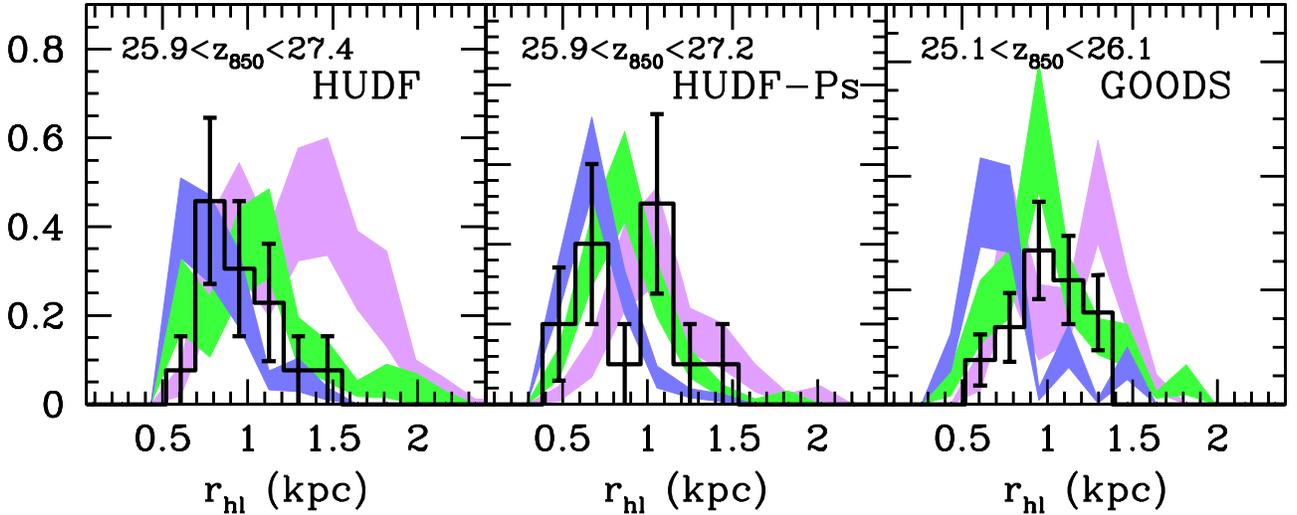}
\end{center}
\caption{Observed half-light radii (\textit{black histogram with}
$1\sigma$ \textit{Poisson errors}) for a bright subset of $z\sim6$
$i$-dropouts from the HUDF, HUDF-Ps, and GOODS fields vs. that predicted
from $(1+z)^0$ (\textit{violet shaded region}), $(1+z)^{-1}$
(\textit{green shaded region}), and $(1+z)^{-2}$ (\textit{blue shaded
region}) size scalings of a $z\sim2.5$ $U$-dropout sample from the
HDF-N + HDF-S fields (B06a).  The normalization is arbitrary.  The
breadth of the shaded regions indicates the $\pm1\sigma$ uncertainties
based on the finite size of our input samples (Bouwens et al.\ 1998a).
Predictions for the GOODS data sets appear ``noisy'' due to the
limited number of bright galaxies in our HDF-N+HDF-S input samples.
The best-fit is obtained for a $(1+z)^{-1.1\pm0.3}$ size scaling (see
\S3.7 for details).  Typical $i$-dropouts at $z_{850,AB}\sim27$ (from
the HUDF-Ps and HUDF) have PSF-corrected half-light radii of $\sim0.8$
kpc.}
\end{figure*}

\subsection{Best-Fit Surface Densities}

It is useful to combine the results from our three data sets into a
single measure of the $i$-dropout surface density as a function of
magnitude.  To derive this, we apply a maximum likelihood procedure.
For all three data sets, the model counts are multiplied by the
transfer functions (Appendix D3: from the HUDF to the relevant field),
multiplied by the normalization factors from Table~10
(from the cosmic average to the normalization of the particular
field), and then compared with the observed counts.  In these fits, we
do not include counts faintward of $z_{850,AB}=27.0$ in the two GOODS
fields and faintward of $z_{850,AB}=28.0$ in the HUDF-Ps to be
conservative.  This allows us to avoid any systematics that may occur
in modeling the selection effects near the completeness limit.  The
resulting surface density of $i$-dropouts is tabulated in the ``I''
column of Table~11 and shown in the bottom panel of
Figure~5.  This surface density spans 5 mag, running
all the way from $z_{850,AB}\sim24.5$ to 29.5.  We remind the reader
that the surface densities quoted here are as measured at HUDF depths
and are not free of the incompleteness/flux biases implicit at these
levels.  Because of this, we have also included a second column in
Table~11 that quotes the surface densities at HUDF
depths corrected for blending with foreground objects (see Appendix
D1).

\begin{deluxetable}{ccc}
\tablewidth{3in}
\tabletypesize{\footnotesize}
\tablecaption{$i$-dropouts surface densities estimated from
HUDF, HUDF-Ps, and GOODS fields, corrected up to the HUDF completeness
levels.\tablenotemark{a}}
\tablehead{
\colhead{} & \multicolumn{2}{c}{Surface Density (arcmin$^{-2}$)}\\
\colhead{Magnitude} & \colhead{I} & \colhead{II}}
\startdata
$24.50<z_{850}<25.00$ & $0.004\pm0.004$ & $0.004\pm0.004$\\
$25.00<z_{850}<25.50$ & $0.024\pm0.011$ & $0.028\pm0.014$\\
$25.50<z_{850}<26.00$ & $0.054\pm0.019$ & $0.066\pm0.022$\\
$26.00<z_{850}<26.50$ & $0.166\pm0.033$ & $0.201\pm0.039$\\
$26.50<z_{850}<27.00$ & $0.551\pm0.071$ & $0.664\pm0.085$\\
$27.00<z_{850}<27.50$ & $1.175\pm0.265$ & $1.416\pm0.320$\\
$27.50<z_{850}<28.00$ & $2.589\pm0.478$ & $3.119\pm0.575$\\
$28.00<z_{850}<28.50$ & $1.643\pm0.456$ & $1.980\pm0.549$\\
$28.50<z_{850}<29.00$ & $4.731\pm0.797$ & $5.701\pm0.960$\\
$29.00<z_{850}<29.50$ & $1.743\pm0.481$ & $2.100\pm0.580$\\
\enddata
\tablenotetext{a}{Because of the modest ($\sim17$\%) incompleteness
due to object blending in the HUDF (Appendix D1), we quote two
different surface densities here.  Column ``I'' gives the equivalent
surface densities at HUDF depths.  Column ``II'' corrects the column
``I'' surface densities for blending (i.e., by multiplying column
``I'' by $1/0.83$).  The results in column ``II'' should be largely
free of selection or measurement biases brightward of
$z_{850,AB}\sim28.5$.  Faintward of this, incompleteness becomes
important.}
\end{deluxetable}

\section{Comparison With Previous Results}

\subsection{Source Lists and Surface Densities}

In \S3, we used $i$-dropouts measured at three different depths
(GOODS, HUDF-Ps, and HUDF) to derive an optimal measure of the surface
density of $i$-dropouts.  Previously, there have been several attempts
to compile the counts from these fields, and so it is useful to make
comparisons with the source lists first before trying to understand
possible differences in the interpretation.  We begin with the
$i$-dropouts from the HUDF, for which several source lists have
already been compiled (BSEM04; Yan \& Windhorst 2004b; Beckwith
et al.\ 2006).  Fortunately, these papers use selection criteria
nearly identical to our sample, facilitating the comparisons.  As far
as the current catalogs are concerned, 48 of the 54 $i$-dropouts
compiled by BSEM04 appear in our primary list
(Table~4), four appear in our blended $z\sim6$
candidate list (Table~D4: see Appendix D1), one
(BSEM04\#49117) was blended with a foreground object in both our
catalogs (Tables~4~and~D4), and one
(BSEM04\#17487) had a $V_{606}$-band flux ($V_{606}-z_{850}=2.4$)
inconsistent with our $i$-dropout selection criteria.  Eighty-four of
the brightest 95 $i$-dropouts ($z_{850,AB}<29.5$) from the Yan \&
Windhorst (2004b) catalog also appear in our primary list
(Table~4), five appear in our blended $z\sim6$
candidate list (Table~D4), three had $V_{606}$-band
fluxes inconsistent with our $i$-dropout criteria, and three were near
the edges of the HUDF image and therefore outside our selection area.
Possible differences in object splitting between catalogs are ignored
in the above comparisons.  As for the previously published catalogs,
35 of the brightest 39 $i$-dropouts from Table~4
($z_{850,AB}<27.9$) appear in the BSEM04 catalog and 34 of these 39
appear in the Yan \& Windhorst (2004b) compilation.  Objects appear to
be missing from the previous catalogs due to their surface brightness
(e.g., as with HUDF-42566566 or HUDF-34998369), proximity to the
$(i_{775}-z_{850})_{AB}=1.3$ color cut, and proximity to the edge of
the HUDF frame, as is the case for HUDF-42209119 which is not given in
the BSEM04 catalog.

In the GOODS fields, the surface densities we derive are less than
those first reported by Giavalisco et al.\ (2004b) and Dickinson et
al.\ (2004) using a similar $(i_{775}-z_{850})_{AB}>1.3$ selection on
the three-epoch data.  We obtain 0.10$\pm$0.02 and 0.30$\pm$0.05
arcmin$^{-2}$ to $z_{850,AB}\sim26$ and 26.5, respectively, versus
their surface densities of 0.17 and 0.37 arcmin$^{-2}$ to the same
magnitude limits, after applying their estimated correction for
contamination from photometric scatter (20\%) and spurious fraction
(23\%).  The disagreement becomes even worse, however, if an account
is made for the fact that their surface densities derive from the
three-epoch data (and would need to be corrected upward to account for
the considerable incompletenesses at these depths).  What is the
source of this disagreement?  A quick investigation suggests that it
has come from a substantial underestimate of the contamination rate in
these previous studies.  Here we can revisit these estimates using the
now deeper imaging data over the GOODS fields and in particular the
HUDF-Ps and HUDF data.  Of the 251 $i$-dropouts in the Dickinson et
al.\ (2004) $i$-dropout catalog, only 12 overlap with the deeper HUDF
(2 mag fainter) and HUDF-Ps (1 mag fainter) data.  Three (25\%) of
these objects appear to be bona-fide $i$-dropouts, two (17\%) are
low-redshift interlopers, and seven objects (58\%) are not found at
all in the deeper data and therefore appear to be spurious.  This
works out to a 75\% contamination rate, which is much higher than the
$\sim$45\% estimated in the Giavalisco et al.\ (2004b) and Dickinson
et al.\ (2004) studies.  To be fair, we note that these studies
stressed the substantial uncertainties in their estimates.  More
striking is the fact that only 94 of the 251 $i$-dropouts in the
Dickinson et al.\ (2004) catalog are even associated with real sources
in our GOODS catalogs (based on data that are $\sim0.7$ mag deeper in
the $z_{850}$ band than that used by Dickinson et al.\ 2004) and just
48 of these appear to be bona-fide $i$-dropouts (Table~6).  This
suggests that the majority of objects in the original Dickinson et
al.\ (2004) compilation were simply spurious sources.  A cursory
examination of these sources in the current ACS GOODS reduction bears
out this supposition.

From the HUDF, BSEM04 made the point that the cumulative surface
density of $i$-dropouts is only 0.1$\pm$0.1 arcmin$^{-2}$ to
$z_{850,AB}\sim26.5$.  While the present results roughly corroborate
this claim, we find a slightly higher density (0.30 arcmin$^{-2}$) to
the same bright limit in our corrected counts
(Table~11).  The current value is a bit lower than the
completeness corrected $0.5\pm0.2$ $i$-dropouts arcmin$^{-2}$ cited in
our earlier study on the RDCS1252-2927 + HDF-N fields (Bouwens et al.\
2003b), but this appears to have been the result of large scale
structure (B06a) and lensing by the prominent foreground cluster in
that study.  This surface density (0.30 arcmin$^{-2}$) also appears to
be consistent with the three-epoch estimate from the GOODS team, if we
assume the 75\% contamination fraction derived earlier (and apply a
small completeness correction).

\subsection{Is the Surface Density of $i$-dropouts in the HUDF Typical?}

The normalization of the $i$-dropout counts in a given field can show
large variations (e.g., 35\% rms for a single ACS field) depending on
the large scale structure (``cosmic variance'').  In \S3.6, we are
able to estimate the normalizations for our fields relative to the
large area GOODS fields.  One field that was of particular concern in
this analysis was the HUDF because (1) it provides our best constraint
on the number of faint $i$-dropouts and (2) it was selected to contain
one particularly bright $z_{850,AB}=25.0$ $i$-dropout.  Since rare
objects are typically associated with overdensities, one might have
expected the $i$-dropouts in the HUDF to be overdense relative to the
cosmic average, compromising any LF we might have determined using its
data.

In \S3.6, we show that this is not likely an important concern, and
that $i$-dropouts in the HUDF have a surface density that is just
$0.76\pm0.20$ times that of the two GOODS fields (and thus the HUDF may
even be \textit{underdense} relative to the cosmic average).
Nevertheless, one might have expected this to be a concern given the
recent findings by Malhotra et al.\ (2005) using the HUDF GRISM data.
Comparing the redshift distribution of $i$-dropouts they observed with
that obtained from their modelling, Malhotra et al.\ (2005) argued
that the HUDF contained a factor of $\sim2$ overdensity in the number
of $i$-dropouts at $z=5.9\pm0.2$ (15 of the total 23 $i$-dropouts).
At first glance, these results may seem contradictory to our own, but
one needs to remember that the Malhotra et al.\ (2005) measurement is
really just a comparison between the volume density of $i$-dropouts
inside the interval $z\sim5.9\pm0.2$ and that outside it.  Since the
comparison was made entirely within the area of the HUDF, it simply
provides us with information on the large-scale structure at $z\sim6$
along that line of sight.

\section{Luminosity Function}

The combined data from the HUDF, HUDF-Ps, and GOODS fields provide a
unique opportunity to derive the luminosity function at $z\sim6$ to
unprecedented depths and accuracy.  Such detail is important for
making accurate inferences about galaxy evolution and the reionization
of the universe.  It allows us to address questions about the
subsequent evolution of $UV$-bright galaxies to $z\sim3$, indicating
whether there has been evolution in $L_{*}$, $\phi^{*}$, or $\alpha$.
It also allows us to make reliable estimates of the UV background
produced by $z\sim6$ galaxies.  The UV background density is crucial
for assessing the impact of $z\sim6$ galaxies on reionization.

Estimating the LF would be straightforward if there were a simple way
of converting the observed fluxes $m$ to an absolute magnitude $M$
that was essentially independent of redshift.  Unfortunately, the
$z_{850}$-band fluxes are heavily attenuated by the forest and thus
conversions to absolute magnitude are highly dependent on the redshift
of the source (see Figure~7).  By contrast, our
infrared fluxes--while not highly affected by the forest--are of much
lower S/N and moreover are not available for many of our fields.  As a
result, our only recourse here is to use the $z_{850}$-band fluxes to
work back to the absolute magnitudes through a modelling of the
$i$-dropout redshift distribution.  To do this, we consider an
integral over the full redshift range in deriving the luminosity
function $\phi(M)$:
\begin{equation}
\int_z \phi(M(m,z)) P(m,z) \frac{dV}{dz} dz = N(m)
\end{equation}
where $\frac{dV}{dz}$ is the cosmological volume element, $m$ is the
apparent $z_{850}$-band magnitude, $N(m)$ is the number counts,
$P(m,z)$ is the selection function, and $M$ is the absolute magnitude
at 1350 $\AA$.  The absolute magnitude $M$ is a function of both the
apparent magnitude $m$ and redshift $z$.

\begin{figure}
\includegraphics[width=3.4in]{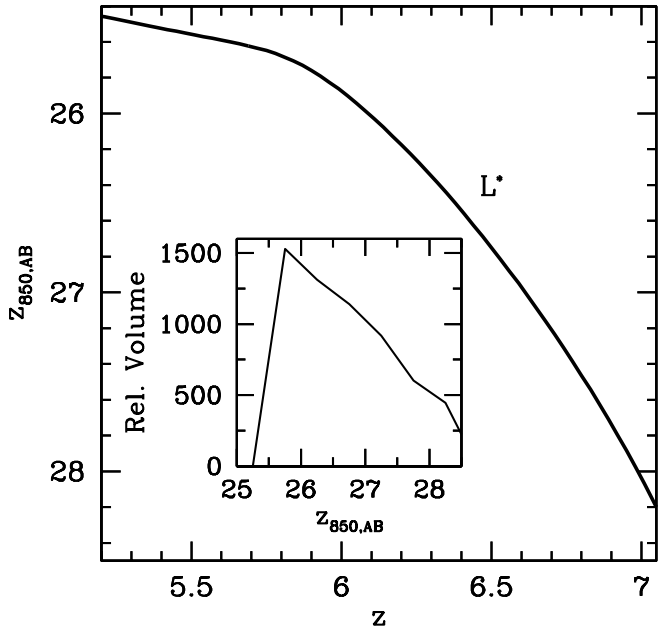}
\caption{The $z_{850}$-band magnitude vs. redshift (\textit{thick
solid line}) for objects of a fixed luminosity (here a $L_{z=3}^{*}$
galaxy).  Consequently, objects at a particular $z_{850}$-band
magnitude can correspond to a wide range of luminosities (e.g., a
$z_{850,AB}\sim27$ $i$-dropout would correspond to a $0.3L_{z=3}^{*}$
object at $z\sim5.5$ and a $2.5L_{z=3}^{*}$ object at $z\sim7$).  To
cope with this issue, we model the redshift distribution and integrate
the LF ($\phi_k$) over the relevant selection volume when fitting the
observed counts $N_m$ (Eq. 5).  One example of the effective kernel
$V_{m,k}$ (Eq. 4) used in these integrations is shown here in the
inset (for an object whose absolute magnitude corresponds to
$L_{z=3}^{*}$).  The effective kernels for other absolute magnitudes
are similar.  The vertical axis for the inset is in units of the
selection volume per unit area per unit magnitude (Mpc$^3$
arcmin$^{-2}$ mag$^{-1}$).}
\end{figure}

The selection function $P(m,z)$ can be estimated by projecting a
complete $B$-dropout sample from the HUDF (Bouwens et al.\ 2004b) to
$z\sim5-7$ and reselecting it using a similar procedure to that
described in \S3.1.  The projected $B$-dropout sample is assumed to
have a $UV$-continuum slope $\beta$ with mean $-2.0$ and $1\sigma$
scatter of 0.5, similar to our fits in \S3.4.  It also makes sense to
adopt a $(1+z)^{-1.1}$ size scaling (for fixed luminosity: \S3.7).
Motivated by the findings of Stanway et al.\ (2004a) and Dow-Hygelund
et al.\ (2006), we also assume that 25\% of the projected $B$-dropouts
have Ly$\alpha$ emission with an equivalent width of $30\AA$.  This
latter assumption provides a rough account for the bias introduced by
the current $(i_{775}-z_{850})_{AB}>1.3$ selection against galaxies
with strong Ly$\alpha$ emission at $z\sim5.5-5.9$ (Malhotra et al.\
2005; Figure 6 of Dow-Hygelund et al.\ 2006).  Since Ly$\alpha$
emission falls in the $i_{775}$ band for objects at these redshifts,
such objects will not readily show up as dropouts.  This reduces the
selection volume for $z\sim6$ galaxies by $\sim3$\%.  The selection
function we derive is shown in Figure~8.

\begin{figure}
\epsscale{0.7}
\includegraphics[width=3.2in]{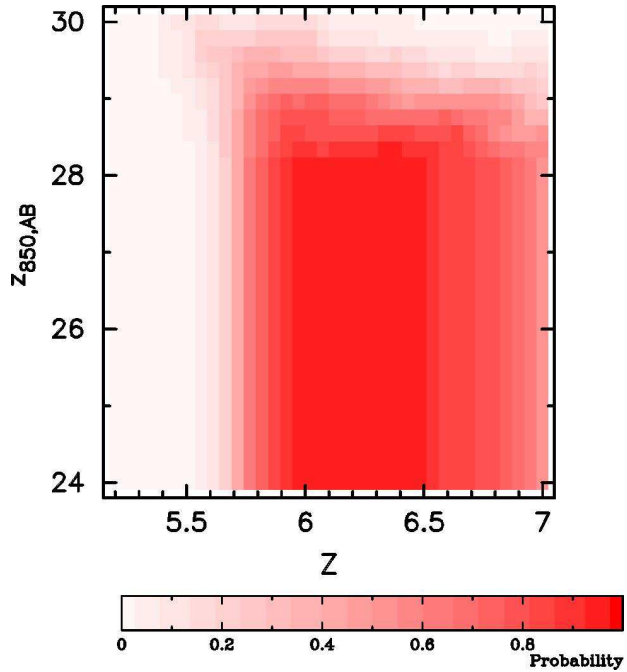}
\caption{Probability $P(m,z)$ that some object of apparent
$z_{850,AB}$-band magnitude and redshift $z$ is included in our HUDF
$i$-dropout sample.  This function was computed by projecting a HUDF
$B_{435}$-dropout sample (Bouwens et al.\ 2004b) to $z\sim6$ assuming
a $(1+z)^{-1.1}$ size scaling (for fixed luminosity: \S3.7).  Other
scalings [e.g., $(1+z)^{-1}$ or $(1+z)^{-1.5}$] yield only modest
differences with respect to the adopted selection function $P(m,z)$
and therefore only have a minor effect on the shape of the LF (e.g.,
$\Delta \alpha = \pm0.1$).  The rest-frame UV slopes $\beta$ of our
input sample are assumed to have a mean of $-2.0$, with a $1\sigma$
scatter of 0.5.  25\% of the objects are assumed to have a Ly$\alpha$
equivalent width of 30$\AA$ (Dow-Hygelund et al.\ 2006).  This
function does not include the small incompleteness ($\sim11-17$\%) due
to blending with foreground sources (Appendix
D1).}
\end{figure}

\subsection{Direct Method}

Here we present our primary determination of the rest-frame UV LF at
$z\sim6$.  We express the LF in terms of a set of stepwise functions
$\phi_k W(M-M_k)$ of half-magnitude width:
\begin{equation}
\phi(M) = \Sigma _k \phi_k W(M-M_k)
\end{equation}
where 
\begin{equation}
W(x) = 
\begin{array}{cc} 
0, & x < -1/4\\
1, & -1/4 < x < 1/4\\
0, & x > 1/4
\end{array}
\end{equation}
We then derive the coefficients on the stepwise function through a
maximum likelihood procedure, from a fit to the observed counts
(Table~11).  To simplify the computation, we derive
kernels $V_{m,k}$ that convert the luminosity function $\phi_k$ to
predicted counts:
\begin{equation}
V_{m,k} = \int _z \int _{m-1/4} ^{m+1/4} W(M(m',z) - M_k) P(m',z)
\frac{dV}{dz} dm' dz 
\end{equation}
With this definition, Eq. (1) reduces to
\begin{equation}
\Sigma _k \phi_k V_{m,k} = N_m
\end{equation}
where $N_m = \int _{m-1/4} ^{m+1/4} N(m') dm'$.  One example of the
kernel $V_{m,k}$ that appears in Eq. (4) is shown in
Figure~7.  Since our procedure here is essentially a
deconvolution of $N_m$ (to obtain $\phi_k$), the LF we derive will
have correlated errors.  The LF will also appear somewhat more
``noisy'' than the original counts.  As a result (and because of the
Poissonian noise in the observed counts at $z_{850,AB}\gtrsim27.5$),
we have enlarged the size of our faintest two bins ($M_{1350,AB}>-19$)
to be 1.0 mag in width.  The resulting LF is shown in
Figure~10 (see also Table~12) and extends
over 2 orders of magnitude: from 4 $L_{z=3}^{*}$ to 0.04
$L_{z=3}^{*}$.  Remarkably, this is fainter than what Steidel et al.\
(1999) was able to obtain at $z\sim3$ (where the limit was
$\approx0.1L_{z=3}^{*}$).  As a check on the current procedure, we
repeated it on the surface density predictions made in
Figure~5 (\textit{bottom}) based on the $z\sim3$ LF
(Steidel et al.\ 1999) and were able to recover the input LF.  For
context, we present the predicted redshift distribution for this LF
(and our HUDF $i$-dropout selection) in Figure~9.

\begin{deluxetable}{cc}
\tablewidth{2.5in}
\tabletypesize{\footnotesize}
\tablecaption{A stepwise determination of the $z\sim6$ rest-frame UV luminosity function (see also Figure~10).\tablenotemark{a}}
\tablehead{
\colhead{$M_{1350,AB}$} & \colhead{$\phi_k$ (Mpc$^{-3}$ mag$^{-1}$)}}
\startdata
$-21.94$ & 0.00001 $\pm$ 0.00001\\
$-21.44$ & 0.00007 $\pm$ 0.00004\\
$-20.94$ & 0.00012 $\pm$ 0.00007\\
$-20.44$ & 0.00033 $\pm$ 0.00012\\
$-19.94$ & 0.00128 $\pm$ 0.00030\\
$-19.44$ & 0.00313 $\pm$ 0.00118\\
$-18.69$ & 0.00332 $\pm$ 0.00115\\
$-17.69$ & 0.00771 $\pm$ 0.00211\\
\enddata
\tablenotetext{a}{Note that adjacent bins in our LF are not
independent [see Eq. (1) and Figure~7], and therefore
the errors on the individual bins include some covariance with their
neighbors.}
\end{deluxetable}

In addition to breaking up the LF in stepwise intervals, it has also
become conventional to parametrize it in terms of a Schechter function
(Schechter 1976).  Because of the degeneracies among the parameters
$\alpha$, $\phi^{*}$, and $M^{*}$, the results are expressed as
likelihood contours (Figure~11, \textit{blue solid lines}).  In
deriving these contours, we allowed $\alpha$ to extend to values as
steep as $-2$ to explore the broadest possible parameter space.  Even
though the luminosity density is formally divergent for such steep
values of the faint-end slope, it seems clear that the LF must cut off
at some physical scale and so the total light will not diverge.  We
evaluate the likelihood of different Schechter parametrizations by
calculating the equivalent values of $\phi_k$ for the parametrization
(by integrating the Schechter function over the full 0.5 mag interval
relevant for the considered $\phi_k$), comparing them with the
observed counts $N_m$ (Table~11) using Eq. (5), and then computing
$\chi^2$.  We have incorporated large-scale structure uncertainties
into these likelihood estimates by smoothing the $\chi^2$ likelihood
contours with a kernel that encapsulates the joint uncertainty in
$\alpha$, $M^{*}$, and $\phi^{*}$ arising from field-to-field
variations ($\sigma_{\alpha}=0.2$, $\sigma_{M^{*}}=0.14$,
$\sigma_{\phi}=0.0006$, and their internal correlations: see Appendix
E).  An additional $\sim20\%/\sqrt{2}\sim14\%$ uncertainty in
$\phi^{*}$ results from the expected field-to-field variations in the
$i$-dropout surface densities over the two GOODS fields (Somerville et
al.\ 2004; \S3.6).  To illustrate the effect of fixing the different
Schechter parameters at the $z\sim3$ values, green contours are
overplotted in Figure~11.  These results can be put in context by
comparing them with the equivalent $z\sim3$ determinations (Steidel et
al.\ 1999).  In the left two plots, we see evidence for lower
characteristic luminosities $M^{*}$ at $z\sim6$, with little change in
$\phi^{*}$ or $\alpha$.  Fainter values of $M^{*}$ are favored at
99.7\% confidence.  If we try to minimize changes in $M^*$, we can see
that our results favor steeper values for $\alpha$ at $z\sim6$.  Note
that scenarios, such as density evolution ($\phi^{*}$) which do not
include these changes in $M^{*}$ (toward fainter values) or $\alpha$
(toward steeper values) are excluded at $>$99.99\% confidence.  Our
most likely values for $\phi^{*}$, $M_{1350,AB}^{*}$, and $\alpha$ are
$2.02_{-0.76}^{+0.86}\times10^{-3}$ Mpc$^{-3}$, $-20.25\pm0.20$, and
$-1.73\pm0.21$, respectively.\footnote{We note that the best-fit
parameters are $M_{1350,AB}^{*} = -20.31\pm0.20$ and $\phi^{*} =
1.80_{-0.68}^{+0.77}\times10^{-3}$ Mpc$^{-3}$ if we express them using
the cosmological parameters
$(\Omega_{M},\Omega_{\Lambda},h)=(0.24,0.76,0.73$) preferred by the
one year WMAP measurements (Spergel et al.\ 2003).}  As illustrated
in Figure~10 (\textit{black line}), this fit is in good agreement with
the stepwise LF determined earlier (Table~12).  Because of the
proximity of the present faint-end slope to $-2$, where the integral
of the total light diverges, extrapolations to zero luminosity can be
somewhat uncertain.  A much more robust number is the total luminosity
density integrated to the approximate faint-end limit of the HUDF
($0.04 L_{z=3} ^{*}$): $1.77\pm0.45 \times 10^{26}\, \textrm{ergs}\,
\textrm{s}^{-1}\, \textrm{Hz}^{-1}\, \textrm{Mpc}^{-3}$.  This is
equal to 0.68 times, 0.50 times, and 0.24 times the luminosity density
integrated to zero assuming faint-end slopes $\alpha$ of $-1.6$,
$-1.7$, and $-1.9$, respectively.

\subsection{STY79 Method}

A more conventional way of deriving the LF (across multiple fields) is
to use the Sandage, Tammann, \& Yahil (1979, hereafter STY79) fitting
procedure.  This procedure has the advantage of being relatively
insensitive to large-scale structure.  Only the shape of the
luminosity function factors into the fits and not the normalization,
allowing one to derive extremely robust measures on the overall shape.
We do not use this procedure as our primary fitting procedure since
our degradation procedure (\S3.6) provides us with a slightly more
direct measure of the field-to-field variance (the STY79 approach may
be more sensitive to errors in our transfer functions).  However, we
show that the two results are in very good agreement, suggesting that
our overall result here is robust.

As with our primary approach, an important complication is the rather
inexact relationship between apparent and absolute magnitudes
(Figure~7).  This makes it more convenient to work in
terms of the apparent rather than absolute magnitudes.  Our procedure
then becomes one in which we are maximizing the likelihood of
producing the observed counts (here distributed over three different
fields) given a LF.  In detail, this approach really is not that
different from what we performed in \S3.6 to match up the counts from
our three different data sets, and so it should not be surprising that
the best-fit parameters we obtained from this procedure (i.e.,
$M^{*}_{1350,AB}=-20.28$ and $\alpha=-1.74$) and their likelihood
contours (Figure~12) are in good agreement with those
obtained with our primary methodology (Figure~11).
The $\phi^{*}$ we derive fixing the shape of the LF and fitting to the
number counts in the two GOODS fields (Figure~5,
\textit{middle}) is $1.94\times10^{-3}$ Mpc$^{-3}$ and also quite
consistent.

\begin{figure}
\epsscale{0.5}
\includegraphics[width=3.4in]{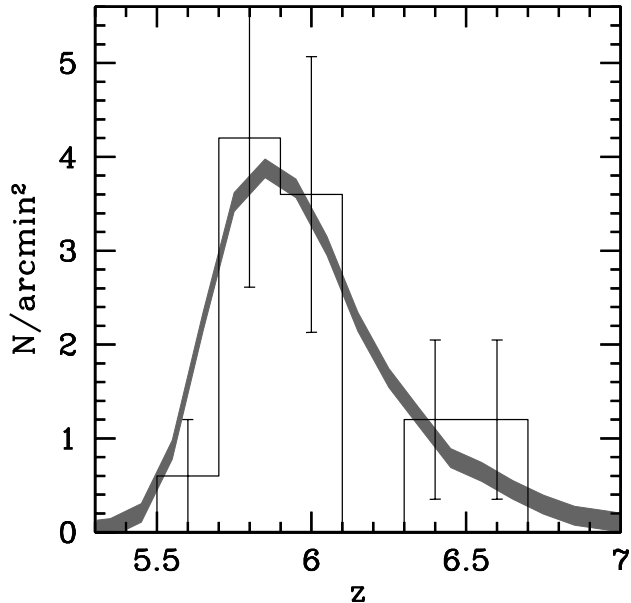}
\caption{Redshift distribution of $i$-dropouts in our HUDF selection
assuming the $z\sim6$ LF from Figure~10 and a mean
rest-frame UV continuum slope $\beta$ of $-2.0$ with a $1\sigma$
scatter of 0.5 (\textit{shaded gray region}).  Object profiles used in
the simulations were drawn from comparable luminosity HUDF
$B$-dropouts (Bouwens et al.\ 2004b) scaled in size as $(1+z)^{-1.1}$
(\S3.7).  Our predicted redshift distribution is in good agreement
with that obtained by Malhotra et al.\ (2005) for $i$-dropouts from
the HUDF [\textit{histogram}: clipped to include only objects with
$(i_{775}-z_{850})>1.3$ and with a vertical normalization scaled to
match our predictions].  This suggests that our model for the
rest-frame $UV$-colors of the $z\sim6$ galaxy population is
reasonable.  It may also indicate that Malhotra et al.\ (2005)
overestimated the size of the overdensity at $z\sim5.9\pm0.2$ in the
HUDF.
}
\end{figure}

\begin{figure}
\epsscale{1.0}
\includegraphics[width=3.4in]{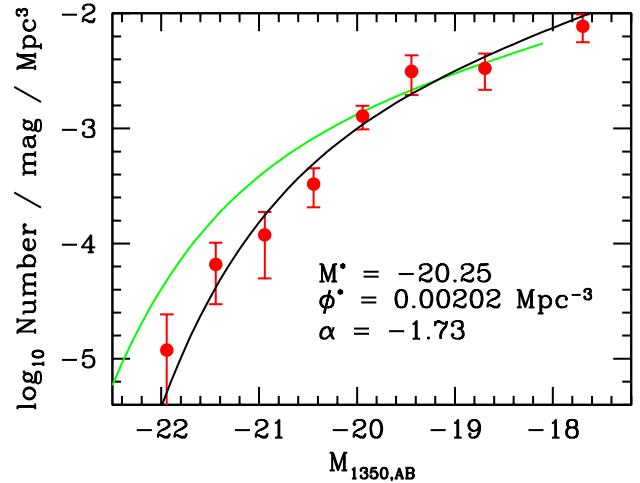}
\caption{Rest-frame continuum $UV$ ($\sim1350\AA$) LF estimated from
the HUDF, the HUDF-Ps, and the GOODS fields, shown in terms of the
best-fit stepwise parameterizations (\textit{red circles with}
$1\sigma$ \textit{errors}; see Table~12) and Schechter
function (\textit{black line}).  Because of the greater noise in the
$i$-dropout counts at $z_{850,AB}\gtrsim27.5$, the LF is binned on 1.0
mag intervals faintward of $M_{1350,AB}=-19$ (otherwise 0.5 mag
intervals are used).  Note that adjacent bins in our LF are not
independent [see Eq. (1) and Figure~7], and therefore
the errors on the individual bins include some covariance with their
neighbors.  The $z\sim3$ Steidel et al.\ (1999) LF (with the
$k$-corrected equivalent $M_{1350,AB}^{*}=-20.87$, see
Table~13) is shown for comparison (\textit{green line}) and
only plotted to its faint-end limit $0.1L_{z=3}^{*}$.  Amazingly, this
is brighter than the faint-end limit we were able to obtain at
$z\sim6$ (0.04 $L_{z=3}^{*}$).  Our $z\sim6$ LF shows a clear turnover
at the bright end relative to the $z\sim3$ LF and suggests that there
has been an evolution in the characteristic luminosity from $z\sim6$
to 3 ($\sim 0.6$ mag of brightening: see also
Figure~11).}
\end{figure}

\begin{figure*}
\includegraphics[width=7.0in]{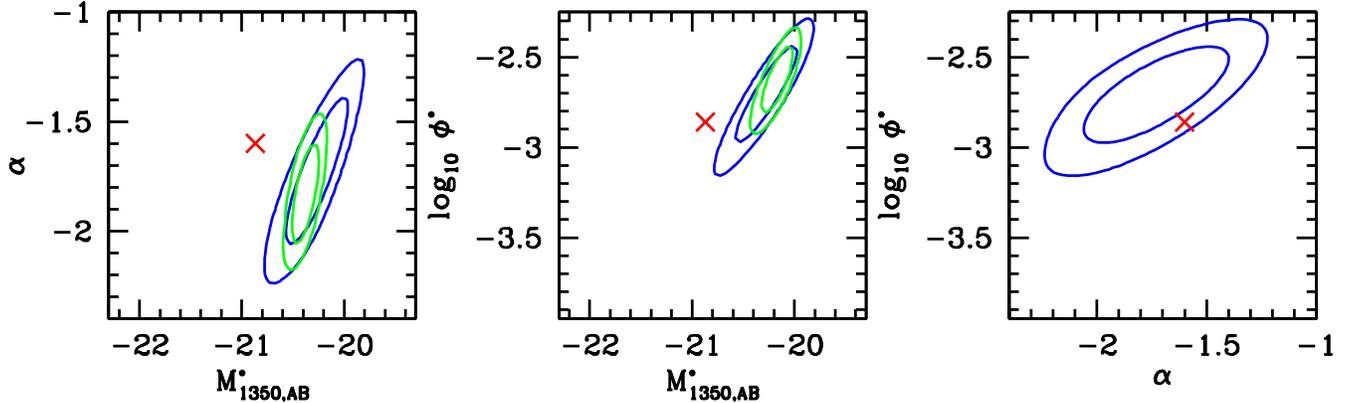}
\caption{Maximum likelihood Schechter parameters ($M^{*}$, $\alpha$,
and $\phi^{*}$) for the $z\sim6$ UV ($\sim1350\AA$) luminosity
function.  The inner and outer contours indicate the 68\% and 95\%
confidence regions, respectively.  The blue contours indicate the
confidence intervals after marginalizing across the third parameter in
the LF.  The green contours show these confidence intervals if no
change is allowed in this third parameter from $z\sim3$ (Steidel et
al.\ 1999).  The red cross indicates the parameters for the $z\sim3$
Steidel et al.\ (1999) LF shifted to 1350 $\AA$ rest frame.  Note that
even though the luminosity density is formally divergent for faint-end
slopes $\alpha<-2$, it seems clear that the LF must cut off at some
physical scale and so the total light will not diverge.  We considered
such steep slopes to explore the broadest possible parameter space.
The simplest way to accommodate the observed evolution is to shift the
characteristic luminosity $M^{*}$ by $\sim0.6$ mags (brightward) from
$z\sim6$ to 3 although an evolution in the faint-end slope $\alpha$
(from $-1.9$ at $z\sim6$ to $-1.6$ at $z\sim3$) can also help.  LFs
that do not include these changes (a fainter $M^*$ or a steeper
$\alpha$ at $z\sim6$) are excluded at $>$99.99\%
confidence.}
\end{figure*}

\begin{figure}
\epsscale{0.35}
\begin{center}\includegraphics[width=2.5in]{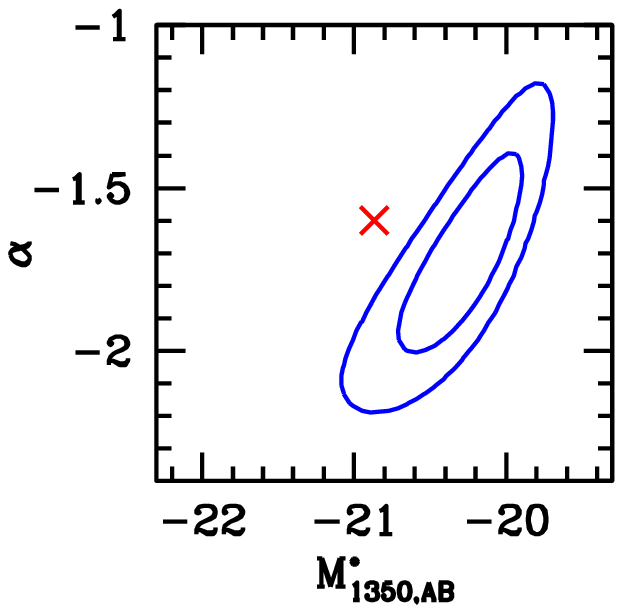}\end{center}
\caption{Maximum likelihood Schechter parameters ($M^{*}$, $\alpha$)
for the $z\sim6$ UV ($\sim1350\AA$) luminosity function.  This figure
is similar to the left-hand panel of Figure~11, but
using the maximum likelihood procedure of Sandage et al.\ (1979).  The
results here are in excellent agreement with those obtained with the
direct method (\S5.1:
Figures~10-11)}
\end{figure}

\subsection{Direct Method (without LSS correction)}

Finally, it is interesting to compute the $z\sim6$ LF but without any
correction for large-scale structure (``cosmic variance'').  Since
field-to-field variations (i.e., 35\% rms for a single ACS field:
\S3.6) are only slightly larger than our measurement errors on these
variations (the uncertainties on the normalization factors for the HUDF
are 26\% rms: see Table~10), the LF we derive ignoring
the normalization altogether (i.e., assuming each field is
representative of the cosmic average) should be fairly competitive
with our primary determination (\S5.1).  Meanwhile, differences we
observe relative to this determination can provide us with a good
sense for the representative errors.  Rederiving the LF with these
assumptions, we obtained the following best-fit Schechter parameters:
$\phi^{*}=1.76 \times 10^{-3}$ Mpc$^{-3}$, $M_{1350,AB}^{*}=-20.28$,
and $\alpha=-1.60$.  It is encouraging that these values are only
slightly different from those obtained from the two previous methods
(Figures~11 and 12).  In retrospect,
we might have expected this level of agreement from some simulations
we ran to assess the impact of cosmic variance on the derived LF
(Appendix E).

\subsection{Luminosity Densities}

Having obtained a basic fit to the observed LF, we can move on to look
at the UV continuum luminosity density and how it compares with
previous determinations at higher and lower redshift.  Because of the
limited sensitivies of the highest redshift probes (e.g., the Bouwens
et al.\ 2004c study at $z\sim7-8$ and the Bouwens et al.\ 2005 study
at $z\approx10$), we make these comparisons to two different
luminosity limits: 0.3 times and 0.04 times the characteristic
luminosity at $z=3$ (Steidel et al.\ 1999).  This is important to
properly account for possible evolution in the characteristic
luminosity $L^*$ or faint-end slope $\alpha$ with redshift.  To a
limiting magnitude of $0.3L_{z=3}^{*}$, the present LF integrates out
to $5.8 \pm0.9 \times 10^{25}\,\textrm{ergs}\, \textrm{s}^{-1}\,
\textrm{Hz}^{-1} \textrm{Mpc}^{-3}$.

To convert these UV luminosity densities into star formation rate
(SFR) densities (uncorrected for extinction), we assume a Salpeter IMF
and use the now somewhat canonical conversion factors of Madau et al.\
(1998):
\begin{equation}
L_{UV} = \textrm{const}\,\, \textrm{x}\,\, \frac{\textrm{SFR}}{M_{\odot} \textrm{yr}^{-1}} \textrm{ergs}\, \textrm{s}^{-1}\, \textrm{Hz}^{-1}
\end{equation}
where const = $8.0 \times 10^{27}$ at 1500 $\AA$.  Both the present
luminosity densities and SFR densities are shown in
Figure~13 relative to many previous determinations
(Steidel et al.\ 1999; Giavalisco et al.\ 2004b; Bouwens et al.\
2004a, 2004c, 2005; BSEM04; Schiminovich et al.\ 2005).  The fall off
in the luminosity density towards high redshift is much sharper at
brighter luminosities [$> 0.3 L_{z=3} ^{*}$;
$\rho(z=6)/\rho(z=3)=0.52\pm0.08$] than it is when integrated to $0.04
L_{z=3}^{*}$ [$\rho(z=6)/\rho(z=3)=0.82\pm0.21$].

\begin{figure}
\epsscale{0.85}
\includegraphics[width=3.7in]{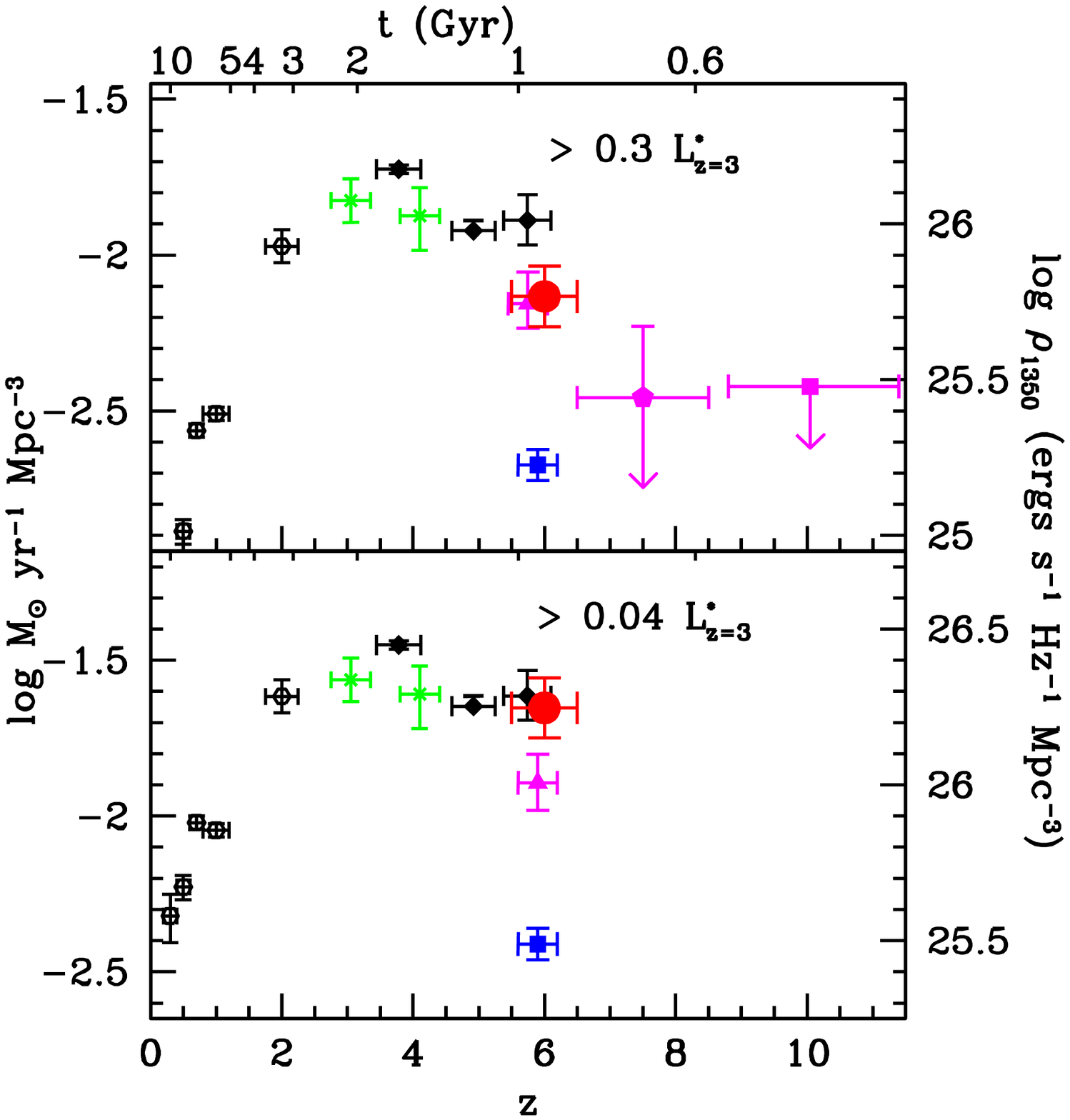}
\caption{Cosmic star formation history (uncorrected for extinction)
integrated to $0.3 L_{z=3} ^{*}$ (\textit{top panel}) and $0.04
L_{z=3} ^{*}$ (\textit{bottom panel}).  These luminosities correspond
to the faint-end limits for $z_{850}$ and $i_{775}$-dropout probes at
$z\sim7-8$ and 6, respectively.  The large red circle denotes the
current determination at $z\sim6$.  A Salpeter IMF was used to convert
the rest-frame continuum $UV$ luminosity density ($\sim1350\AA$) to a
SFR density.  For comparison, the previous determinations by
Schiminovich et al.\ (2005; \textit{open hexagons}), Steidel et al.\
(1999; \textit{green crosses}), Giavalisco et al.\ (2004b;
\textit{black diamonds}), Bouwens et al.\ (2004a; magenta triangle),
BSEM04 (\textit{blue square}), Bouwens et al.\ (2004c; \textit{magenta
pentagon}: shifted slightly to the left in the upper panel to avoid
confusion), and Bouwens et al.\ (2005; \textit{magenta square}) are
also included.  The age of the universe is plotted along the top.  The
plotted position of the BSEM04 $z\sim6$ determination is as quoted in
their paper (although our own fits to the BSEM04 counts yield values
$\sim2.7$ times higher; \S6.1).  The Giavalisco et al.\ (2004b)
$z\sim6$ determination appears to have been significantly affected by
contamination (\S4.1).  The figure is divided into two panels to
illustrate how much stronger the evolution is at the bright end of the
LF ($\gtrsim 0.3 L_{z=3}^{*}$) than it is when integrated to the
faint-end limit of the current probe
($0.04L_{z=3}^{*}$).}
\end{figure}

\section{Discussion}

The combination of the HUDF, HUDF-Ps, and GOODS datasets, especially the
very deep HUDF data, provides a unique opportunity to explore a number
of issues for $z\sim6$ galaxies.  These include refining our knowledge
of the rest-frame $UV$-continuum luminosity function, assessing the
impact of $z\sim6$ galaxies on the reionization of the universe, and
using $z\sim6$ as a baseline for assessing evolution to even higher
redshift.  Our analysis of the rest-frame $UV$-colors also permits us
to revisit the issue of a possible evolution in the $UV$-continuum
slope $\beta$.

\subsection{UV Continuum Luminosity Function}

One of the principal goals of this paper is to obtain an optimal
determination of the luminosity function in the rest-frame continuum
UV ($\sim1350\AA$).  The present approach has several important
advantages over several previous derivations (Dickinson et al.\ 2004;
Yan \& Windhorst 2004a, 2004b; Bouwens et al.\ 2004a; BSEM04; Malhotra
et al.\ 2005).  These include obtaining a self-consistent selection of
$i$-dropouts from three of the deepest, widest area data sets (GOODS,
HUDF-Ps, and HUDF); systematic use of the deeper ACS and infrared data
to derive completeness, flux, and contamination corrections; use of
the average UV continuum colors in our selection volume estimates; an
inclusion of the selection biases against strong Ly$\alpha$ emitters
in these same selection volume estimates; and a detailed matching-up
of the surface density of $i$-dropouts in our deeper fields with that
obtained in shallower, wider area fields to ensure a proper
normalization of the overall LF.

The current refinement to the $z\sim6$ LF puts us in a good position
to examine several previous determinations of this LF and the
associated claims for evolution from $z\sim3$ (Dickinson et al.\ 2004;
Yan \& Windhorst 2004a, 2004b; Bouwens et al.\ 2004a; BSEM04; Malhotra
et al.\ 2005).  A summary of many previous Schechter parameterizations
are given in Table~13 and plotted relative to the current
determination in Figure~14.  We divide this discussion between the
bright and faint ends of the LF.  At the bright end
($M_{1350,AB}\lesssim-21$), we find a substantial (factor of $\sim6$)
deficit relative to the $z\sim3$ LF.  This supports the initial
findings of Stanway et al.\ (2003, 2004b) and Dickinson et al.\
(2004).  Our current estimate for the number density of $i$-dropouts
at the bright end is slightly smaller than what we reported in two
previous studies (Bouwens et al.\ 2003b, 2004a).  In the first case
this was because of a substantial (factor of $\approx2$) overdensity
in the RDCS1252-2927 field relative to the cosmic average (\S4.1;
B06a) and in the second case it was because of slight ($\sim20$\%)
overestimates of the surface densities and completeness present in the
GOODS fields (Bouwens et al.\ 2004a).  The number density is also less
than reported by Yan \& Windhorst (2004b).  This appears to have been
due to their reliance on the three-epoch GOODS $i$-dropout catalog
(Dickinson et al.\ 2004) which, as we discuss earlier (\S4.1),
overestimates the surface density of $i$-dropouts.  Recent searches at
bright magnitudes ($z_{R,AB}<25.6$) with Subaru also find strong
($\approx11$ times) deficits at $z\sim6$ relative to $z\sim3$ values
(Shimasaku et al.\ 2005).

\begin{deluxetable*}{lccc}
\tabletypesize{\footnotesize}
\tablecaption{Determinations of the best-fit parameters for the rest-frame UV ($\sim1350\AA$) LF at $z\sim6$.\tablenotemark{a}}
\tablehead{
\colhead{Study} & \colhead{$M_{1350,AB} ^{*}$} & \colhead{$\phi^{*}$ (Mpc$^{-3}$)} & \colhead{$\alpha$}}
\startdata
This work & $-20.25\pm0.20$ & $0.00202_{-0.00076}^{+0.00086}$ & $-1.73\pm0.21$\\
Dickinson et al.\ 2004 & $-19.87$\tablenotemark{b} & 0.00527 & $-1.6$ (fixed)\\
Bouwens et al.\ 2004a & $-20.26$ & 0.00173 & $-1.15$\\
Bunker et al.\ 2004 & $-20.87$\tablenotemark{b} & 0.00023 & $-1.6$\\
Yan \& Windhorst 2004b & $-21.03$ & 0.00046 & $-1.8$\\
Malhotra et al.\ 2005 & $-20.83$ & 0.0004 & $-1.8$ (assumed)\\
\enddata
\tablenotetext{a}{Figure~14 provides a visual comparison
of these LFs.}
\tablenotetext{b}{Since the quoted LF was expressed in terms of the
$z\sim3$ LF (Steidel et al.\ 1999) which is at rest-frame $1700\AA$,
it was necessary to apply a k-correction (0.20 mag) to obtain the
equivalent luminosity at 1350 $\AA$ (calculated using the typical
colors of $z\sim3$ LBGs).}
\end{deluxetable*}

\begin{figure}
\epsscale{1.0}
\includegraphics[width=3.4in]{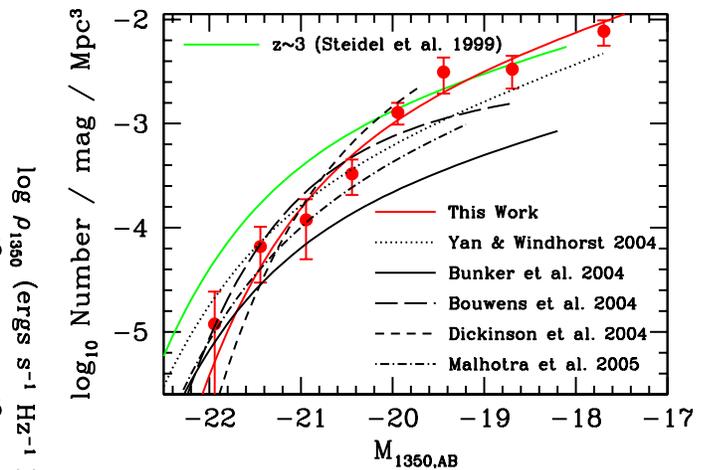}
\caption{Comparison of our rest-frame continuum UV ($\sim1350\AA$)
luminosity function (Figure~10; \textit{red line} and
\textit{red circles}) with that of others.  Included are the $z\sim6$
LFs of Yan \& Windhorst (2004; \textit{dotted black line}), BSEM04
(\textit{thin solid black line}), Bouwens et al.\ (2004a;
\textit{thick dashed black line}), Dickinson et al.\ (2004;
\textit{thin dashed black line}), and Malhotra et al.\ (2005;
\textit{dashed-dotted line}).  The $z\sim3$ Steidel et al.\ (1999) LF
shifted to 1350 $\AA$ rest-frame is shown for context (\textit{green
line}).  All the LF determinations are only plotted to their nominal
faint-end limits.  A compilation of the Schechter parameterizations of
the plotted LFs is provided in Table~13.  The careful and
comprehensive nature of the current analysis should make the present
determination of the $z\sim6$ LF the most robust
(\S6.1).}
\end{figure}

At fainter luminosities, the $z\sim6$ LF shows much better agreement
with $z\sim3$ than at the bright end.  This suggests evolution.  As
discussed earlier (\S5), the simplest way to accommodate these changes
is through an evolution of the characteristic luminosity (99.7\%
confidence).  Our best-fit is a $0.6\pm0.2$ mag brightening in
$M^{*}$.  An evolution of the faint-end slope $\alpha$ to $-1.9$ can
also help (from $-1.6$ at $z\sim3$: Steidel et al.\ 1999).  The latter
option echoes earlier claims made by Yan \& Windhorst (2004b) for a
steep faint-end slope ($\alpha=-1.8$) using data from the HUDF.
However such faint-end slopes do not appear to be required
(Figure~11).  The faint-end slope is nevertheless
steeper than the $\alpha=-1.15$ determined in our earlier work using
the HUDF-Ps (Bouwens et al.\ 2004a).  The shallower slope from that
study appears to have derived from the significantly lower surface
density of $i$-dropouts present in the HUDF-Ps ($\sim0.6$ times the
cosmic average: see Table~10).  Contrary to this work
(\S3.6), no attempt was made there to treat possible field-to-field
variations, and therefore the shape of the LF was affected.  The
Dickinson et al.\ (2004) determination, by contrast, was too high at
lower luminosities.  This appears to have been a consequence of their
substantial underestimate of the contamination rate (\S4.1).

Our determination also differs substantially from the best-fit LF of
BSEM04 (Figure~14), particularly at the faint end where
our LF is nearly a factor of $\sim$10 higher.  Since the derived
counts from BSEM04 are only slightly lower than those in our study
(Figure~5), how can the differences in the LF be so
large?  The volume element does not appear to be the culprit since the
BSEM04 no-evolution predictions from $z\sim3$ (Steidel et al.\ 1999)
closely match our own.  The only possible explanation appears to be
due to some peculiarity in the way that BSEM04 derived their best-fit
parameters.  From their figures 10 and 11, it would appear that BSEM04
conducted their fits ($\chi_r ^2$) on the \textit{cumulative} counts,
not the \textit{differential} counts.  If so, this would not be
appropriate as the data points in the \textit{cumulative} counts are
not independent.  Our own fits to their \textit{differential} counts
(Figure~5, \textit{blue circles}) yield
$M_{1350,AB}^{*}=-20.49$ and $\phi^{*}=0.00097$ assuming a fixed
$\alpha=-1.6$.  This fit gives a cumulative luminosity density to
their faint-end limit ($z_{850,AB}=28.5$) which is $\sim2.7$ times
higher than their optimal fit (a factor of $\approx6$ drop in
$\phi^{*}$ from $z\sim3$).

The Bunker et al.\ (2004) work excepted, there has been a growing
consensus among $z\sim6$ studies that the evolution in the UV LF at
high redshift occurs primarily at the bright end.  Shimasaku et al.\
(2005) made a similar argument based on a comparison of their bright
$i$-dropout search with those obtained at fainter magnitudes (Bouwens
et al.\ 2004a; Bunker et al.\ 2004; Yan \& Windhorst 2004b).  Such
luminosity-dependent trends would also partially explain the supposed
discrepancy (e.g., Trimble \& Schwanden 2005; Stanway et al.\ 2004b)
between several early $z\sim6$ results, in which different
evolutionary factors were quoted relative to no-evolution $z\sim3$
expectations (e.g., $\approx6$ by Stanway et al.\ 2003 vs. $\approx2$
by Bouwens et al.\ 2003b).  Although it was previously believed that
these differences might be due to uncertainties in the completeness
and contamination rates (Bouwens et al.\ 2003b; Stanway et al.\
2004b), it now appears that differences in the flux limit may have
played an equally important role.\footnote{In principle, comparisons
between the UV LF at $z\sim4-5$ and $z\sim3$ also inform our
understanding of high-redshift galaxy evolution.  Unfortunately,
studies have come to different conclusions.  Iwata et al.\ (2003) at
$z\sim5$ and Sawicki \& Thompson (2005) at $z\sim4$ found the
predominant evolution at the faint-end of their LFs, while Ouchi et
al.\ (2004) found this evolution at the bright end.}

It seems relevant to step back and look at the observed evolution in
the larger context of galaxy evolution.  What is remarkable about the
evolution we observe is that the characteristic luminosity of galaxies
in the UV shows a significant \textit{increase} over the range
$z\sim6$ to 3.  This is in contrast to the strong \textit{decrease}
observed from $z\sim2$ to 0 (Arnouts et al.\ 2005; Gabasch et al.\
2004) and suggests that galaxy formation is a very different process
early on than it is at much later times.  At early times, it seems
reasonable to imagine that this increase in luminosity we observe is
just a simple consequence of the merging and coalescence of galaxies
expected in hierarchical scenarios.  The fact that this does not occur
at later times suggests that something must halt this growth and even
turn it around.  Although we discuss it no further, two promising
explanations for this turn-around include active galactic nucleus
(AGN) feedback (e.g., Scannapieco et al.\ 2005; Croton et al.\ 2005;
Granato et al.\ 2004; Scannapieco \& Oh 2004; Binney 2004; Di Matteo
et al.\ 2005) and the transition from cold to hot flows (e.g.,
Birnboim \& Dekel 2003).

In light of the likely relationship between the luminosity evolution
observed and the evolution of the mass function, it makes sense to
examine this connection briefly.  Figure~15 presents
the mass function at $z\sim3$ and 6 calculated from the Sheth \&
Tormen (1999) formalism and a $\Lambda$CDM power spectrum (Bardeen et
al.\ 1986) with $\sigma_8=0.9$, $\Omega_b = 0.045$, $\Omega_{M}=0.3$,
$\Omega_{\Lambda}=0.7$, and $H_{0}=70\,\textrm{km/s/Mpc}$.  A
horizontal blue arrow is overplotted to indicate the approximate mass
range of dropouts which make up the current LF (e.g., Cooray 2005).
Two aspects are evident in the evolution of the mass function: (1)
halos of a given density are $\sim2-3$ times more massive at $z\sim3$
as at $z\sim6$ and (2) the slope of the mass function becomes
shallower with time ($\Delta \alpha = 0.27$ from $z\sim6$ to
$z\sim3$).  The first change is very similar to $\sim0.6$ mag (factor
of $\approx2$) brightening of the LF observed here.  The second
change--this trend toward shallower faint-end slopes--is less clear
from current data (cf. Yan \& Windhorst 2004b), but will almost
certainly be tested in the near future.  Similarities between the
observed evolution and predictions for the mass function suggests that
we are actually observing hierarchical growth over the range $z\sim6$
to 3 (see e.g., Cooray 2005 and Night et al.\ 2005 for more
sophisticated treatments).

\begin{figure}
\epsscale{0.95}
\includegraphics[width=3.4in]{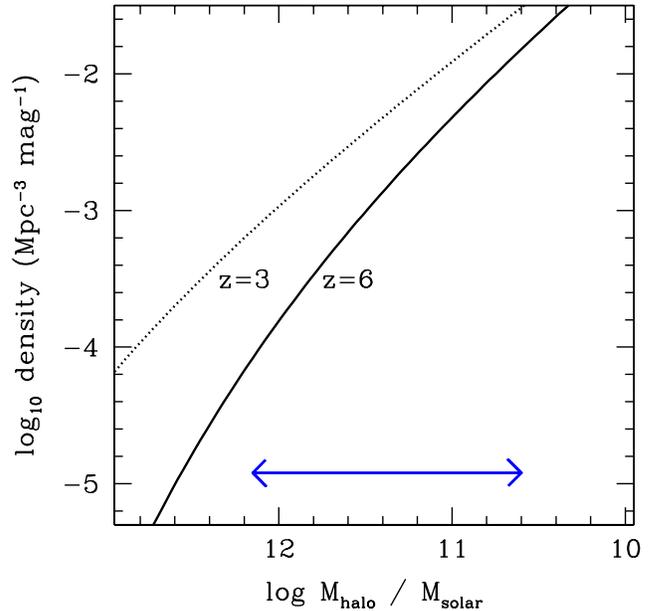}
\caption{Mass function (comoving volume density) at $z\sim3$
(\textit{dotted line}) and $z\sim6$ (\textit{solid line}) calculated
using the Sheth \& Tormen (1999) formalism and a $\Lambda$CDM power
spectrum (Bardeen et al.\ 1986) with $\sigma_8 = 0.9$, $\Omega_b =
0.048$, $\Omega_M = 0.3$, $\Omega_{\Lambda}=0.7$, and $H_0 =
70\,\textrm{km/s/Mpc}$.  The horizontal blue arrow provides the likely
mass range for $i$-dropouts in our sample (e.g., Cooray 2005).
Besides an obvious evolution toward higher masses at later times
(factor of $\approx3$ change), the mass function is also expected to
flatten ($\Delta \alpha = +0.27$ from $z\sim6$ to
3).}
\end{figure}

\subsection{Rest-frame UV colors}

The present sample also allowed us to place constraints on the mean
redshift and rest-frame UV slope $\beta$.  We obtained these
constraints using the measured optical-infrared colors for specific
$i_{775}$-dropouts from the HUDF (Table~4).  A
comparison of our measured colors with those obtained in two previous
studies (Stanway et al.\ 2005; Yan \& Windhorst 2004b) shows no large
systematic differences, but considerable scatter ($\pm0.15$ mag) for
individual objects.  The scatter becomes even larger ($>0.4$ mag) in
cases of possible blending with foreground objects.  Relative to
previous measurements, we would expect our measurements to represent a
modest improvement given our use of more optimized scalable apertures
(thus avoiding most blending problems) and careful aperture
corrections.

Despite no large systematics relative to previous measurements of the
colors, the mean $\beta$ inferred in this study is $-2.0$, which is
redder than the $\beta=-2.2$ inferred in the Stanway et al.\ (2005)
study based on the same data.  The principal reason for the difference
here is that current inferences are based on the $J-H$ colors while
previous inferences were based on the $z-J$ colors.  Since the $z-J$
colors are highly influenced by the redshift of a source and moreover
can be quite insensitive to rest-frame UV color (see B06a), it is
better to use the $J-H$ colors to determine the rest-frame $UV$ slope.
The $z-J$ colors are also more sensitive to errors in image alignment,
errors in the aperture corrections, and uncertainties in the optical
to infrared zero points.  Therefore, we consider the present
determination to be an improvement on the Stanway et al.\ (2005)
estimate (though current uncertainties in the zero points may make all
present measures somewhat uncertain, i.e., $\Delta \beta = \pm0.3$:
\S2.1).

Irrespective of the exact $\beta$, the mean rest-frame UV slope
observed at $z\sim6$ is bluer than that observed at $z\sim3$.  This
evolution is consistent with a number of recent studies (Lehnert \&
Bremer al.\ 2003; Kneib et al.\ 2004; B06a; Bouwens et al.\ 2004c;
Schaerer \& Pell\'{o} 2005; Yan et al.\ 2005; cf. Ouchi et al.\ 2004)
and point towards a lower mean dust extinction at higher redshift.
Changes in age, metallicity, and the IMF have a much smaller effect on
the rest-frame UV slope (e.g., Schaerer 2003; Leitherer et al.\ 1999).
Moreover, a significant contribution from Ly$\alpha$ to the $J_{110}$
flux seems unlikely given constraints from emission line searches at
$z\sim6$ (Ajiki et al.\ 2003; Kodaira et al.\ 2003; Hu et al.\ 2004;
BSEM04; Stanway et al.\ 2004a; Nagao et al.\ 2004; Dow-Hygelund et
al.\ 2006).  This leaves an evolution in the dust content as the most
natural way of explaining this change (see also the discussion in
B06a).

One obvious consequence of this lower dust extinction is an evolution
in the correction factor applied to the SFR densities inferred
directly from the UV luminosity function (see also B06a; B06b; Stanway
et al.\ 2005).  A convenient way of estimating the effect of this
change is through the Meurer et al.\ (1999) fit relating the
extinction $A_{1600}$ to the UV slope $\beta$: $A_{1600} = 4.43 +
1.99\beta$.  Although there is some uncertainty in the exact value of
$\beta$ at $z\sim6$ (and $z\sim3$), it is useful to adopt some
fiducial value of $\beta$ to estimate the size of the effect.  Taking
$\beta$ to equal $-2.0$ at $z\sim6$ and $-1.5$ at $z\sim3$ (Adelberger
\& Steidel 2000) suggests a doubling of the attenuation factor at
$1600\AA$ from $z\sim6$ to $z\sim3$.  Since this is the same direction
as the evolution of the UV LF, it appears that the real evolution
(after correction for extinction) may be large indeed.  So, instead of
the factor of 2 increase in the characteristic luminosity from
$z\sim6$ to 3 inferred in Figure~11, the real evolution in this
quantity may be as large as a factor of $\sim4$ increase after
correction for extinction.  It may also suggest that the total SFR
(and UV luminosity) density at $z\sim6$ (after correction for
extinction and integrated to $0.04L_{z=3}^{*}$) is just $\sim0.3$
times the value at $z\sim3$ (instead of the 0.82 factor given in
\S5.4).  We have included a simple illustration of this effect in
Figure~16 using several representative determinations of the UV
luminosity density (Steidel et al.\ 1999; Schiminovich et al.\ 2005).
The dust corrections we have applied here are $\sim2$ times (0.4 mag)
at $z\sim6$ and otherwise from Schiminovich et al.\ (2005).  As is
apparent from the figure, such changes have wide-range implications
and indicate a much more rapid rise in the corrected star formation
history from $z\sim6$ to $z\sim3$ than in the uncorrected SFR density.
Clearly, it will be important to confirm this change with other
methods (e.g., by using stacked X-ray fluxes: Reddy \& Steidel 2004;
Lehmer et al.\ 2005).

\begin{figure}
\epsscale{0.95}
\includegraphics[width=3.4in]{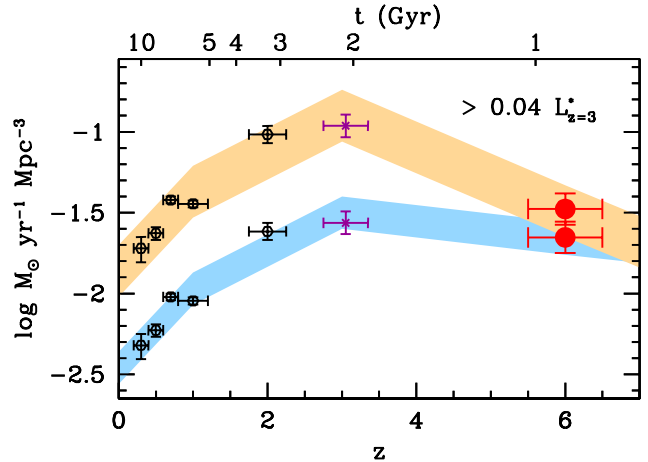}
\caption{Cosmic star formation history integrated to $0.04 L_{z=3}
^{*}$.  This history is shown both with and without extinction
corrections (\textit{upper and lower points}, respectively).  This is
also indicated by the orange and light blue shaded regions,
respectively, where the width here gives the suggested uncertainties
(see Schiminovich et al.\ 2005).  Similar to Figure~13,
we use the Madau et al. (1998) prescription to convert luminosity
densities into SFR densities.  Included are the determinations of
Schiminovich et al.\ (2005; \textit{hexagons}), Steidel et al.\ (1999;
\textit{dark magenta crosses}), and the present $z\sim6$
determinations (\textit{red circles}).  The extinction corrections we
apply at low redshift ($z\lesssim3$) are $\sim1.4$ mag and are
intermediate between the high and low estimates given in Schiminovich
et al.\ (2004; i.e., 1.8 and 1.0, respectively).  The extinction
correction we infer at $z\sim6$ is a significantly smaller $\sim0.4$
mag (from the Meurer et al.\ 1999 prescription).  Evolution in the
extinction correction over the range $z\sim3-6$ appears to have a
substantial impact on the cosmic star formation history
(\S6.2).}
\end{figure}

As we conclude this section, perhaps we should not be surprised by
this evolution in the rest-frame UV slope or the dust extinction.
Given the strong correlation between the total SFRs and the dust
extinction (Wang \& Heckman 1996; Martin et al.\ 2005; Adelberger \&
Steidel 2000), we might have expected the extinction to be lower at
the highest redshifts.  The mass scales are expected to be lower
there, and as we have observed, so are the typical UV luminosities and
apparent SFRs.

\subsection{Reionization of the Universe}

In light of the observational evidence that $z\sim6$ marks the end of
the reionization epoch (Becker et al.\ 2001; Fan et al.\ 2002; White
et al.\ 2003), it has become common to use the observed $i$-dropouts
to comment on the possible reionization of the universe by photons
arising from galaxies (e.g., Stanway et al.\ 2003, 2004b; Lehnert \&
Bremer 2003; Bouwens et al.\ 2003b; Giavalisco et al.\ 2004b;
Dickinson et al.\ 2004; BSEM04; Stiavelli et al.\ 2004b; Yan et al.\
2004a,b).  An estimate of the star formation rate necessary to produce
this reionizing flux can be made using the convenient formulation of
Madau et al.\ (1999) modified to match the baryon density derived from
the one-year \textit{WMAP} results (Spergel et al.\ 2003) and
shifted to $z\sim6$ (Bouwens et al.\ 2003; BSEM04):
\begin{equation}
\dot{\rho_{*}} \approx (0.052 \,\sfrd)\,\left(\frac{0.5}{f_{\rm
esc,rel}}\right)\,C_{30} \left(\frac{1+z}{7}\right)^{3}.
\end{equation}
where $\dot{\rho_{*}}$ is the SFR density, $C_{30}$ is the $\HI$
clumping factor $\left\langle \rho_{\HI}^2
\right\rangle$/$\left\langle \rho_{\HI}\right\rangle^2$ / 30, and
$f_{\rm esc,rel}$ is the relative fraction of ionizing radiation
escaping into the intergalactic medium to that escaping in the
$UV$-continuum ($\sim1500\AA$).  Unfortunately, current constraints on
the total SFR $\dot{\rho_{*}}$ still remain poor.  Although an
integration of our best-fit LF to our faint-end limit and zero
luminosity yields 0.022 and 0.043 $\,\sfrd$, respectively (somewhat
smaller than the fiducial SFR needed), current constraints also allow
for substantially steeper values of the faint-end slope (e.g.,
$\alpha\sim-1.9$: Figure~11).  Such slopes would
nearly double the value of $\dot{\rho_{*}}$ and hence be sufficient to
reionize the universe in this formulation.  Of course, it is also true
that physical constraints become important for some faint-end slope
(given limits on the total stellar mass or metals produced, e.g.,
Madau et al.\ 1998, Stiavelli et al.\ 2004a).

Despite current refinements to the $z\sim6$ UV continuum LF, there
continue to be substantial uncertainties in the role that $z\sim6$
galaxies play in reionizing the universe.  Indeed, we should not
forget that we still do not have a direct measure of the ionizing
radiation escaping into the intergalactic medium (IGM) and are forced
to rely on a proportionality factor, called the relative escape
fraction, to convert the observed rest-frame continuum-UV flux into an
ionizing flux.  While most attempts to measure this escape fraction at
$z\lesssim3$ have thusfar only obtained upper limits (i.e., $<0.1$ to
$<0.4$) (Leitherer et al.\ 1995; Hurwitz et al.\ 1997; Deharveng et
al.\ 2001; Giallongo et al.\ 2002; Fern{\' a}ndez-Soto et al.\ 2003;
Malkan et al.\ 2003; Inoue et al.\ 2005), there have been other
notable efforts (e.g., Steidel et al.\ 2001) that have obtained much
larger values ($\gtrsim0.5$).  The situation remains somewhat
controversial.  As a result of these and other uncertainties (e.g.,
Stiavelli et al.\ 2004b), there has been a wide range of different
claims regarding the capacity of galaxies to reionize the universe.
Some authors have claimed that the observed galaxies are not
sufficient to reionize the universe (BSEM04) while others have claimed
that they are, either because of a higher ionizing efficiency
(Stiavelli et al.\ 2004b) or because of a large contribution from
lower luminosity galaxies at the faint end of the luminosity function
(Yan \& Windhorst 2004a, 2004b).  This study (with its more rigorous
and detailed matching up of the different surveys) provides an
important confirmation and extension of this latter result, although
it is not yet clear that the faint-end slope is unusually steep (i.e.,
$\alpha\lesssim-1.8$: as argued by Yan \& Windhorst 2004a, 2004b).
This being said, we would like to reemphasize the considerable
uncertainties present at this stage and how little knowledge we have
about how the escape fraction might behave, both in its redshift and
in its luminosity dependence.  Better constraints will be available
when we are able (1) to better characterize the escape fraction and
(2) to look at the ionizing flux of $z\gtrsim3$ objects more directly
(as one might obtain through proximity studies).

\subsection{Implications for $z_{850}$-dropout Samples}

Our redetermination of the $z\sim6$ LF allows us to remark on recent
$z\sim7.5$ $z_{850}$-dropout samples selected from the HUDF.  There
are two recent samples that are relevant: the Yan \& Windhorst (2004b)
sample and the Bouwens et al.\ (2004c) sample.  Yan \& Windhorst
(2004b) performed a shallow search for $z$-dropouts
($J_{110,AB}\lesssim26.6$) and found only one candidate, which was
just on the edge of their selection window.  Since they predicted 2.9
candidates to a similar limit from their $i$-dropout LF assuming
no-evolution, they interpreted this as a tentative indication for the
onset of galaxy formation at $z\sim6$.  Performing a much deeper
search ($H_{160,AB}<27.5$) on the same field, Bouwens et al.\ (2004)
found five such $z_{850}$-dropout candidates, four of which they
assume to be real in their fiducial estimates (at least one was
considered to be spurious due to the aggressive nature of the search).
Comparing this with the 14 $z_{850}$-dropouts predicted assuming
no-evolution from $z\sim4$, this appeared consistent with a factor of
$\sim3-5$ drop in the number (luminosity density) of $UV$-bright
objects from $z\sim4$ to 7.5.

It is relevant to revisit these issues using the $z\sim6$ LF.
Comparisons can be made by projecting the present $i$-dropout sample
to $z\sim6-9$ using our cloning software, adding it to the data and
then reselecting it.  In simulating the profiles of specific
$i$-dropouts in our LF, we use scaled versions of specific
$i$-dropouts from the HUDF matched to the current $i$-dropout LF
(Appendix F of B06a details an anologous modelling of $U$-dropouts
using the HDF profiles).  Running through this procedure, 0.8
$z_{850}$-dropouts are expected to $J_{110,AB}\sim26.6$ versus the one
found, and 6.6 $z_{850}$-dropouts are expected to $H_{160,AB}\sim27.5$
versus the four fiducial candidates.  This suggests that there has
only been a modest increase in the number of bright objects from
$z\sim7-8$ to 6, although the uncertainties are large due to small
number statistics, cosmic variance, and some questions about the
$z_{850}$-dropout candidates themselves (items 2b, 2c, and 2f from
Bouwens et al.\ 2004c).

\section{Conclusions}

We have compiled a sample of 506 $i$-dropouts ($z\sim6$ galaxies) from
the HUDF, the HUDF parallel ACS fields (HUDF--Ps), and the GOODS fields
(316 arcmin$^2$), the latter enhanced by the ACS supernova search data
(extending the depth of the ACS $i$ and $z$-data by 0.2 and 0.4 mag,
respectively, to the depth of the ver. 2.0 GOODS release).  This
statistically robust sample consists of 122, 68, and 332 galaxies,
respectively, from the three aforementioned fields and includes
objects as faint as $z_{850,AB}\sim29.5$.  Note that 16 of these
$i$-dropouts appear in more than one of the above samples.  The
current sample of 506 galaxies represents the most comprehensive and
robust compilation to date, and is a significant advance over the
$\sim$50 and $\sim$100 object sample assembled by Bunker et al.\
(2004) and Yan \& Windhorst (2004b) over the HUDF, the 30 object
sample obtained by Bouwens et al.\ (2004a) over the HUDF-Ps, and the
251 object sample that Dickinson et al.\ (2004) compiled from GOODS,
although the latter sample is largely composed of contaminants
($\sim$75\%: see \S4.1).

\begin{deluxetable}{cc}
\tablewidth{2.6in}
\tablecolumns{2}
\tabletypesize{\footnotesize}
\tablecaption{Properties of $z\sim6$ Galaxies.}
\tablehead{
\colhead{Parameter} & {Value}}
\startdata
$r_{hl}$ (at $z_{850,AB}\sim27$) & $\sim$0.8 kpc ($\sim0.14''$)\\
Size-Redshift Scaling & $(1+z)^{-1.1\pm0.3}$\\
UV slope $\beta$ & $-2.0 \pm 0.3$ \\
$\phi^{*}$ & $0.00202_{-0.00076}^{+0.00086}$ Mpc$^{-3}$ \\
$M_{1350,AB}^{*}$ & $-20.25\pm0.20$ \\
$\alpha$ & $-1.73\pm0.21$ \\
${\cal{L}}_{1350} (>0.3L_{z=3}^{*})$ & $5.8\pm0.9\times10^{25}\,\textrm{ergs/s/Hz/Mpc}^3$\\
${\cal{L}}_{1350} (>0.04L_{z=3}^{*})$ & $1.77\pm0.45\times10^{26}\,\textrm{ergs/s/Hz/Mpc}^3$\\
\enddata
\end{deluxetable}

We select these galaxies using the well--established dropout
technique, with an $i$--dropout criterion
[$(i_{775}-z_{850})_{AB}>1.3$, $(V_{606}-z_{850})_{AB}>2.8$] and
demonstrate that the contamination levels on our selection are
$\lesssim8$\% (i.e., $\gtrsim92$\% are real: Appendix D4).

Contamination is a potentially serious concern for dropout samples.
We gave particular attention to four sources of contamination:
intrinsically red low--redshift galaxies, stars, spurious sources, and
low--redshift galaxies scattering into the selection region
(photometric scatter).  We established the contamination levels by
performing object selection in degraded versions of the deepest
fields, and used the deep (NICMOS) and wide-area (ISAAC) infrared
images of these fields.  As we discuss in Appendix D4.1, red galaxies
only appear to be a significant source of contamination
($18_{-9}^{+13}$\% of our $i$--dropout candidates) at bright
magnitudes ($25<z_{850,AB}<26$) and again possibly at the faintest
magnitudes ($10_{-5}^{+8}$\%: $z_{850,AB}\gtrsim28$).  Contamination
from photometric scatter is also small ($<$10\%) and only important
near the faint-end limit.  Contamination from stars is uniformly low
at all magnitudes ($\lesssim3$\%; after filtering out the few obvious
bright cases), while that from spurious sources is insignificant.
Overall, the present $i$--dropout catalogs are extremely clean
($\lesssim8$\% contamination).

An optimal measure of the $i$-dropout surface densities over a 5 mag
range ($24.5<z_{850,AB}<29.5$) is determined from our three samples
(HUDF, HUDF--Ps, enhanced GOODS).  Detailed degradation experiments are
made on our deeper data sets in order to understand object selection
and photometry in our shallower fields and to derive completeness,
flux and contamination corrections.  These corrections are applied to
establish a common baseline across all data sets.  To remove the
effects of large-scale structure (we expect $\sim$35\% rms variations
in the surface density of $i$-dropouts over 11.3 arcmin$^2$ ACS
fields) when combining our three $i$-dropout samples, we carefully
match up the surface density of $i$-dropouts in the deeper HUDF and
HUDF-Ps probes to that found in the two enhanced wide-area GOODS fields
(\S3.6).  The HUDF and HUDF--Ps fields are underdense ($0.77\pm0.20$
times and $0.67\pm0.16$ times) relative to the cosmic average defined
by the two GOODS fields (Table~10).

Finally, we use our derived surface densities to calculate a
rest-frame UV LF at $z\sim6$, and compared this LF with lower redshift
($z\sim3$) LFs.  Quantitative estimates of both the sizes and UV
colors of objects in our samples are used to estimate accurate
selection volumes for the derived LF.  Our principal findings are
summarized in Table~14 and are as follows:

\textit{Galaxy sizes}: Typical $i$-dropouts at $z_{850,AB}\sim27$
(from the HUDF-Ps and HUDF) have PSF-corrected half-light radii of
$\sim0.8$ kpc or $\sim0.14''$ (Figure~6: \S3.7).  By
comparing the observed sizes of $i$-dropouts from our three different
data sets with that predicted using different scalings of a $z\sim2.5$
$U$-dropout sample (B06a), we make inferences about how the physical
sizes of galaxies depend on redshift (for fixed luminosity).  Our
best-fit is a $(1+z)^{-1.1\pm0.3}$ scaling, which is in good agreement
with several previous determinations (Ferguson et al.\ 2004; Bouwens
et al.\ 2004a,b; B06b).

\textit{Rest-frame UV colors / UV-to-total SFR correction:} By
modelling the $z-J_{110}$ and $J_{110}-H_{160}$ colors of $i$-dropouts
from the NICMOS HUDF (see \S3.3), we construct models for the
rest-frame UV colors of the $i$-dropout population and the redshift
distribution (which we find peaks at $z\lesssim6$ -- see also Malhotra
et al.\ 2005).  The mean rest-frame UV spectral slope $\beta$ we infer
is $-2.0$.  This is bluer than the $-1.5$ observed at $z\sim3$
(Adelberger \& Steidel 2000), but redder than the Stanway et al.\
(2005) estimate of $\beta=-2.2$ at $z\sim6$ using the same data.  A
similar evolution from bluer spectral slopes has already been noted in
a number of other high-redshift ($z>3$) studies (Lehnert \& Bremer
2003; Kneib et al.\ 2004; B06a; Bouwens et al.\ 2004c; Schaerer \&
Pell\'{o} 2005; Yan et al.\ 2005; cf. Ouchi et al.\ 2004).  The most
natural explanation for this evolution is an increase in the dust
content at later cosmic times.  The most salient implication of such
an evolution is the effect it would have on the inferred SFR
densities.  Using the Meurer et al.\ (1999) relation between the $UV$
slope $\beta$ and the extinction $A_{1600}$, we estimate this factor
(at $1600\AA$) to change (in linear units) from $\sim2$ at $z\sim6$ to
$\sim4$ at $z\sim3$.

\textit{Luminosity Function:} Using the surface densities of
$i$-dropouts from our data sets and a computed selection function
$P(m,z)$, we derive the rest-frame continuum UV ($\sim1350\AA$) LF at
$z\sim6$ from $4 L_{z=3}^{*}$ to $0.04 L_{z=3}^{*}$
($M_{1350,AB}\sim-17.5$: see \S5).  This is fainter than Steidel et
al.\ (1999) was able to obtain at $z\sim3$ ($0.1L_{z=3}^{*}$).  The
likelihood parameters we derive for different Schechter
parameterizations suggest that there has been an increase in the
characteristic luminosity $M_{1350,AB} ^{*}$ from $z\sim6$ to 3 (99.7\%
confidence).  The best-fit is a $0.6\pm0.2$ mag brightening.  This
evolution in $M^{*}$ can be partially offset by changes in the
faint-end slope $\alpha$ (from $-1.9$ at $z\sim6$ to $-1.6$ at
$z\sim3$; Steidel et al.\ 1999).  Scenarios, such as density evolution
($\phi^{*}$), which do not include this evolution in $M^{*}$ or
$\alpha$ are excluded at $>$99.99\% confidence, demonstrating quite
significantly that galaxies at $z\sim6$ have lower luminosities (on
average) than galaxies at $z\sim3$.  The best-fit Schechter parameters
are $M_{1350,AB}^{*} = -20.25\pm0.20$, $\alpha = -1.73\pm0.21$, and
$\phi^{*} = 2.02_{-0.76}^{+0.86}\times10^{-3}$ Mpc$^{-3}$.  We note
that the best-fit parameters are $M_{1350,AB}^{*} = -20.31\pm0.20$ and
$\phi^{*} = 1.80_{-0.68}^{+0.77}\times10^{-3}$ Mpc$^{-3}$ if we
express them using the cosmological parameters
$(\Omega_{M},\Omega_{\Lambda},h)=(0.24,0.76,0.73$) preferred by the
one year WMAP measurements (Spergel et al. 2003).

\textit{Luminosity/SFR density:} The rest-frame continuum UV
($\sim1350\AA$) luminosity density at $z\sim6$ is
$5.8\pm0.9\times10^{25}\,\textrm{ergs/s/Hz/Mpc}^3$ integrated to
$0.3L_{z=3}^{*}$ and
$1.77\pm0.45\times10^{26}\,\textrm{ergs/s/Hz/Mpc}^3$ integrated to
$0.04L_{z=3}^{*}$ (\S5).  This is $0.52\pm0.08$ times and
$0.82\pm0.21$ times, respectively, the luminosity density at $z\sim3$
(Steidel et al.\ 1999) to comparable faint-end limits.  The large
dispersion in previous results at $z\sim6$ seems at least in part to
have been due to a dependence on the faint-end limit (e.g., compare
the panels in Figure~13).  Adopting the evolution in
the UV-to-total correction factors quoted earlier (and thus dust
content as the reason for the change in the rest-frame UV colors), we
infer a much stronger evolution in the SFR density over the range
$z\sim6$ to 3 than is found in the luminosity density.  Using the
Meurer et al.\ (1999) prescription, we estimate that the SFR density
at $z\sim6$ is only $\sim0.3$ times that at $z\sim3$ (to
$0.04L_{z=3}^{*}$), quite different from the change in the luminosity
density (i.e., $\sim0.82$ times).

\textit{Reionization of the universe:} Assuming an escape fraction of
0.5 and $\HI$ clumping factor of 30, we estimate that a SFR
density of 0.052 $\,\sfrd$ is needed to reionize the universe using
the Madau et al.\ (1999) formulation (\S6.3).  This is to be compared
to the 0.022 $\,\sfrd$ observed to the limit of our probe and the
0.043 $\,\sfrd$ obtained by extrapolating our best-fit LF to zero
luminosity.  Despite being lower than the fiducial SFR densities
required, there are sufficient uncertainties at present (particularly
in the escape fraction, ionizing efficiency, and faint-end slope) that
this factor of 2 difference is not significant.  $z\sim6$ galaxies
seem capable of reionizing the universe (see also Stiavelli et al.\
2004b; Yan \& Windhorst 2004a,b).

\textit{$z\sim7-8$ galaxies}: Projecting the present $i$-dropout LF to
$z\sim6-9$, we make an estimate of the number of $z_{850}$-dropouts
that would have been found in a number of recent work (Yan \&
Windhorst 2004b; Bouwens et al.\ 2004c).  We estimate 0.8
$z_{850}$-dropouts to $J_{110,AB}\sim26.6$ versus the one found by Yan
\& Windhorst (2004) and 6.6 $z_{850}$-dropouts to $H_{160,AB}\sim27.5$
versus the four fiducial candidates found in the Bouwens et al.\
(2004c) study.  Despite substantial uncertainties, this suggests that
the rest-frame UV LF only shows a slight change from $z\sim6$ to 7.5
(\S6.4).

The HST ACS data from the HUDF, HUDF-Ps and GOODS fields (enhanced by
the extensive supernova search data) enabled us to detect 506 $z\sim
6$ galaxies.  This significant sample has been used to derive a
rest-frame $UV$ luminosity function at $z\sim 6$ that extends 3 mag
below $L^*$ (to 0.04$L^*$), as well as to provide improved constraints
on size and color evolution, clearly establishing that galaxies are
smaller and bluer at earlier times.  The $z\sim 6$ LF demonstrates
that the brightest galaxies are less luminous at $z\sim 6$, i.e., that
luminosity evolution is the dominant characteristic of the evolving
galaxy population between $z\sim 6$ (0.9 Gyr) and $z\sim3$ (2 Gyr).
The broad consistency of these results with the expectations of
hierarchical models is encouraging.  However, it is the quantitative
constraints made possible with current data sets that are really
important.  Like $z\sim3$, $z\sim6$ seems destined to become an
important reference point in our studies of galaxy evolution, marking
the end of the reionization epoch and providing a useful baseline for
theoretical exploration to even earlier times.

\acknowledgements

We are appreciative of the many individuals who contributed to our
cloning software with their thoughts, ideas, or other suggestions.  We
acknowledge useful discussions with Brandon Allgood, Tom Broadhurst,
Andy Bunker, Daniel Eisenstein, Mauro Giavalisco, Akio Inoue, Sangeeta
Malhotra, James Rhoads, Evan Scannapieco, Daniel Schaerer, Rodger
Thompson, and Jason Tumlinson.  We are indebted to Adam Riess for use
of the ACS images from his SNe search program over the two GOODS
fields, allowing us to make a much deeper reduction.  We thank Dan
Magee for helping to install Apsis on our computer systems, Wei Zheng
for checking the infrared fluxes on our $i$-dropout candidates over
the ACS HUDF-Ps, Dan Coe, Roderik Overzier, and Ruben Salvaterra for a
careful reading of this manuscript, and our referees for many helpful
comments.  ACS was developed under NASA contract NAS5-32865, and this
research was supported under NASA grant HST-GO09803.05-A and
NAG5-7697.

\appendix

\section{$V-z$ Color Cut}

While $i-z$ color criterion is quite effective at isolating $z>5.5$
galaxies, a more robust selection is possible using the $V$-band
fluxes.  This takes advantage of our expectation that the $V$-band
fluxes of galaxies at $z>4$ will be highly attenuated by the
Ly$\alpha$ forest.  However, we cannot simply demand that our $z\sim6$
candidates have no detectable ($<2\sigma$) $V$-band flux.  This is
because of the residual transmission at $\lambda\sim912-1216\AA$
(incomplete Gunn-Peterson trough) which allows several of the brighter
$z\sim6$ galaxies in the HUDF to show faint $V$-band detections.
Therefore, we must arrive at some criterion which allows for some
$V$-band flux in our $z\sim6$ selection, but not too much.  The
criterion we settled upon was $(V_{606}-z_{850})_{AB}>2.8$ color cut
(Figure~A1) for our high redshift selection.  Although
ideally this criterion would have provided a clean separation between
the low-redshift interlopers and high redshift objects (i.e.,
$z>5.5$), such a separation isn't entirely possible.  Excluding all
low-redshift objects would require the cut to be 3.5
(Figure~A1), while including the bluest objects to
$z\sim5.5$ would nominally require the cut to be 2.9 (although a
consideration of the photometric scatter, typically $\gtrsim0.4$ mag,
suggests that 2.5 would be better).  Therefore, it was necessary for
us to settle on some compromise between 2.5 and 3.5.  After some
experimentation, we chose 2.8.

We note that interlopers not identified with our
$(V_{606}-z_{850})_{AB}<2.8$ criterion should be identified using the
red-galaxy criterion $(z_{850}-K_s)_{AB}>1.6$ in Appendix D4.1 or
using the photometric scatter experiments (Appendix D4.2) where the
$V-z$ colors of the undegraded galaxies in those experiments are known
very well.

\begin{figure}
\begin{center}\includegraphics[width=3.5in]{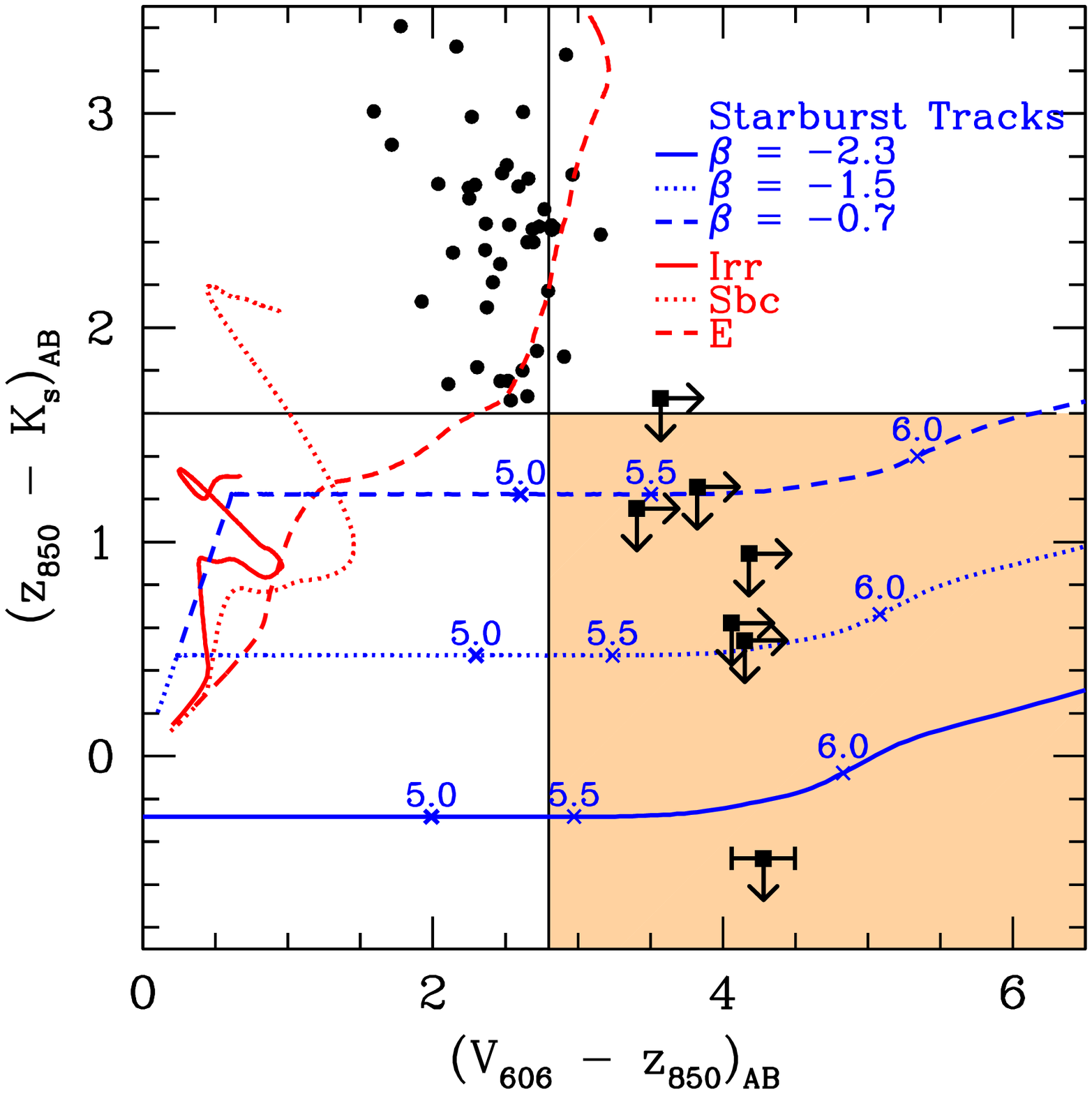}\end{center}
\caption{Motivation for our $(V_{606}-z_{850})_{AB}>2.8$ cut used for
selecting $i$-dropouts.  The measured $(V_{606}-z_{850})_{AB}$ /
$(z_{850}-K_s)_{AB}$ colors for spectroscopically confirmed
$i_{775}$-dropouts (\textit{squares}: Dickinson et al.\ 2004; Malhotra
et al.\ 2005) are contrasted with those obtained from a set of lower
redshift interlopers selected in the CDF-S GOODS ISAAC footprint
(\textit{circles}; Table~2).  The latter objects were
selected to have $(i_{775}-z_{850})_{AB}$ colors redder than 1.0 and
$(z_{850}-J)_{AB}$ colors redder than 0.8.  The model colors are shown
for three different UV spectral slopes $\beta$ and three different
low-z interlopers (Coleman et al.\ 1980).  Redshifts are marked on the
diagram alongside the tracks.  The $(V_{606}-z_{850})_{AB}=2.8$ color
cut shown here (\textit{vertical line}) is used to discriminate
against low-$z$ early types and all later spectral types that enter
into our sample (see Appendix A).  Early-types not caught by the
$(V_{606}-z_{850})$ cut will be included in our estimates of the
contamination fraction using the $z_{850}-K_s$ colors of objects from
the CDF South GOODS ISAAC and HUDF NICMOS data (see
Figure~D3, Table~D7, and Appendix D4.1).
Red $(i_{775}-z_{850})_{AB}>1.3$ objects with $(z_{850}-K_s)_{AB}>1.6$
colors (\textit{horizontal line}) are included in this contamination
fraction.}
\end{figure}

\section{Degradation Procedure}

At several points in our analysis, we found it convenient to degrade
our deeper data to some shallower S/N level.  In \S2.3, we used this
procedure to obtain a uniform S/N level across the two GOODS fields,
and in \S3.4, \S3.6, Appendix C, and Appendix D, we used this
procedure to quantify the effect of S/N on object selection and
photometry.

Before discussing our degradation procedure, it is helpful to provide
some background on both our noise models and weight maps, which are
expressed in units of the inverse variance (equal to what they would
be without any correlation in noise).  To determine the noise model
for each of our images, we measure the rms variance in apertures of
different sizes, and then find an rms noise level and noise kernel
that reproduced the observed variation in rms noise as a function of
aperture size.  The best fit noise levels were then used to scale the
weight (inverse variance) maps provided with the HUDF (Beckwith et
al.\ 2006) or obtained from our ``apsis'' software (Blakeslee et al.\
2003).

By comparing the weight maps on our deeper data with that desired for
our degradation experiments, we determined how much noise needed to be
added to each pixel.  We then made simple realizations of this noise
and smoothed these realizations with the appropriate noise kernels to
obtain the correct correlation properties.  Finally, we added this
noise to our deeper data and updated the pixel-by-pixel weights to
reflect the lower S/N levels.

\section{Degradation Experiments}

To assess the completeness, contamination rate, and flux measurements
in our shallower fields relative to our deeper fields, we degraded our
deeper fields (HUDF and HUDF-Ps) to the depths of our shallower fields
(HUDF-Ps and GOODS) in a series of experiments.  These degradations
provide a very natural way of estimating the effect that photometric
scatter has on both our selection and measurement process.
Experiments included degrading the UDF to the depth of the first HUDF
parallel (HUDFP1), degrading the UDF to the depth of the second HUDF
parallel (HUDFP2), degrading the UDF to the depth of the GOODS fields,
degrading HUDFP1 to the depth of the GOODS fields, and degrading HUDFP2
to the depth of the GOODS fields.  Each degradation was repeated 10
times to minimize the dependence on any particular noise realization.
To maximize realism, we ensured that the pixel-by-pixel weight maps of
the degraded images were identical to those of the shallower fields.
This was of particular interest for the HUDF-Ps (\S2.2) because the
depth in these fields varies by $\sim0.4$ mag across the field of
view.  Then, $i$-dropouts were selected using the selection criteria
of our shallower fields.  Our degradation procedure is detailed in
Appendix B.

Objects obviously associated with the diffraction wings of stars or
bright elliptical galaxies were eliminated to mimic the selection
procedure used for the main catalog (where similar spurious sources
were eliminated).  Objects selected by this procedure were divided
into two categories: contaminants and $z\sim6$ objects.  Objects with
$(V_{606}-z_{850})_{AB}$ colors bluer than 2.8 were classified as
contaminants and objects with $(V_{606}-z_{850})_{AB}$ colors redder
than 2.8 were classified as $z\sim6$ objects (see Appendix A).  In a
few cases, where it was clear that the $V$-band photometry was
contaminated by a nearby foreground object, we reclassified what would
otherwise be labeled a contaminant as a $z\sim6$ $i$-dropout.  Despite
some ambiguity regarding the exact split between the two categories,
our results are not expected to depend on the exact split chosen.
More stringent $(V_{606}-z_{850})_{AB}$ cuts will result in a higher
contamination rate for the shallower field, but this will be offset by
a lower selection volume.

\section{Corrections Applied to Our Data}

This section describes the corrections we applied to the surface
densities of $i$-dropouts derived from our shallower data in order to
put them on a similar footing to our deeper HUDF data.  These
corrections compensate for the greater incompleteness levels, flux
biases, and contamination expected to be present in the shallower
data.

We start by looking at what can be said about the completeness levels
and flux biases by degrading the available data.  Although these
issues are best treated with transfer functions (Appendix D3), our
initial analyses here provide some valuable benchmarks that we can use
later to assess the validity of the transfer functions we determine.

\subsection{Completeness Corrections}

A generic consequence of S/N thresholds and standard detection
algorithms is an overall incompleteness at faint magnitudes and large
sizes.  An illustration of this is provided in
Figure~D1 for the three fields under study.  It is
immediately apparent that the distribution of $i$-dropouts in the HUDF,
HUDF-Ps, and GOODS fields do not extend much to the upper-right of the
three 50\% completeness contours shown.  As a result, significant
incompleteness is not expected until at least $z_{850,AB}\sim26.8$ in
the GOODS fields, $z_{850,AB}\sim27.5$ in the HUDF-Ps fields, and
$z_{850,AB}\sim29$ in the HUDF.

\begin{figure}
\begin{center}\includegraphics[width=4.0in]{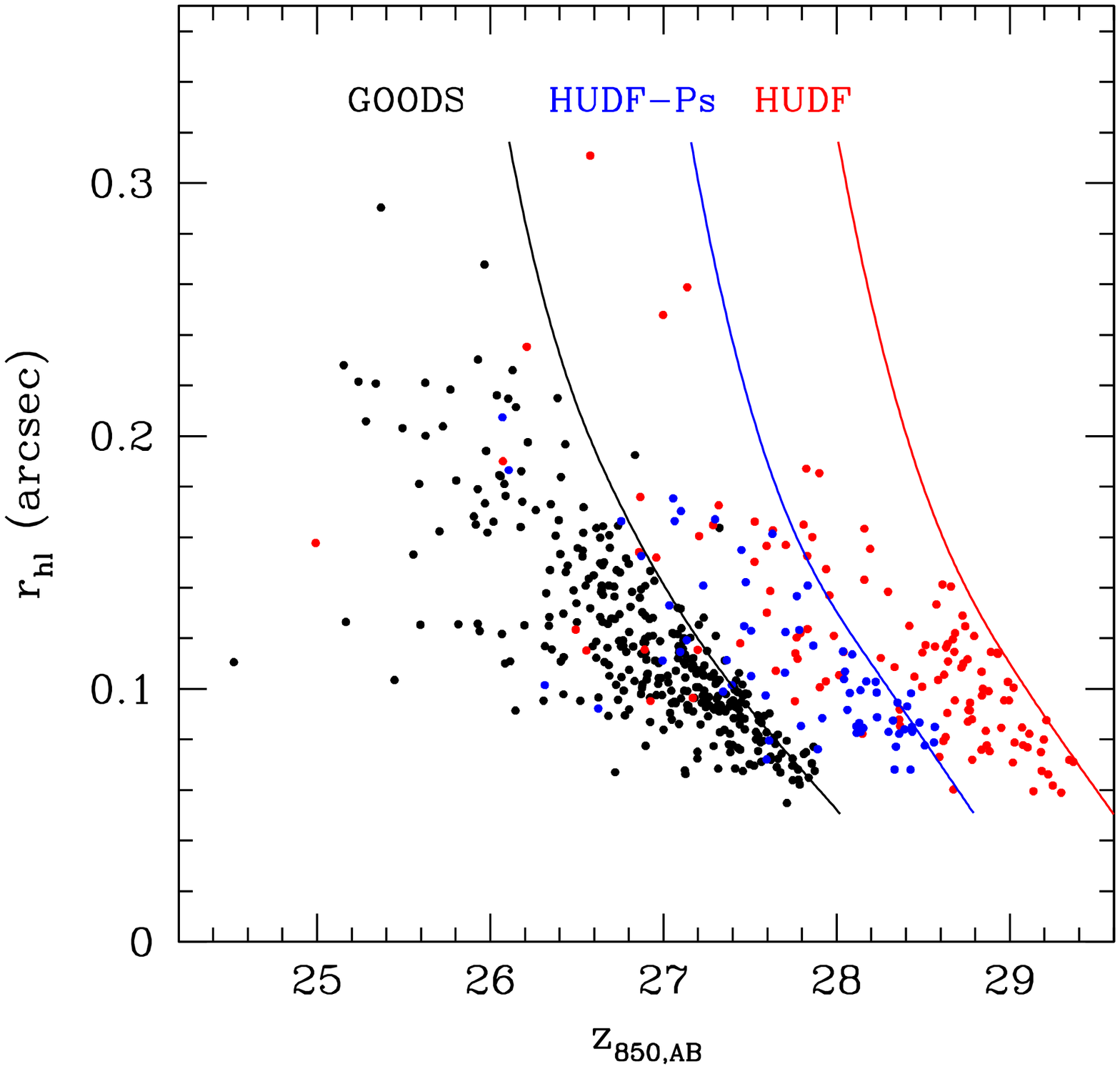}\end{center}
\caption{Size-magnitude diagram for $i_{775}$-dropouts from the HUDF
(\textit{red dots}), HUDF-Ps (\textit{blue dots}), and GOODS fields
(\textit{black dots}).  Size is presented here in terms of the
half-light radius.  The 50\% completeness limits (\textit{solid
lines}) are overplotted for the three fields and assume an $r^{1/4}$
surface brightness profile.  These limits were determined by laying
down galaxies of different sizes and total magnitudes on a noise frame
and then attempting to recover them with our selection procedure
(\S3.1).  By comparing the size-magnitude distribution of objects from
our deeper surveys with our shallower surveys, it is obvious that
significant incompleteness only sets in beyond $z_{850,AB}\sim26.8$ in
the GOODS fields and $z_{850,AB}\sim27.5$ in the HUDF-Ps.  A more
detailed quantification of these biases is provided in the text and
Table~D3 (see also Bouwens et al.\ 2004b).}
\end{figure}

Perhaps the most model-independent way of estimating the
incompleteness in our shallower fields relative to the HUDF is to
degrade the deeper data sets to the same S/N as these shallower fields
and then repeat the selection procedure.  The very similar PSFs, pixel
sizes, and passbands for all data sets considered here make this a
straightforward process.  The deeper data also provide a natural way
of determining the fraction of objects on the degraded frames that are
high redshift objects and the fraction of objects that are likely
contaminants or noise.  These simulations are described in Appendix C.

By comparing the surface density of the sources recovered in the
deeper images with that recovered at the shallower depths (while
excluding those objects whose $V_{606}$-band fluxes indicate they
might be contaminants, Appendix D4.2), we are able to compute the
completeness for the different fields under study.  The results of the
simulations are given in
Tables~D1-D2 and can be put
together to obtain an estimate of the completeness relative to the
HUDF.

\begin{deluxetable}{lrrr}
\tablewidth{300pt}
\tablecolumns{4}
\tabletypesize{\footnotesize}
\tablecaption{Number of $i$-dropouts from the HUDF (11 arcmin$^2$)
recovered at the depths of our shallower two data
sets.\tablenotemark{a}}
\tablehead{
\colhead{Magnitude Interval} & \colhead{GOODS\tablenotemark{b}} &
\colhead{HUDF-Ps\tablenotemark{b}} & \colhead{HUDF}}
\startdata
$24.5<z_{850,AB}<25.0$ & \textbf{1.0} & \textbf{1.0} & \textbf{1} \\
$25.0<z_{850,AB}<25.5$ & \textbf{0.0} & \textbf{0.0} & \textbf{0} \\
$25.5<z_{850,AB}<26.0$ & \textbf{0.0} & \textbf{0.0} & \textbf{0} \\
$26.0<z_{850,AB}<26.5$ & \textbf{2.7} & \textbf{2.9} & \textbf{3} \\
$26.5<z_{850,AB}<27.0$ & 3.1 & \textbf{7.0} & \textbf{7} \\
$27.0<z_{850,AB}<27.5$ & 1.8 & \textbf{7.9} & \textbf{8} \\
$27.5<z_{850,AB}<28.0$ & 0.6 & 13.7 & \textbf{24} \\
$28.0<z_{850,AB}<28.5$ & 0.0 & 4.9 & \textbf{16} \\
$28.5<z_{850,AB}<29.0$ & 0.0 & 2.8 & 46 \\
\enddata

\tablenotetext{a}{The figures here correspond to the number of
$i$-dropouts found in degradations of the HUDF (Appendix C and
Appendix D1) to different depths (or as found in the original data).
The degradation experiments were repeated 10 times, which is why the
quoted values are often non-integer.  This demonstrates how
completeness can depend on depth.  Magnitude intervals in which the
selection is largely complete are shown in bold.}
\tablenotetext{b}{All of the dropouts listed in these columns were
identified as objects in our HUDF catalogs.  This ensures that
differences in the deblending with foreground galaxies do not have a
large effect on these results.}
\end{deluxetable}

\begin{deluxetable}{lrr}
\tablewidth{250pt}
\tablecolumns{3}
\tabletypesize{\footnotesize}
\tablecaption{Number of $i$-dropouts from the HUDF-Ps (17 arcmin$^2$)
recovered at the depths of the GOODS
fields.\tablenotemark{a}}
\tablehead{
\colhead{Magnitude Interval} & \colhead{GOODS\tablenotemark{b}} & \colhead{HUDF-Ps}}
\startdata 
$24.5<z_{850,AB}<25.0$ & \textbf{0.0} & \textbf{0} \\
$25.0<z_{850,AB}<25.5$ & \textbf{0.0} & \textbf{0} \\
$25.5<z_{850,AB}<26.0$ & \textbf{0.0} & \textbf{0} \\
$26.0<z_{850,AB}<26.5$ & \textbf{2.0} & \textbf{3}\\
$26.5<z_{850,AB}<27.0$ & 2.4 & \textbf{3}\\
$27.0<z_{850,AB}<27.5$ & 2.9 & \textbf{15}\\
$27.5<z_{850,AB}<28.0$ & 1.1 & 15\\ 
$28.0<z_{850,AB}<28.5$ & 0.1 & 29\\
$28.5<z_{850,AB}<29.0$ & 0.0 & 3\\
\enddata 
\tablenotetext{a}{The figures here correspond to the number of
$i$-dropouts found in degradations of the HUDF-Ps (Appendix C and
Appendix D1) to the depth of the two GOODS fields (or as found in the
original data).  The degradation experiments were repeated 10 times,
which is why the quoted values are often non-integer.  This
demonstrates how completeness can depend upon depth.  Magnitude
intervals in which the selection is largely complete are shown in
bold.}
\tablenotetext{b}{Each dropout listed in this column was identified as
an object in our HUDF-Ps catalogs.  This ensures that differences in
the deblending with foreground galaxies do not have a large effect on
these results.}
\end{deluxetable}

An application of binomial statistics to the results of
Table~D1 enables a fairly straightforward
determination of the magnitude-dependent completeness of the HUDF-Ps
relative to the HUDF.  While a similar procedure can be used to
calculate the completeness of the GOODS probe relative to the HUDF,
tighter constaints can be obtained by using objects from both the HUDF
(Table~D1) and HUDF-Ps
(Table~D2).  This takes advantage of the fact
that the HUDF-Ps are significantly more complete than the GOODS fields
are.  However, to use the results from the HUDF-Ps, we need to make a
small correction for the small differences in the completeness between
the HUDF and HUDF-Ps selections (based on the results in
Table~D1).  The $1\sigma$ confidence intervals on
the incompleteness of both fields are tabulated in
Table~D3.

\begin{deluxetable}{lcc|cc}
\tablewidth{380pt}
\tablecolumns{4}
\tabletypesize{\footnotesize}
\tablecaption{The Relative Completeness of the shallower data sets to
the HUDF.}
\tablehead{
\colhead{Magnitude Interval} & \colhead{GOODS (OBS)\tablenotemark{a}}
& \colhead{GOODS(SIM)\tablenotemark{b}} & \colhead{HUDF-Ps
(OBS)\tablenotemark{a}} & \colhead{HUDF-Ps (SIM)\tablenotemark{b}}}
\startdata 
$24.5<z_{850,AB}<25.0$ & $>0.56$ & 0.98 & $>0.56$ & 1.00 \\ 
$25.0<z_{850,AB}<25.5$ & --   & 0.92 & -- & 0.99 \\ 
$25.5<z_{850,AB}<26.0$ & --   & 0.86 & -- & 0.98 \\
$26.0<z_{850,AB}<26.5$ & $0.77_{-0.17}^{+0.13}$ & 0.79 & $0.97_{-0.24}^{+0.03}$ & 0.98 \\
$26.5<z_{850,AB}<27.0$ & $0.55_{-0.14}^{+0.14}$ & 0.63 & $>0.87$ & 0.90 \\
$27.0<z_{850,AB}<27.5$ & $0.22_{-0.07}^{+0.10}$ & 0.32 & $0.99_{-0.12}^{+0.01}$ & 0.86 \\
$27.5<z_{850,AB}<28.0$ & $0.04_{-0.02}^{+0.04}$ & 0.05 & $0.57_{-0.10}^{+0.10}$ & 0.61 \\
$28.0<z_{850,AB}<28.5$ & $<0.03$ & 0.00 & $0.31_{-0.10}^{+0.11}$ & 0.28 \\
$28.5<z_{850,AB}<29.0$ & $<0.02$ & 0.00 & $0.06_{-0.03}^{+0.04}$ & 0.04 \\
\enddata 
\tablenotetext{a}{The relative completeness here depends on the numbers
obtained from the degraded data
(Tables~D1-D2).  $1\sigma$
errors are calculated assuming binomial statistics (Appendix D1).
Lower limits are $1\sigma$.}
\tablenotetext{b}{The relative completeness here is based on the
simulations we use to compute the transfer functions (Appendix D3).
Good agreement is observed relative to those extracted from the data,
suggesting that the transfer functions we derive from these
simulations are accurate.}
\end{deluxetable}

Finally, we discuss issues of incompleteness due to blending with
foreground sources (object overlap).  Although not generally
considered to be an important source of incompleteness ($\lesssim
10$\%) for HST studies, here blending played a slightly larger role.
This was due to our choice of blending parameters
(DEBLEND\_MINCONT=0.15), which we adopted to ensure that SExtractor
kept many of the more lumpy dropouts in our sample in a single piece
(see \S3).

To compute the incompleteness from blending, we included $i$-dropouts
from our samples onto the image frames, and then attempted to recover
them with our selection procedure.  We used analytic versions (i.e.,
best-fit exponential profiles) of these dropouts in the simulations to
avoid introducing additional noise onto the image frames.  To control
for possible incompleteness from photometric scatter and surface
brightness selection effects, we also laid down dropouts on empty
frames.  The net increase in incompleteness due to the presence of
foreground objects is approximately 17\%, 10\%, and 8\% for
$i$-dropouts in our HUDF, HUDF-P, and GOODS fields, respectively.  These
numbers appear to be relatively insensitive to the flux of the
source.\footnote{Note that the incompleteness is slightly larger for
our deeper fields than for the shallower fields.  This can be
attributed to our choice of deblending parameters (i.e.,
DEBLEND\_MINCONT=0.15).  For such large values of DEBLEND\_MINCONT,
SExtractor rarely deblends sources.  As a result, objects whose
profiles overlap at or above some minimum surface brightness threshold
will be blended together.  Since lower thresholds are accessed in our
deeper fields, the blending will also be larger there.}

As a basic check on these results and to see how much our
incompleteness determinations were affected by our choice of
deblending parameters, we experimented with a smaller value for the
deblending parameter (DEBLEND\_MINCONT=0.0001) using the HUDF data.
With this choice, we calculated an incompleteness of 11\%, again using
the above procedure.  Since this is smaller than what we calculated
for our fiducial parameters (17\%), we should find more dropouts in
the HUDF with these parameters.  In fact, nine additional $i$-dropouts
(Table~D4) were found.  This increase (from 122 to
131) is almost exactly what we would have expected by comparing the
incompleteness results for the two different values of
DEBLEND\_MINCONT (i.e., 11\% vs. 17\%).

\begin{deluxetable}{ccccccccc}
\tablewidth{5.5in}
\tablecolumns{9}
\tabletypesize{\footnotesize}
\tablecaption{$i$-dropouts in the HUDF which were blended with
foreground sources in our main HUDF catalog (Table~4).
\tablenotemark{a}}
\tablehead{
\colhead{} & \colhead{} & \colhead{} & \colhead{} & \colhead{} & \colhead{} & \colhead{} & \colhead{} & \colhead{$r_{hl}$}\\
\colhead{Object ID} &
\colhead{R.A.} & \colhead{Decl.} &
\colhead{$z_{850}$} & \colhead{$i - z$} & \colhead{$z - J$} & \colhead{$J - H$} & \colhead{S/G} & \colhead{(arcsec)}}
\startdata
HUDF-38397588 & 03:32:38.39 & -27:47:58.8 & 26.94$\pm$0.04 & 1.3 & -0.2 & 0.0 & 0.02 & 0.14 \\
HUDF-36458342 & 03:32:36.45 & -27:48:34.2 & 27.25$\pm$0.06 & 1.3 & $<$-0.1 & faint & 0.01 & 0.17 \\
HUDF-33556441 & 03:32:33.55 & -27:46:44.1 & 27.27$\pm$0.07 & 1.3 & --- & --- & 0.00 & 0.19 \\
HUDF-37278545 & 03:32:37.27 & -27:48:54.5 & 27.48$\pm$0.05 & 3.0 & --- & --- & 0.22 & 0.12 \\
HUDF-42548398 & 03:32:42.54 & -27:48:39.8 & 27.74$\pm$0.08 & 1.5 & --- & --- & 0.01 & 0.15 \\
HUDF-33556440 & 03:32:33.55 & -27:46:44.0 & 27.86$\pm$0.08 & 2.2 & --- & --- & 0.01 & 0.14 \\
\enddata
\tablenotetext{a}{Table~D4 is published in its
entirety in the electronic version of the Astrophysical Journal.  A
portion is shown here for guidance regarding its form and content.
Similar comments to Table~4 apply.  Objects in this
catalog were found (Appendix D1) using a more aggressive splitting
parameter (DEBLEND\_MINCONT=0.0001) than used in the main catalog
(DEBLEND\_MINCONT=0.15: see \S3).  Adding these sources to our main
catalogs would increase the total of $i$-dropouts in the HUDF by
$\sim7$\%.  Units of right ascension are hours, minutes, and seconds,
and units of declination are degrees, arcminutes, and arcseconds.}
\end{deluxetable}

As one final check to test the plausibility of $\sim10$\%
incompletenesses estimated here, we computed the fractional area
covered by sources in the HUDF (from a $V_{606}$-band selected
catalog).  We took the covering area of each object to be equal to 1.5
times the Kron (1980) radii (this closely corresponded with the
apparent visual boundaries of each object).  Summing over all objects
in the HUDF, we obtained a total covering area of 1.4 arcmin$^2$, which
is $\sim13$\% of our total selection area (11.2 arcmin$^2$).  This
estimate is very close to the incompleteness computed above for the
HUDF-Ps and GOODS fields and for the HUDF using our smaller deblending
parameters (DEBLEND\_MINCONT=0.0001).

\subsection{Flux Corrections}

Depth can also have an impact on the measured magnitudes.  This is
particularly true for scalable aperture magnitudes (MAG\_AUTO) as used
by SExtractor, for which both the shape and size of the aperture are
set by the light above some isophote.  Fainter lower surface
brightness objects tend to have significantly smaller isophotal areas,
and this can bias the size of the aperture derived for flux
measurements.  To estimate the extent of this bias, we compared the
$z_{850}$-band magnitudes measured for specific $i$-dropouts in the
HUDF with measurements made on the same objects degraded to GOODS and
HUDF-Ps depths and plotted these differences as a function of magnitude
(Figure~D2).  Again, we considered the results of 10
different degradation experiments in constructing this plot (see
Appendix C or Appendix D1 for a description).  Despite considerable
amounts of scatter, magnitudes measured in the HUDF were found to be
$\sim0.1$ and $\sim0.2$ mag brighter than that measured at HUDF-Ps and
GOODS depth, respectively.  Near the selection limit, there was a
noticeable decrease in the mean flux bias.  This appears to be the
result of a Malmquist-like selection effect (i.e., near the magnitude
limit, brightward-scattering objects make it into our selection while
faintward-scattering objects do not).  We compiled the results of
these experiments into an average offset versus magnitude
(Figure~D2, \textit{red vertical bars}).  The 68\%
confidence limits on these offsets were derived from the
object-to-object scatter.

\begin{figure}
\epsscale{0.8}
\begin{center}\includegraphics[width=4.4in]{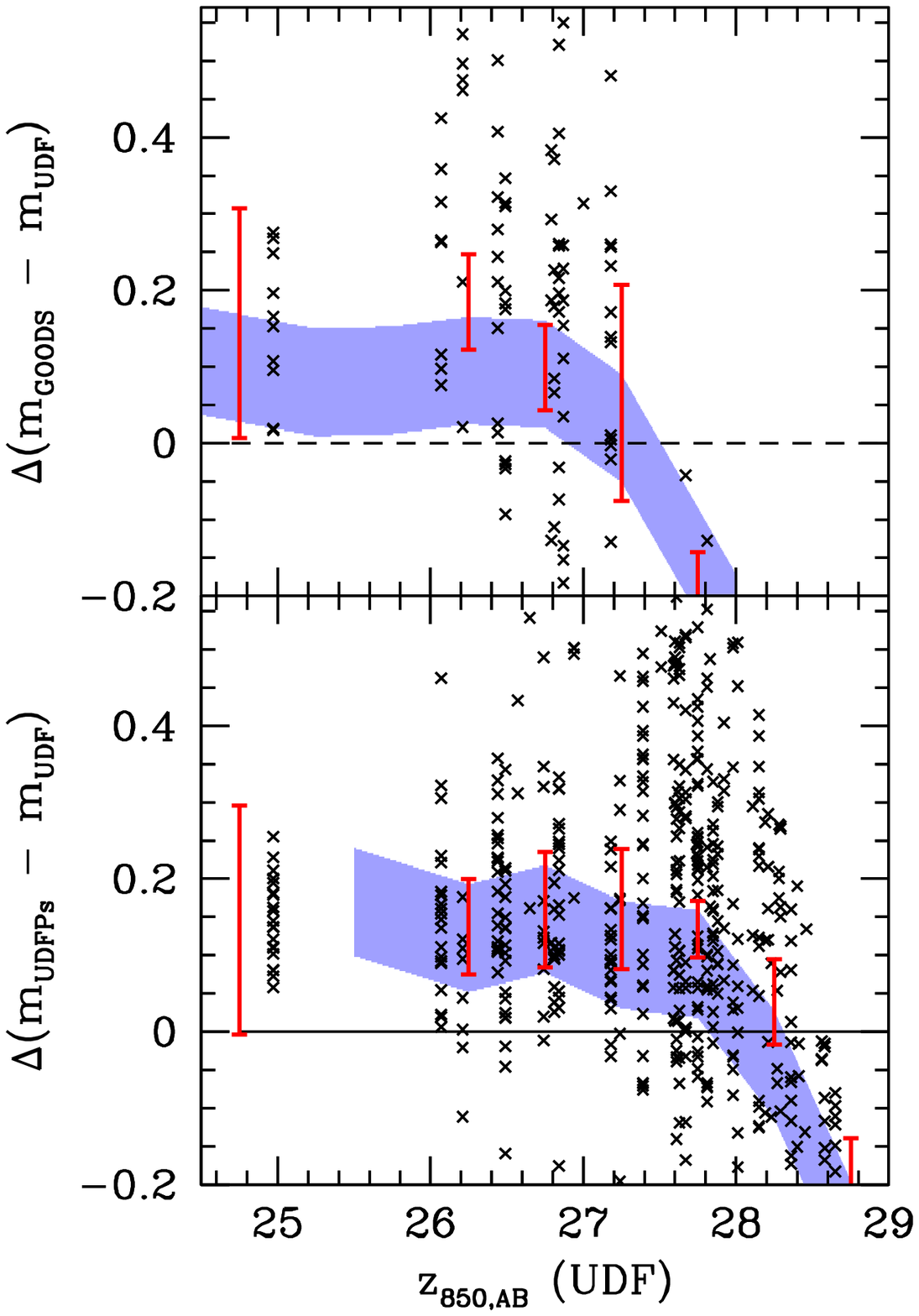}\end{center}
\caption{Differences between the $z_{850,AB}$-band magnitudes measured
for $i$-dropouts in the HUDF and those measured after degrading these
data to the depths of our shallower fields and reselecting these
objects (see Appendix C for a description of these simulations).  The
results of 10 different degradations of the HUDF to GOODS depth are
shown (\textit{top}).  Twenty different degradations of the HUDF to
HUDF-Ps depth are included in the bottom panel (10 for HUDFP1 and 10 for
HUDFP2).  The red vertical bars (positioned at half-magnitude
intervals) denote the 68\% confidence intervals on the mean magnitude
shift (clipped to exclude large $>0.45$ mag deviations).  The blue
shaded region shows the shifts obtained when repeating these
experiments on several simulated fields used to extract our transfer
functions (Appendix D3).  The width of the shaded region ($\pm0.07$
mag) provides suggested error bars on the derived shifts due to
systematic (different deblending effects with nearby objects) and
random errors (limited number of input templates in the simulations).
Although there is some indication, particularly in the top panel, that
the derived shifts from the observations (\textit{red error bars}) are
larger than those obtained from the simulations (\textit{blue shaded
region}), this may simply be an artefact of the objects we use to make
these estimates (three objects from the HUDF were used to derive the
mean shifts for the brightest two magnitude intervals).  The shifts
are similar enough to give us confidence in using the simulations to
determine our transfer functions.}
\end{figure}

\subsection{Transfer Functions}

The completeness and flux corrections detailed in Appendix D1 and D2
can be more properly implemented using transfer functions.  Transfer
functions take surface densities observed at one depth and convert
them to their equivalent densities if measured at another.  In this
formulation, incompleteness is incorporated as a decrease in the
surface density from the input to output stage.  Magnitude biases are
included by effecting a shift from one magnitude interval to another.

Ideally, we would determine the transfer functions in the same way as
we estimate the completeness and flux biases in the previous sections
(e.g., by performing degradation experiments on the real data).
Unfortunately, the available data are simply not sufficient to
adequately determine these functions.  Without a large number of input
objects, the computed transfer functions would be overly dependent on
the position of particular objects within the different magnitude bins
(and their morphologies), compromising the overall accuracy of the
simulations.  This is particularly true at bright magnitudes
($z_{850,AB}\lesssim26$) where there is only one object in our deeper
fields.

As such, it appeared that our best option was simply to rely on
simulations--again using our cloning software to generate the mock
fields.  The inputs to the simulations consisted of $B$-dropout
samples from both the HUDF-Ps (Bouwens et al.\ 2004b) and the HUDF
(Bouwens et al.\ 2004b).  Our use of $z\sim3.8$ $B$-dropout samples
was motivated by the much higher surface brightness sensitivites
available for $B$-dropouts than for $i$-dropouts in the same data [due
to $(1+z)^{4}$ cosmic surface brightness dimming].  Moreover, objects
from these samples should be fairly similar to the $i$-dropout sample
in both size and morphology, minimizing the importance of different
assumptions regarding their evolution over cosmic time ($\sim$700
Myr).  Objects were projected over the range $z\sim5.2-7.0$ in
accordance with their volume density and then added to the artificial
HUDF frames.  Object sizes were scaled as $(1+z)^{-1.1}$ (for fixed
luminosity) to match the observed scalings (\S3.7).

Our transfer functions were calculated by degrading the above
simulations and then comparing the magnitudes of objects selected on
the original frames (at HUDF depths) with those selected on the
degraded frames.  The transfer functions are initially binned on 0.1
mag scales to form familiar two-dimensional matrices, and then
smoothed along the diagonals (to improve the statistics while
preserving flux biases).  The smoothing length is set so that at least
30 different objects from our simulated images contribute to each
element in this matrix (this is equivalent to a smoothing length of
$\Delta m \sim 0.5$ at $z_{850,AB}<25.5$, but $\Delta m \sim 0.1$ at
$z_{850,AB}\gtrsim26.5$).  After smoothing, the results are rebinned
on 0.5 mag intervals to match the binning for the number counts
(Figure~5).  A tabulation of our two transfer functions
is provided in Tables~D5 and
D6.  They are expressed in such a way that one
can use matrix multiplication procedures to go from surface densities
selected in HUDF-type data to the equivalent surface densities
measured in the shallower data.  Note that since object blending is
not properly included in these simulations (the surface density of
objects is comparably low), we corrected our transfer functions
upwards to account for the greater incompleteness in the HUDF due to
object blending (see Appendix D1).

\begin{deluxetable}{cccccccccccc}
\tablewidth{5.5in}
\tabletypesize{\footnotesize}
\tablecolumns{12}
\tablecaption{Transfer Function to take the surface densities measured
at HUDF depths to their equivalent surface densities at GOODS and HUDF-Ps depths
(Appendix D3).\tablenotemark{a}}
\tablehead{\colhead{} & \multicolumn{10}{c}{HUDF}\\
\colhead{GOODS} & \multicolumn{10}{c}{$z_{850}$ Band (mag)}\\
\colhead{$z_{850}$ band} & \colhead{24.25} & \colhead{24.75} & \colhead{25.25} & \colhead{25.75} & \colhead{26.25} & \colhead{26.75} & \colhead{27.25} & \colhead{27.75} & \colhead{28.25} & \colhead{28.75} & \colhead{29.25} }
\startdata
24.25 & \textbf{0.776}& 0.011& 0.000& 0.000& 0.000& 0.000& 0.000& 0.000& 0.000& 0.000& 0.000\\
24.75 & 0.326& \textbf{0.817}& 0.008& 0.000& 0.000& 0.000& 0.000& 0.000& 0.000& 0.000& 0.000\\
25.25 & 0.000& 0.255& \textbf{0.711}& 0.018& 0.000& 0.000& 0.000& 0.000& 0.000& 0.000& 0.000\\
25.75 & 0.000& 0.000& 0.279& \textbf{0.658}& 0.024& 0.000& 0.000& 0.000& 0.000& 0.000& 0.000\\
26.25 & 0.000& 0.000& 0.017& 0.256& \textbf{0.600}& 0.041& 0.000& 0.000& 0.000& 0.000& 0.000\\
26.75 & 0.000& 0.000& 0.002& 0.018& 0.219& \textbf{0.415}& 0.044& 0.000& 0.000& 0.000& 0.000\\
27.25 & 0.000& 0.000& 0.000& 0.002& 0.030& 0.229& \textbf{0.241}& 0.028& 0.000& 0.000& 0.000\\
27.75 & 0.000& 0.000& 0.000& 0.000& 0.002& 0.014& 0.073& \textbf{0.030}& 0.001& 0.000& 0.000\\
28.25 & 0.000& 0.000& 0.000& 0.000& 0.000& 0.000& 0.000& 0.000& \textbf{0.000}& 0.000& 0.000\\
28.75 & 0.000& 0.000& 0.000& 0.000& 0.000& 0.000& 0.000& 0.000& 0.000& \textbf{0.000}& 0.000\\
29.25 & 0.000& 0.000& 0.000& 0.000& 0.000& 0.000& 0.000& 0.000& 0.000& 0.000& \textbf{0.000}\\
\enddata
\tablenotetext{a}{The diagonal elements are shown in bold.}
\end{deluxetable}

\begin{deluxetable}{cccccccccccc}
\tablewidth{5.5in}
\tabletypesize{\footnotesize}
\tablecolumns{12}
\tablecaption{Transfer Function to take the surface densities measured
at HUDF depths to their equivalent surface densities at HUDF-Ps depths
(Appendix D3).\tablenotemark{a}}
\tablehead{\colhead{} & \multicolumn{10}{c}{HUDF}\\
\colhead{HUDF-Ps} & \multicolumn{10}{c}{$z_{850}$ Band (mag)}\\
\colhead{$z_{850}$ band} & \colhead{24.25} & \colhead{24.75} & \colhead{25.25} & \colhead{25.75} & \colhead{26.25} & \colhead{26.75} & \colhead{27.25} & \colhead{27.75} & \colhead{28.25} & \colhead{28.75} & \colhead{29.25}}
\startdata
24.25 & \textbf{0.899} & 0.006& 0.000& 0.000& 0.000& 0.000& 0.000& 0.000& 0.000& 0.000& 0.000\\
24.75 & 0.180& \textbf{0.922}& 0.008& 0.000& 0.000& 0.000& 0.000& 0.000& 0.000& 0.000& 0.000\\
25.25 & 0.000& 0.153& \textbf{0.862}& 0.013& 0.000& 0.000& 0.000& 0.000& 0.000& 0.000& 0.000\\
25.75 & 0.000& 0.000& 0.202& \textbf{0.800}& 0.013& 0.000& 0.000& 0.000& 0.000& 0.000& 0.000\\
26.25 & 0.000& 0.000& 0.000& 0.248& \textbf{0.755}& 0.022& 0.000& 0.000& 0.000& 0.000& 0.000\\
26.75 & 0.000& 0.000& 0.000& 0.001& 0.272& \textbf{0.644}& 0.035& 0.000& 0.000& 0.000& 0.000\\
27.25 & 0.000& 0.000& 0.000& 0.000& 0.016& 0.277& \textbf{0.633}& 0.040& 0.001& 0.000& 0.000\\
27.75 & 0.000& 0.000& 0.000& 0.000& 0.003& 0.026& 0.247& \textbf{0.408}& 0.055& 0.001& 0.001\\
28.25 & 0.000& 0.000& 0.000& 0.000& 0.000& 0.005& 0.018& 0.208& \textbf{0.223}& 0.028& 0.001\\
28.75 & 0.000& 0.000& 0.000& 0.000& 0.000& 0.001& 0.000& 0.009& 0.021& \textbf{0.014}& 0.000\\
29.25 & 0.000& 0.000& 0.000& 0.000& 0.000& 0.000& 0.000& 0.000& 0.000& 0.000& \textbf{0.000}\\
\enddata
\tablenotetext{a}{The diagonal elements are shown in bold.}
\end{deluxetable}

It is possible to obtain a useful check on the results obtained from
these simulated fields by estimating the completeness levels and flux
biases on these same fields.  Our estimates of these quantities were
computed in a way very similar to how they were computed on the actual data
(i.e., Appendix D1 and D2) to ensure consistency.  The results are
shown in the ``SIM'' columns of Table~D3 and in
Figure~D2 (blue shaded regions) and appear to be in
broad agreement with those obtained from our degradation
experiments.\footnote{Although there is some indication that the flux
biases we derive from the simulations may underestimate those obtained
from the observations (Figure~D2, \textit{top}), this
may simply be an artifact of the objects we use to make these
estimates (only three objects from the HUDF were used to derive the
mean flux biases in the brightest two magnitude bins).  Since possible
systematics are much smaller in size than the uncertainties due to
large-scale structure [i.e., $\sigma(M_{1350}^{*})\sim0.15$; Appendix
E], we ignore this issue for the rest of this analysis.}  This
provides us with confidence in the transfer functions we determine
from the simulations.

\subsection{Contamination Corrections}

In principle, the availability of $B_{435}$ and $V_{606}$-band imaging
provides an effective means of eliminating contaminants directly.
Lower redshift interlopers are expected to be significantly brighter
in the $B_{435}$ and $V_{606}$-bands than genuine high-redshift
objects and therefore our requirement that objects be redder than 2.8
in $V_{606}-z_{850}$ (Figure~A1, Appendix A) should prove
to be an effective means of eliminating such objects.  Unfortunately,
near the magnitude limit of each field, only limited constraints can
be set on the $V_{606}$-band fluxes and therefore it is difficult to
effectively filter out all contaminants.

We can however estimate this contamination statistically, using the
deeper optical and infrared data available for some of our fields.  We
break these contamination estimates into four different components:
(1) contamination from intrinsically red objects, (2) contamination
from photometric scatter, (3) contamination from low mass stars, and
(4) contamination from spurious sources.  Explicit effort is made to
ensure that the contribution from each component is independent (and
thus no contaminant is subtracted twice).

\subsubsection{Contamination from Intrinsically Red Objects} 

A small fraction of low-redshift ($z\sim1-3$) galaxies have colors
that are red enough to satisfy our $(i_{775}-z_{850})_{AB}>1.3$
selection.  Since such objects have very different optical-infrared
colors from bona-fide $z\sim6$ objects (Figures 3 and D3),
we can use the deep infrared data available to make an estimate of the
approximate contamination rate.  We already provided a preliminary
estimate of this contamination rate from the HUDF in \S3.2, but we can
obtain a much better estimate of this contamination rate at bright
magnitudes ($25<z_{850,AB}<27$) using the ISAAC data available over
the CDF-S GOODS field.  Similar to the procedures outlined at the
beginning of \S3, $z_{850}-J$ and $z_{850}-K_{s}$ colors for
$i$-dropouts in CDF-South GOODS were measured by smoothing the
$z_{850}$-band data to the same PSF as in the infrared images and
measuring the flux in an aperture whose diameter was 2 times the FWHM
of the object.  Compiling galaxies from the entire 131 arcmin$^2$
CDF-S ISAAC mosaic, candidate low-$z$ interlopers were identified with
the criteria: $(i_{775}-z_{850})_{AB}>1.3, (z_{850}-K_s)_{AB}>1.6$.
Only two such objects were found (Figure~D3): one at
$z_{850,AB}\sim25.4$ and one at $z_{850,AB}\sim26.0$.  The majority of
objects with $i_{775}-z_{850}>1.3$ colors had $(z_{850}-K_s)_{AB}$
colors bluer than 1.6.  Over the interval $25.0<z_{850,AB}<26.0$, this
works out to $18_{-9}^{+13}$\% contamination rate from intrinsically
red objects and over the interval $26.0<z_{850,AB}<27.0$, the
contamination rate is $\lesssim2$\% ($1\sigma$).  These results are
combined with similar estimates from the HUDF IR data (\S3.2) and
summarized in Table~D7.

\begin{figure}
\begin{center}\includegraphics[width=4.0in]{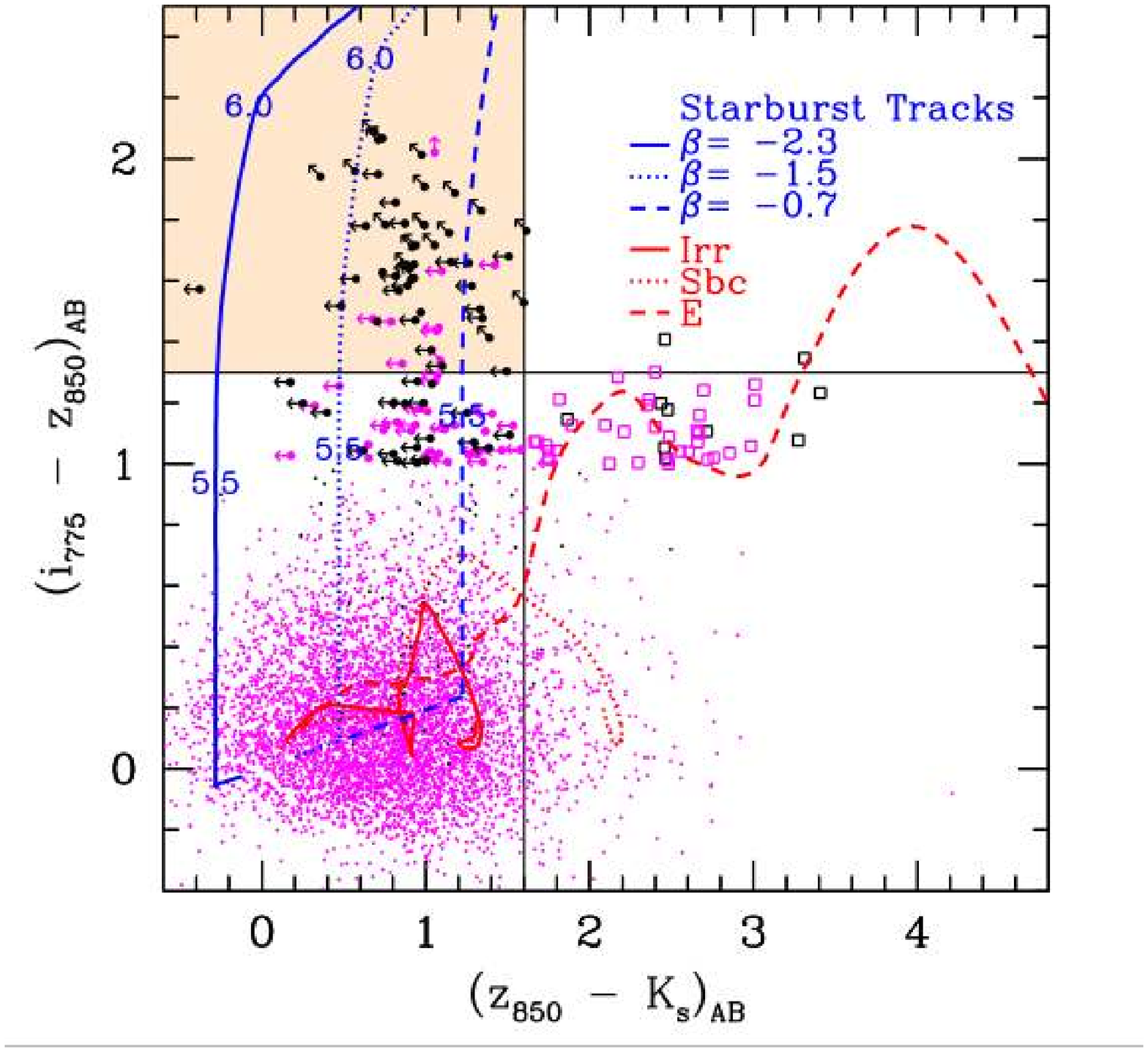}\end{center}
\caption{The $(i_{775}-z_{850})_{AB}$/$(z_{850}-K_s)_{AB}$ colors of
objects in the CDF-S GOODS field (Appendix D4.1) with
$z_{850,AB}<26.8$.  Objects that are undetected ($<2\sigma$) in the
$V_{606}$-band are shown in black while objects which are detected at
the $2\sigma$ level are shown in magenta.  Objects that made it into
our low-redshift interloper selection [Table~2;
$(i_{775}-z_{850})_{AB}>1$ and $(z_{850}-J)_{AB}>0.8$] are shown as
enlarged open squares (see also Figure~A1).  Otherwise
objects are shown as small filled circles.  Color-color tracks of
low-redshift templates and high-redshift starbursts with different
reddenings are as in Figure~2.  Arrows denote $2\sigma$
limits on the $(i_{775}-z_{850})_{AB}$ and $(z_{850}-K_s)_{AB}$
colors.  The solid horizontal line shows our $(i_{775}-z_{850})_{AB}$
cut for selecting $i$-dropouts while the solid vertical line shows our
$(z_{850}-K_s)_{AB}$ cut which serves to separate dropouts from
intrinsically red objects (Figure~A1).  The majority
($\lesssim2$\%) of objects with $i_{775}-z_{850}>1.3$ colors had
$(z_{850}-K_s)_{AB}$ colors bluer than 1.6.  This suggests that
contamination from intrinsically red objects is very small
($\lesssim2$\%: Table~D7).}
\end{figure}

\begin{deluxetable}{ccc}
\tablewidth{350pt}
\tablecolumns{3}
\tabletypesize{\footnotesize}
\tablecaption{The estimated number of intrinsic red contaminants using the CDF-S GOODS + HUDF data}
\tablehead{
\colhead{} & \colhead{GOODS\tablenotemark{a}} & 
\colhead{HUDF\tablenotemark{a}}\\
\colhead{Magnitude Interval} & \colhead{(arcmin$^{-2}$)} & \colhead{(arcmin$^{-2}$)}}
\startdata
$24.0<z_{850,AB}<25.0$ & 0.000 ($<43$\%) & 0.000 ($<43$\%) \\
$25.0<z_{850,AB}<26.0$ & 0.014 ($18_{-9}^{+13}$\%) & -- \\
$26.0<z_{850,AB}<27.0$ & 0.000 ($<2$\%) & 0.000 ($<17$\%) \\
$27.0<z_{850,AB}<28.0$ & -- & 0.000 ($<6$\%) \\
$28.0<z_{850,AB}<29.0$ & -- & 0.767 ($10_{-5}^{+8}$\%) \\
$29.0<z_{850,AB}<29.5$ & -- & -- 
\enddata
\tablenotetext{a}{The number in parentheses
indicates the fraction of $i$-dropout candidates with optical-infrared
colors suggesting that they are intrinsically red low redshift
contaminants (Appendix D4.1, \S3.2).  Uncertainties are $1\sigma$ and
were determined from binomial statistics.}
\end{deluxetable}

\subsubsection{Contamination from Photometric Scatter}

Here we estimate the contamination rate from photometric scatter.  As
with our estimates of the completeness levels and flux biases, perhaps
the most model-independent procedure is to use the results of our
degradation experiments described earlier (Appendix C).  Objects that
are selected as $i$-dropouts can be compared with the original source
catalogs available for the HUDF and HUDF-Ps fields and contaminants
identified.\footnote{Note that objects are only classified as
contaminants if the deeper photometry suggests that their redshifts
are likely well below 5.0, i.e., significantly below our nominal lower
redshift limit of $z\sim5.5$ (Figure~9).  This will
happen for $(V_{606}-z_{850})_{AB}$ colors bluer than 2.8 (Appendix
A).  Having $(i_{775}-z_{850})_{AB}$ colors bluer than 1.3 (in the
deeper photometry) is not sufficient to label an object a contaminant.
This avoids classifying as contaminants objects that are just below
our nominal low redshift limit ($z\sim5.5$: see Figure~A1)
and thus readily scatter into our selection.}  The results of these
simulations are compiled in
Tables~D8-D9 as a function of
magnitude, and again this source of contamination is small
($\lesssim10$\%) and only of significance within $\approx1$ mag of the
faint-end limit.

In our shallower fields, contamination from photometric scatter can
effectively be controlled for using degradations of the HUDF.  However,
the HUDF itself has no deeper field that can serve as a control (which
is an issue faintward of $z_{850,AB}\sim28.5$ where the HUDF
$V_{606}$-band fluxes are no longer of sufficient S/N to filter out
contaminants).  Therefore, we needed an alternative procedure, and so
we elected to model the faint objects in our catalog with the colors
of intermediate magnitude $25.9<z_{850,AB}<27.4$ objects and then add
photometric scatter.  To ensure that the intermediate magnitude
objects were really at low redshift, we required the objects to have
$(i_{775}-z_{850})_{AB}$ colors bluer than 0.9 and
$(V_{606}-z_{850})_{AB}$ colors bluer than 2.5.  These criteria
explicitly excluded objects that were close to qualifying as
$i_{775}$-dropouts (see Figure~D4).  In performing the
simulations, we iterated over all faint $z_{850,AB}>27.9$ objects in
the HUDF (2908 objects), randomly picking an intermediate magnitude
object and then perturbing this object's photometry to match the S/N
of the faint object we were iterating over.  After repeating this
scattering experiment on all faint objects in the HUDF in four separate
trials, we found only one contaminant, or just 0.25 contaminant per 11
arcmin$^2$ field.  This is a smaller fraction than what we found in
our simulations of the HUDF-Ps and GOODS fields
(Tables~D8-D9) and may owe to
the depth of the HUDF $i_{775}$-band imaging.  In our other fields, the
$i_{775}$-band depths only exceeded the $z_{850}$-band depths by
$\sim$0.4 mag, but in the HUDF this difference is 0.7 mag.  Also note
that because at faint magnitudes almost all objects are blue
$(i_{775}-z_{850})\lesssim0.6$, $(V_{606}-z_{850})_{AB}<1.3$, most
objects would still be quite significant detections in the bluer bands
at the limits of the HUDF $i_{775}$-dropout probe
($z_{850,AB}\sim29.5$).

\begin{deluxetable}{lrr}
\tablewidth{300pt}
\tablecolumns{3}
\tabletypesize{\footnotesize}
\tablecaption{The estimated number of contaminants in our GOODS and
HUDF-Ps $i$-dropout samples resulting from photometric
scatter.\tablenotemark{a}}
\tablehead{\colhead{} &
\colhead{GOODS\tablenotemark{b}} & \colhead{HUDF-Ps\tablenotemark{b}}\\
\colhead{Magnitude Interval} & \colhead{(arcmin$^{-2}$)} & \colhead{(arcmin$^{-2}$)}}
\startdata
$24.0<z_{850,AB}<24.5$ & 0.00 & 0.00\\
$24.5<z_{850,AB}<25.0$ & 0.00 & 0.00\\
$25.0<z_{850,AB}<25.5$ & 0.00 & 0.00\\
$25.5<z_{850,AB}<26.0$ & 0.00 & 0.00\\
$26.0<z_{850,AB}<26.5$ & 0.01 & 0.00\\
$26.5<z_{850,AB}<27.0$ & 0.01 & 0.00\\
$27.0<z_{850,AB}<27.5$ & 0.04 & 0.01\\
$27.5<z_{850,AB}<28.0$ & 0.01 & 0.18\\
$28.0<z_{850,AB}<28.5$ & -- & 0.12\\
$28.5<z_{850,AB}<29.0$ & -- & 0.08\\
\enddata
\tablenotetext{a}{Based upon degradations of the HUDF.  These
degradation experiments are described in Appendix C and Appendix D4.2.
They only include the contamination from photometric scatter and do
not include the contamination from intrinsically red objects (\S3.2,
Appendix D4.1: Figure~2, Figure~D3, and
Table~D7).}
\tablenotetext{b}{Errors arise from small number statistics and are
typically half the size of the quoted values.}
\end{deluxetable}

\begin{deluxetable}{lr}
\tablewidth{200pt}
\tablecolumns{2}
\tabletypesize{\footnotesize}
\tablecaption{The estimated number of contaminants in our GOODS
$i$-dropout sample resulting from photometric scatter.\tablenotemark{a}}
\tablehead{\colhead{} & \colhead{GOODS}\\
\colhead{Magnitude Interval} & \colhead{(arcmin$^{-2}$)}}
\startdata
$24.0<z_{850,AB}<24.5$ & 0.00 \\
$24.5<z_{850,AB}<25.0$ & 0.00 \\
$25.0<z_{850,AB}<25.5$ & 0.00 \\
$25.5<z_{850,AB}<26.0$ & 0.00 \\
$26.0<z_{850,AB}<26.5$ & 0.02 \\
$26.5<z_{850,AB}<27.0$ & 0.01 \\
$27.0<z_{850,AB}<27.5$ & 0.01 \\
$27.5<z_{850,AB}<28.0$ & 0.02 \\
\enddata
\tablenotetext{a}{Based upon degradations of the HUDF-Ps.  These
degradation experiments are described in Appendix C and D4.2.  They
only include the contamination from photometric scatter and do not
include the contamination from intrinsically red objects (\S3.2 and
D4.1; Figure~2, Figure~D3, and
Table~D7).}

\end{deluxetable}

\begin{figure}
\epsscale{1.0}
\begin{center}\includegraphics[width=7.0in]{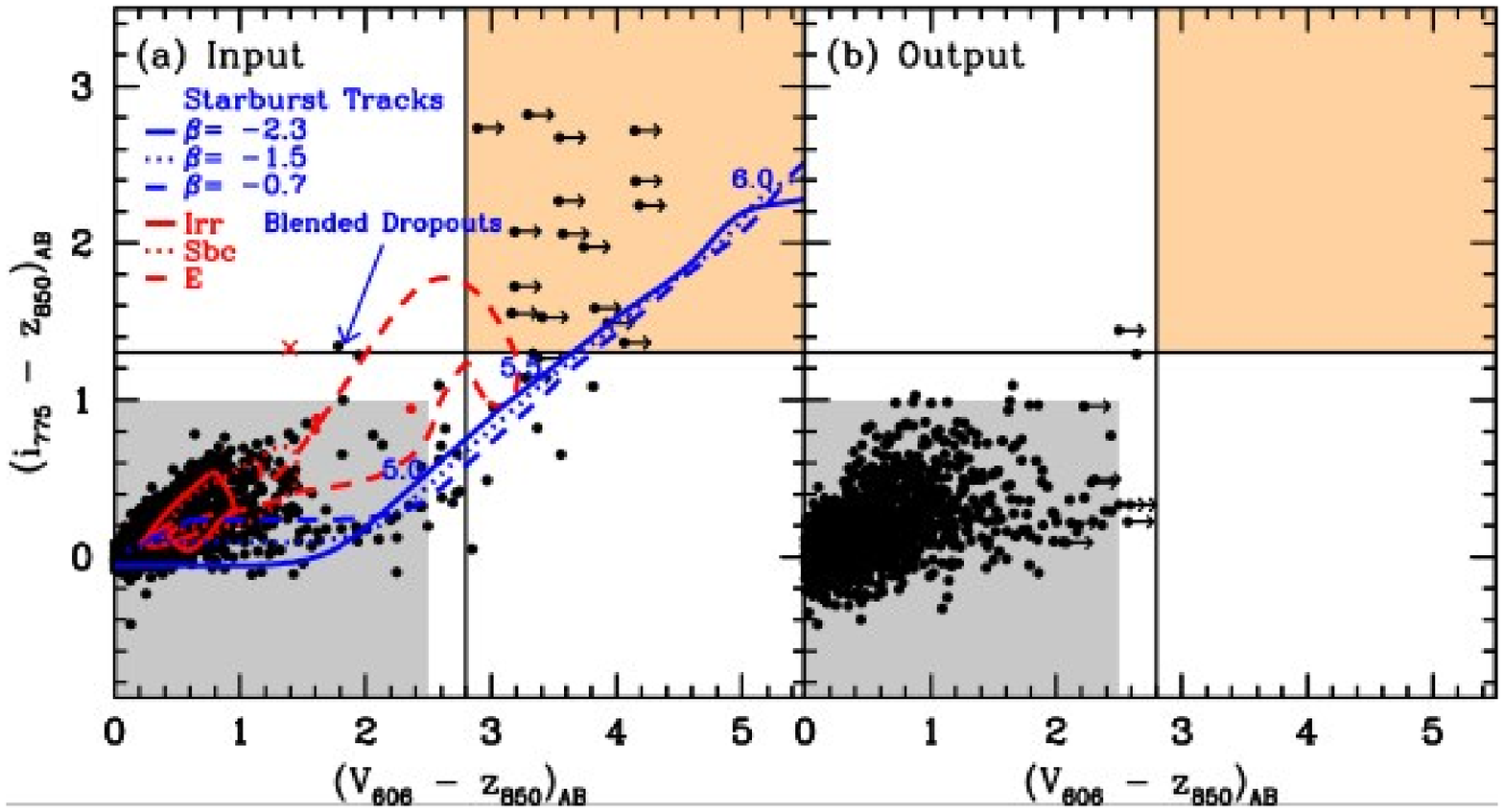}\end{center}
\caption{(\textit{a}) The intermediate magnitude photometric sample
(\textit{gray shaded region in the bottom left-hand corner}) used to
estimate the susceptibility of our faint ($z_{850,AB}>27.9$) HUDF
$i$-dropout sample to contamination from low-redshift interlopers due
to photometric scatter.  Shown are the
$(i_{775}-z_{850})_{AB}/(V_{606}-z_{850})_{AB}$ colors of faint
$25.9<z_{850,AB}<27.4$ objects from the HUDF (\textit{black circles}).
The horizontal and vertical lines show the $(i_{775}-z_{850})_{AB}$
and $(V_{606}-z_{850})_{AB}$ selection cuts used for selecting
$i$-dropouts.  Objects that are particularly red ($\gtrsim0.7$) in
$(z_{850}-J_{110})_{AB}$ and therefore likely low-redshift early-types
are shown as red circles.  Arrorws indicate $2\sigma$ lower limits.
The color-color tracks for low-redshift interlopers are also included
(in red) along with the position of high-redshift starbursts with
various amounts of reddening (in blue).  Objects in the top right-hand
corner (\textit{orange shaded region}) are $i$-dropouts, and objects
in the lower right-hand corner are objects that are likely just below
our redshift cut.  The position of two $i$-dropouts that is partially
blended with foreground objects are indicated by the blue arrow
(Table~D4) while the position of one point-like
star is indicated by the red cross.  \textit{(b)} The
$(i_{775}-z_{850})_{AB}/(V_{606}-z_{850})_{AB}$ colors for objects
from the HUDF input sample (\textit{bottom left-hand corner with gray
shading}) scattered to match the photometric errors of faint
($z_{850,AB}>27.9$) objects in the HUDF.  The output of the
simulations indicates that contamination from photometric scatter at
faint magnitudes is negligible ($<1$ object) (see Appendix
D4.2).}
\end{figure}

\subsubsection{Contamination from Low-mass Stars} 

Low mass stars have similar $(i_{775}-z_{850})_{AB}$ colors to
$z\sim6$ objects, and therefore can act as contaminants to our
samples.  Fortunately, this has not proven to be an important concern,
mostly because the majority of $i$-dropouts ($\gtrsim90\%$) are
clearly resolved at ACS resolution (0.10\arcs$\,$FWHM) and therefore
it is possible to distinguish these objects (which have typical
half-light radii of $\sim0.1-0.2$\arcs) from stellar contaminants.  We
have found that the SExtractor stellarity parameter works particularly
well in this regard, especially for sources with significant
($>10\sigma$) detections in the $z_{850}$ band.  Such S/Ns are
achieved at $z_{850,AB}\lesssim26.8$ for the GOODS fields,
$z_{850,AB}\lesssim27.5$ for the HUDF-Ps, and $z_{850,AB}\lesssim28.4$
for the HUDF.

Unfortunately, beyond these limits, the SExtractor stellarity
parameter no longer gives reliable results -- making it difficult to
use this statistic to identify and remove stellar sources.  So, the
question becomes: how shall we deal with contamination from stars at
such magnitudes?  We think the best approach is a statistical one: (1)
determine the fraction of low-mass stars in $i$-dropout samples as a
function of magnitude using the deeper ACS data and then (2) apply
that contamination fraction to the shallower data.  An estimate of
this contamination fraction can be obtained by examining the data at
all three depths and plotting the fraction of point-like objects as a
function of the $z_{850}$-band magnitude in all three fields.  As is
clear in Figure~D5, there is a monotonic decrease in the
fraction of these objects with magnitude, from $\sim80-100$\% at
bright magnitudes ($z_{850,AB}\sim23-25$) to a mere $\sim$1-2\% at
fainter magnitudes ($z_{850,AB}\sim26-27$).

\begin{figure}
\begin{center}\includegraphics[width=3.4in]{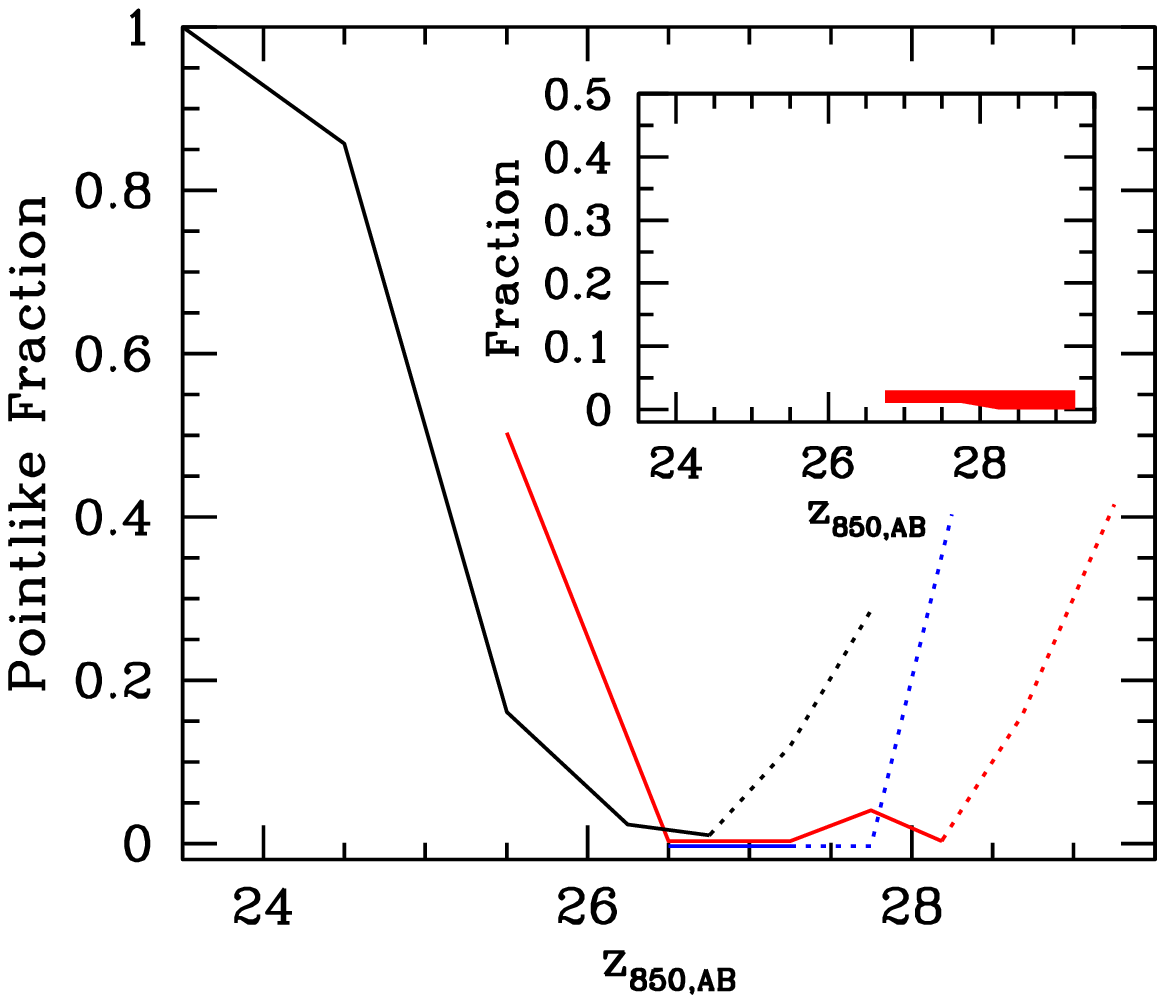}\end{center}
\caption{Fraction of $i$-dropout candidates that are point-like
(SExtractor stellarity $>$0.75, where 0 is an extended object and 1 is
a point source) and thus likely stellar contaminants
vs. $z_{850}$-band magnitude.  The main panel shows the observed
fraction in the HUDF (\textit{red lines}), HUDF-Ps (\textit{blue
lines}), and GOODS fields (\textit{black lines}).  The red and blue
lines are offset slightly from zero for clarity.  The lines become
dotted at the point where the S/N is too low to discriminate between
extended and point-like objects.  \textit{Inset;} The assumed fraction
of stellar contaminants (the shaded red region shows the assumed
$1\sigma$ uncertainties).  Stellar contaminants are rejected using the
measured stellarities brightward of $z_{850,AB}$ equal to 28.4, 27.5,
and 26.8, for the HUDF, HUDF-Ps, and GOODS fields, respectively.
Faintward of this, no such attempt is made and a contamination
fraction is assumed, based on an extrapolation from bright magnitudes
(see Appendix D4.3 for more details).}
\end{figure}

Connecting the surveys up and extrapolating the trends beyond
$z_{850,AB}\sim28.4$, we can arrive at an approximate contamination
fraction as a function of magnitude (Figure~D5,
\textit{inset}).  By multiplying these fractions by the observed
surface densities (Table~11), the contamination rate
from low mass stars can be derived (see Ryan et al. 2005 for
independent estimates).

\subsubsection{Contamination from Spurious Sources}

In principle, our samples were also sensitive to contamination from
spurious objects resulting from noise spikes or other non-Gaussian
features.  If present, such spurious sources would easily qualify as
dropouts given the unlikelihood that similar spikes would occur in the
other passbands.  Therefore, analogous to the simulations described in
B06a, Dickinson et al.\ (2004), and Yan \& Windhorst (2004b), we
repeated our selection procedure on the negative images, and similar
to the results in B06a and Yan \& Windhorst (2004b), no objects were
found in our data sets at all three depths.  Therefore, it seems
unlikely that spurious objects represent a significant source of
contamination for our samples ($\lesssim1$\%).

\subsubsection{Summary}

Table~D10 shows the sum of all three sources of
contamination for the samples considered here (spurious sources do not
appear to be a concern).  Totaling up these results for all three
samples and all magnitude intervals, we can arrive at an approximate
contamination rate for our cumulative sample.  This result is
$\lesssim8$\% (i.e., $\gtrsim92$\% of $i$-dropouts are at $z\sim6$).

\begin{deluxetable}{lrrr}
\tablewidth{300pt}
\tabletypesize{\footnotesize}
\tablecolumns{4}
\tablecaption{Total Contamination Rate (Intrinsically Red + Photometric Scatter + Stars).\tablenotemark{a}}
\tablehead{
\colhead{} & \multicolumn{3}{c}{Field}\\
\colhead{} & \colhead{GOODS} & \colhead{HUDF-Ps} & \colhead{HUDF}\\
\colhead{Magnitude Range} & \colhead{(arcmin$^{-2}$)} & \colhead{(arcmin$^{-2}$)} & \colhead{(arcmin$^{-2}$)}}
\startdata
$24.0<z_{850}<24.5$ & 0.000 & 0.000 & 0.000\\
$24.5<z_{850}<25.0$ & 0.000 & 0.000 & 0.000\\
$25.0<z_{850}<25.5$ & 0.003 & 0.000 & 0.000\\
$25.5<z_{850}<26.0$ & 0.010 & 0.000 & 0.000\\
$26.0<z_{850}<26.5$ & 0.015 & 0.000 & 0.000\\
$26.5<z_{850}<27.0$ & 0.010 & 0.000 & 0.000\\
$27.0<z_{850}<27.5$ & 0.022 & 0.009 & 0.000\\
$27.5<z_{850}<28.0$ & 0.016 & 0.198 & 0.000\\
$28.0<z_{850}<28.5$ & -- & 0.310 & 0.147\\
$28.5<z_{850}<29.0$ & -- & 0.094 & 0.467\\
$29.0<z_{850}<29.5$ & -- & -- & 0.177\\
\enddata 
\tablenotetext{a}{Since the brighter stars ($z_{850}<26.8$, $27.5$,
and $28.4$ for the GOODS fields, HUDF-Ps, and HUDF, respectively) are
explicitly filtered out using the measured stellarities (\S3.1:
Table~3), we assume no contribution to the contamination
rate from stellar objects at these magnitudes.  Faintward of these
limits, the stellar contamination is assumed to be a declining
fraction of the total surface density (Appendix D4.3,
Figure~D5).}
\end{deluxetable}

\section{Uncertainties in the LF due to Field-to-Field
Variations}

In deriving the rest-frame continuum $UV$ luminosity function at
$z\sim6$, we make use of $i$-dropouts from three different fields.
One possibly significant concern is that since the surface density of
$i$-dropouts can show significant differences in normalization from
one field to another (we expect $\sim$35\% rms for a 11.3 arcmin$^2$ ACS
field: \S3.6), these differences may have an effect on our derived LF.
To quantify the size of this effect, we ran a series of Monte-Carlo
simulations.  Using the normalization
$\phi^*=0.00202\,\textrm{Mpc}^{-3}$, faint-end slope $\alpha=-1.73$
from our best-fit LF (\S5), and an ensemble of different
characteristic luminosities $M_{1350,AB}^{*}$ (i.e., $-19.75$,
$-19.85$, $-19.95$, ..., $-20.65$) scattered around our preferred
value of $M_{1350,AB}^{*} = -20.25$ (\S5), we generated number count
predictions for each of our fields (i.e., the HUDF, the HUDF-Ps, and the
GOODS fields).  Our computed counts included the relevant selection
and measurement biases as shown in
Tables~D5-D6 and
Figure~8.  We varied the normalization on our counts
for our deepest two fields (i.e., the HUDF and HUDF-Ps) by 30\% rms (the
approximate uncertainties on the relative normalization of our
different fields), combined the counts from all our fields (\S3.8),
and then fit them to a Schechter function (\S5).  Repeating this
experiment 100 times using different normalizations for our three
fields, we derived rms errors on our three Schechter parameters that
result from the uncertain normalizations.  The rms errors on $\alpha$
were consistently $\sim0.20$ for all input values of $M_{1350}^{*}$,
while the rms errors on $M_{1350}^{*}$ and $\phi^{*}$ increased from
0.10 and 0.00041, respectively, for fainter values of $M_{1350}^{*}$
(i.e., $-19.65$) to 0.17 and 0.00065, respectively, for brighter
values of $M_{1350}^{*}$ (i.e., $-20.65$). This suggests that it is
currently not possible to determine the normalization of the
luminosity function $\phi^{*}$ to better than 30\% and the faint-end
slope $\alpha$ to better than 0.2.


\begin{thebibliography}{}
\bibitem[Adelberger \& Steidel(2000)]{2000ApJ...544..218A} Adelberger, 
K.~L.~\& Steidel, C.~C.\ 2000, \apj, 544, 218
\bibitem[Ajiki et al.(2003)]{2003AJ....126.2091A} Ajiki, M., et al.\ 2003, 
\aj, 126, 2091
\bibitem[Arnouts et al.(2001)]{2001A&A...379..740A} Arnouts, S., et al.\ 
2001, \aap, 379, 740
\bibitem[Arn]{2005arn} Arnouts, S., et al.\  2005, \apjl, 619, L43
\bibitem[Bardeen, Bond, Kaiser, \& Szalay(1986)]{1986ApJ...304...15B} 
Bardeen, J.~M., Bond, J.~R., Kaiser, N., \& Szalay, A.~S.\ 1986, \apj, 304, 
15
\bibitem[Barkana \& Loeb (1999)]{1999ApJ...523...54B} Barkana, R.~\& Loeb, 
A.\ 1999, \apj, 523, 54
\bibitem[Becker et al.(2001)]{2001AJ....122.2850B} Becker, R.~H.~et al.\ 
2001, \aj, 122, 2850
\bibitem[Beckwith et al.(2006)]{2006AJ....132.1729B} Beckwith, S.~V.~W., et 
al.\ 2006, \aj, 132, 1729 
\bibitem[Bertin and Arnouts (1996)]{1996A&AS..117..393B} Bertin, E.\ and 
Arnouts, S.\ 1996, \aaps, 117, 393
\bibitem[Binney(2004)]{2004MNRAS.347.1093B} Binney, J.\ 2004, \mnras, 347, 
1093
\bibitem[Birnboim]{birnboim} Birnboim, Y. \& Dekel, A.  2003, \mnras,
345, 349
\bibitem[Blakeslee et al.(2003)]{2003adass..12..257B} Blakeslee,
J.~P., Anderson, K.~R., Meurer, G.~R., Ben{\'{\i}}tez, N., \& Magee,
D.\ 2003a, ASP Conf.~Ser.~295: Astronomical Data Analysis Software and
Systems XII, 12, 257
\bibitem[Blakeslee et al.(2004)]{2004ApJ...602L...9B} Blakeslee, J.~P., et 
al.\ 2004, \apjl, 602, L9
\bibitem[Bouwens, Broadhurst and Silk (1998)]{1998ApJ...506..557B} Bouwens, 
R., Broadhurst, T.\ and Silk, J.\ 1998a, \apj, 506, 557
\bibitem[Bouwens, Broadhurst and Silk (1998)]{1998ApJ...506..579B}
Bouwens, R., Broadhurst, T.\ and Silk, J.\ 1998b, \apj, 506, 579.
\bibitem[Bouwens, Broadhurst, \&
Illingworth(2003)]{2003ApJ...593..640B} Bouwens, R., Broadhurst, T.,
\& Illingworth, G.\ 2003a, \apj, 593, 640
\bibitem[Bouwens et al.(2003)]{2003ApJ...595..589B} Bouwens, R.~J.~et
al.\ 2003b, \apj, 595, 589
\bibitem[Bouwens et al.(2004)]{2004ApJ...606L..25B} Bouwens, R.~J., et al.\ 
2004a, \apjl, 606, L25
\bibitem[Bouwens et al.\ 2004]{b2004b} Bouwens, R.~J., Illingworth,
G.D., Blakeslee, J.P., Broadhurst, T.J., \& Franx, M.  2004b, \apjl,
611, L1
\bibitem[Bouwens et al.\ 2004]{b2004c} Bouwens, R.~J., et al.\  2004c, \apjl,
616, L79
\bibitem[Bouwens et al.\ 2005]{b2005} Bouwens, R.~J., Illingworth,
G.D., Thompson, R.I., \& Franx, M.  2005, \apj, 624, L5
\bibitem[Bouwens et al.\ 2006]{bbi2006a} Bouwens, R.~J., Illingworth,
G.D., Broadhurst, T.J., Meurer, G.R., Blakeslee, J.P., Franx, M., \&
Ford, H.C.  2006a, \apj, submitted (B06a)
\bibitem[Bouwens et al.\ 2006]{bbi2006b} Bouwens, R.~J., Broadhurst,
T.J., Illingworth, G.D., Meurer, G.R., Blakeslee, J.P., Franx, M., \&
Ford, H.C.  2006b, \apj, submitted (B06b)
\bibitem[Bunker, Stanway, Ellis, \& McMahon(2004)]{2004MNRAS.355..374B} 
Bunker, A.~J., Stanway, E.~R., Ellis, R.~S., \& McMahon, R.~G.\ 2004, 
\mnras, 355, 374
\bibitem[Cimatti et al.(2002)]{2002A&A...381L..68C} Cimatti, A., et al.\ 
2002, \aap, 381, L68
\bibitem[Coe et al. 2006]{coe2006} Coe, D., Ben\'{i}tez, N.,
S\'{a}nchez, S.F., Jee, M., Bouwens, R., Ford, H.  2006, \aj, in press,
astro-ph/0605262
\bibitem[Coleman, Wu, \& Weedman(1980)]{1980ApJS...43..393C} Coleman, G.\ 
D., Wu, C.\ -., \& Weedman, D.\ W.\ 1980, \apjs, 43, 393
\bibitem[Cooray(2005)]{2005MNRAS.364..303C} Cooray, A.\ 2005, \mnras, 364, 
303
\bibitem[Croton et al.\ (2005)]{2005croton} Croton, D., et al.\ 2005,
\mnras, 365, 11
\bibitem[Daddi et al.(2003)]{2003ApJ...588...50D} Daddi, E., et al.\ 2003, 
\apj, 588, 50
\bibitem[Deharveng et al.(2001)]{2001A&A...375..805D} Deharveng, J.-M., 
Buat, V., Le Brun, V., Milliard, B., Kunth, D., Shull, J.~M., \& Gry, C.\ 
2001, \aap, 375, 805
\bibitem[dejong]{dejong} de Jong, R.S., et al.\ 2006, The 2005 HST
Calibration Workshop, 121.
\bibitem[Dickinson et al.(2004)]{2004ApJ...600L..99D} Dickinson, M.~et al.\ 
2004, \apjl, 600, L99
\bibitem[Di Matteo et al.(2005)]{2005Natur.433..604D} Di Matteo, T., 
Springel, V., \& Hernquist, L.\ 2005, \nat, 433, 604
\bibitem[Dow-Hygelund]{dowhygelund} Dow-Hygelund, C.~et al.\ 2006, submitted
\bibitem[Fan et al.(2002)]{2002AJ....123.1247F} Fan, X., Narayanan, V.~K., 
Strauss, M.~A., White, R.~L., Becker, R.~H., Pentericci, L., \& Rix, H.\ 
2002, \aj, 123, 1247
\bibitem[Ferguson et al.(2004)]{2004ApJ...600L.107F} Ferguson, H.~C.~et 
al.\ 2004, \apjl, 600, L107
\bibitem[Fern{\' a}ndez-Soto, Lanzetta, \& Chen(2003)]{2003MNRAS.342.1215F} 
Fern{\' a}ndez-Soto, A., Lanzetta, K.~M., \& Chen, H.-W.\ 2003, \mnras, 
342, 1215
\bibitem[Gabasch]{gabasch} Gabasch, A., et al.\ 2004, \aap, 421, 41.
\bibitem[Giallongo, Cristiani, D'Odorico, \& 
Fontana(2002)]{2002ApJ...568L...9G} Giallongo, E., Cristiani, S., 
D'Odorico, S., \& Fontana, A.\ 2002, \apjl, 568, L9
\bibitem[Geballe et al.(2002)]{2002ApJ...564..466G} Geballe, T.~R., et al.\ 
2002, \apj, 564, 466
\bibitem[Giavalisco et al.(2004)]{2004ApJ...600L..93G} Giavalisco, M., et 
al.\ 2004a, \apjl, 600, L93
\bibitem[Giavalisco et al.(2004)]{2004ApJ...600L.103G} Giavalisco, M., et 
al.\ 2004b, \apjl, 600, L103
\bibitem[Gnedin \& Ostriker(1997)]{1997ApJ...486..581G} Gnedin, N.~Y.~\& 
Ostriker, J.~P.\ 1997, \apj, 486, 581
\bibitem[Granato et al. (2004)]{gran} Granato, G.~L., De Zotti, G.,
Silva, L., Bressan, A., \& Danese, L.\ 2004, \apj, 600, 580
\bibitem[Hu et al.(2004)]{2004AJ....127..563H} Hu, E.~M., Cowie, L.~L., 
Capak, P., McMahon, R.~G., Hayashino, T., \& Komiyama, Y.\ 2004, \aj, 127, 
563
\bibitem[Hurwitz, Jelinsky, \& Dixon(1997)]{1997ApJ...481L..31H} Hurwitz, 
M., Jelinsky, P., \& Dixon, W.~V.~D.\ 1997, \apjl, 481, L31
\bibitem[Inoue et al.(2005)]{2005A&A...435..471I} Inoue, A.~K., Iwata, I., 
Deharveng, J.-M., Buat, V., \& Burgarella, D.\ 2005, \aap, 435, 471
\bibitem[Iwata et al.(2003)]{2003PASJ...55..415I} Iwata, I., Ohta, K., 
Tamura, N., Ando, M., Wada, S., Watanabe, C., Akiyama, M., \& Aoki, K.\ 
2003, \pasj, 55, 415
\bibitem[Knapp et al.(2004)]{2004AJ....127.3553K} Knapp, G.~R., et al.\ 
2004, \aj, 127, 3553 
\bibitem[Kneib, Ellis, Santos, \& Richard(2004)]{2004ApJ...607..697K} 
Kneib, J., Ellis, R.~S., Santos, M.~R., \& Richard, J.\ 2004, \apj, 607, 
697
\bibitem[Kodaira et al.(2003)]{2003PASJ...55L..17K} Kodaira, K., et al.\ 
2003, \pasj, 55, L17
\bibitem[Kron(1980)]{1980ApJS...43..305K} Kron, R.\ G.\ 1980, \apjs, 43, 
305
\bibitem[Leggett et al.(2002)]{2002ApJ...564..452L} Leggett, S.~K., et al.\ 
2002, \apj, 564, 452
\bibitem[Lehmer et al.(2005)]{2005AJ....129....1L} Lehmer, B.~D., et al.\ 
2005, \aj, 129, 1
\bibitem[Lehnert \& Bremer(2003)]{2003ApJ...593..630L} Lehnert, M.~D.~\& 
Bremer, M.\ 2003, \apj, 593, 630
\bibitem[Leitherer, Ferguson, Heckman, \& 
Lowenthal(1995)]{1995ApJ...454L..19L} Leitherer, C., Ferguson, H.~C., 
Heckman, T.~M., \& Lowenthal, J.~D.\ 1995, \apjl, 454, L19
\bibitem[Leitherer et al.(1999)]{1999ApJS..123....3L} Leitherer, C., et 
al.\ 1999, \apjs, 123, 3
\bibitem[Madau et al.\ 1998]{mad98} Madau, P., Pozzetti, L. \&
Dickinson, M. 1998, \apj, 498, 106
\bibitem[Madau, Haardt, \& Rees(1999)]{1999ApJ...514..648M} Madau, P.,
Haardt, F., \& Rees, M.~J.\ 1999, \apj, 514, 648
\bibitem[Malkan, Webb, \& Konopacky(2003)]{2003ApJ...598..878M} Malkan, M., 
Webb, W., \& Konopacky, Q.\ 2003, \apj, 598, 878
\bibitem[Malhotra et al.(2005)]{2005ApJ...626..666M} Malhotra, S., et al.\ 2005, \apj, 626, 666
\bibitem[Martin et al.(2005)]{2005ApJ...619L..59M} Martin, D.~C., et al.\ 2005, \apjl, 619, L59
\bibitem[Mo \& White(1996)]{1996MNRAS.282..347M} Mo, H.~J., \& White, 
S.~D.~M.\ 1996, \mnras, 282, 347
\bibitem[Nagao et al.(2004)]{2004ApJ...613L...9N} Nagao, T., et al.\ 2004, 
\apjl, 613, L9
\bibitem[Night et al.(2006)]{2006MNRAS.366..705N} Night, C., Nagamine, K., 
Springel, V., \& Hernquist, L.\ 2006, \mnras, 366, 705 
\bibitem[Ouchi et al.(2004)]{2004ApJ...611..660O} Ouchi, M., et al.\ 2004, 
\apj, 611, 660
\bibitem[Pirzkal et al.(2004)]{2004ApJS..154..501P} Pirzkal, N., et al.\ 
2004, \apjs, 154, 501
\bibitem[Pirzkal et al.(2005)]{2005ApJ...622..319P} Pirzkal, N., et al.\ 
2005, \apj, 622, 319
\bibitem[Press \& Schechter(1974)]{1974ApJ...187..425P} Press, W.~H.~\& 
Schechter, P.\ 1974, \apj, 187, 425
\bibitem[Reddy \& Steidel(2004)]{2004ApJ...603L..13R} Reddy, N.~A., \& 
Steidel, C.~C.\ 2004, \apjl, 603, L13
\bibitem[Ryan et al.(2005)]{2005ApJ...631L.159R} Ryan, R.~E., Jr., Hathi, 
N.~P., Cohen, S.~H., \& Windhorst, R.~A.\ 2005, \apjl, 631, L159
\bibitem[Sandage, Tammann, \& Yahil(1979)]{1979ApJ...232..352S} Sandage, 
A., Tammann, G.~A., \& Yahil, A.\ 1979, \apj, 232, 352
\bibitem[Sawicki \& Thompson(2005)]{2005ApJ...635..100S} Sawicki, M., \& 
Thompson, D.\ 2005, \apj, 635, 100
\bibitem[Scannapieco \& Oh(2004)]{2004ApJ...608...62S} Scannapieco, E., \& 
Oh, S.~P.\ 2004, \apj, 608, 62
\bibitem[Scannapieco et al.(2005)]{2005ApJ...635L..13S} Scannapieco, E., 
Silk, J., \& Bouwens, R.\ 2005, \apjl, 635, L13
\bibitem[Schaerer \& Pell{\'o}(2005)]{2005MNRAS.362.1054S} Schaerer, D., \& 
Pell{\'o}, R.\ 2005, \mnras, 362, 1054
\bibitem[Schechter(1976)]{1976ApJ...203..297S} Schechter, P.\ 1976, \apj, 
203, 297
\bibitem[Schiminovich et al.(2005)]{2005ApJ...619L..47S} Schiminovich, D., 
et al.\ 2005, \apjl, 619, L47
\bibitem[Schlegel, Finkbeiner, \& Davis(1998)]{1998ApJ...500..525S} 
Schlegel, D.~J., Finkbeiner, D.~P., \& Davis, M.\ 1998, \apj, 500, 525
\bibitem[Sheth \& Tormen(1999)]{1999MNRAS.308..119S} Sheth, R.~K.~\& 
Tormen, G.\ 1999, \mnras, 308, 119
\bibitem[Shimasaku et al.(2005)]{2005PASJ...57..447S} Shimasaku, K., Ouchi, 
M., Furusawa, H., Yoshida, M., Kashikawa, N., \& Okamura, S.\ 2005, \pasj, 
57, 447
\bibitem[Sirianni et al.(2005)]{2005PASP..117.1049S} Sirianni, M., et al.\ 
2005, \pasp, 117, 1049
\bibitem[Skrutskie et al.(1997)]{1997ilsn.proc...25S} Skrutskie, M.~F., et 
al.\ 1997, ASSL Vol.~210: The Impact of Large Scale Near-IR Sky Surveys, 25
\bibitem[Somerville et al.(2004)]{2004ApJ...600L.171S} Somerville, R.~S., 
Lee, K., Ferguson, H.~C., Gardner, J.~P., Moustakas, L.~A., \& Giavalisco, 
M.\ 2004, \apjl, 600, L171
\bibitem[Spergel et al.(2003)]{2003ApJS..148..175S} Spergel, D.~N., et al.\ 
2003, \apjs, 148, 175
\bibitem[Stanway, Bunker, \& McMahon(2003)]{2003MNRAS.342..439S}
Stanway, E.~R., Bunker, A.~J., \& McMahon, R.~G.\ 2003, \mnras, 342,
439
\bibitem[Stanway et al.(2004)]{2004ApJ...604L..13S} Stanway, E.~R., et al.\ 
2004a, \apjl, 604, L13
\bibitem[Stanway et al.(2004)]{2004ApJ...607..704S} Stanway, E.~R., Bunker, 
A.~J., McMahon, R.~G., Ellis, R.~S., Treu, T., \& McCarthy, P.~J.\ 2004b, 
\apj, 607, 704
\bibitem[Stanway et al.(2005)]{2005MNRAS.359.1184S} Stanway, E.~R., 
McMahon, R.~G., \& Bunker, A.~J.\ 2005, \mnras, 359, 1184
\bibitem[Steidel et al.\ (1999)]{1999ApJ...519....1S} Steidel, C.\ C., 
Adelberger, K.\ L., Giavalisco, M., Dickinson, M.\ and Pettini, M.\ 1999, 
\apj, 519, 1
\bibitem[Steidel, Pettini, \& Adelberger(2001)]{2001ApJ...546..665S} 
Steidel, C.~C., Pettini, M., \& Adelberger, K.~L.\ 2001, \apj, 546, 665
\bibitem[Stiavelli, Fall, \& Panagia(2004)]{2004ApJ...600..508S} Stiavelli, 
M., Fall, S.~M., \& Panagia, N.\ 2004a, \apj, 600, 508
\bibitem[Stiavelli, Fall, \& Panagia(2004)]{2004ApJ...610L...1S} Stiavelli, 
M., Fall, S.~M., \& Panagia, N.\ 2004b, \apjl, 610, L1
\bibitem[Thompson et al.(2005)]{2005AJ....130....1T} Thompson, R.~I., et 
al.\ 2005, \aj, 130, 1
\bibitem[Trimble \& Aschwanden(2005)]{2005PASP..117..311T} Trimble, V., \& 
Aschwanden, M.\ 2005, \pasp, 117, 311
\bibitem[Wang \& Heckman(1996)]{1996ApJ...457..645W} Wang, B., \& Heckman, 
T.~M.\ 1996, \apj, 457, 645
\bibitem[White, Becker, Fan, \& Strauss(2003)]{2003AJ....126....1W} White, 
R.~L., Becker, R.~H., Fan, X., \& Strauss, M.~A.\ 2003, \aj, 126, 1
\bibitem[Yan, Windhorst, \& Cohen(2003)]{2003ApJ...585L..93Y} Yan, H., 
Windhorst, R.~A., \& Cohen, S.~H.\ 2003, \apjl, 585, L93
\bibitem[Yan \& Windhorst(2004)]{2004ApJ...600L...1Y} Yan, H.~\& Windhorst, 
R.~A.\ 2004a, \apjl, 600, L1
\bibitem[Yan \& Windhorst(2004)]{2004ApJ...612L..93Y} Yan, H.~\& Windhorst, 
R.~A.\ 2004b, \apjl, 612, L93
\bibitem[Yan et al.(2005)]{2005ApJ...634..109Y} Yan, H., et al.\ 2005, 
\apj, 634, 109
\end{thebibliography}
\end{document}